%% file: prd.tex
\RequirePackage{lineno}
\documentclass[aps,prd,twocolumn,showpacs,superscriptaddress,groupedaddress]{revtex4}  
\usepackage{graphicx}  
\usepackage{dcolumn}   
\usepackage{bm}        
\usepackage{amssymb}   
\usepackage{amsmath}
\usepackage{sidecap}

\input{Commands.tex}

\def\sigmaZ{255.8}
\def\sigmaZErrUP{5.1}
\def\sigmaZErrDN{12.0}

\def\ZZRatVal{0.216}
\def\ZZRatStat{0.058}
\def\ZZRatSyst{0.017}
\def\WZRatVal{0.593}
\def\WZRatStat{0.080}
\def\WZRatSyst{0.017}

\def\ZZtheoryVAL{1.30}
\def\ZZtheoryERR{0.10}
\def\WZtheoryVAL{3.21}
\def\WZtheoryERR{0.19}

\def\ZZValue{1.64}
\def\ZZStat{0.44}
\newcommand{\ZZSystUP}{0.13}
\newcommand{\ZZSystDN}{0.15}

\def\ZZTot{0.46} 

\newcommand{\ZZCombVal}{1.44}
\newcommand{\ZZCombStatUP}{0.31}
\newcommand{\ZZCombStatDN}{0.28}
\newcommand{\ZZCombSystUP}{0.17}
\newcommand{\ZZCombSystDN}{0.19}
\newcommand{\ZZCombTotUP}{0.35}
\newcommand{\ZZCombTotDN}{0.34}

\def\WZValue{4.50}
\def\WZStat{0.61}
\newcommand{\WZSystUP}{0.16}
\newcommand{\WZSystDN}{0.25}
\newcommand{\WZTotUP}{0.63}
\newcommand{\WZTotDN}{0.66}

\def\plotdir{plots/}
\begin{document}


\hspace{5.2in} \mbox{FERMILAB-PUB-12-022-E}

\title{A measurement of the \boldmath{$WZ$} and \boldmath{$ZZ$} production cross sections
using leptonic final states in 8.6 fb\boldmath{$^{-1}$} of \boldmath{$p\bar{p}$} collisions}
\input author_list.tex       
\date{\today}

\begin{abstract}
  We study the processes $p\bar{p}$ $\rightarrow$ \wzlnull\ and 
  $p\bar{p}$ $\rightarrow$ \zzllnunu, where $\ell = e$ or $\mu$.
  Using 8.6~\invfb\ of integrated luminosity collected by the D0 experiment at the Fermilab Tevatron collider,
  we measure the \WZ\ production cross section to be 
  $\WZValue^{+\WZTotUP}_{-\WZTotDN}$~pb which is consistent with, but slightly
  above a prediction of the standard model. 
  The \ZZ\ cross section is measured to be 
  \ZZValue~$\pm$~\ZZTot~pb,
  in agreement with a prediction of the standard model.
  Combination with an earlier analysis of the \zzllll\ channel yields a
  $ZZ$ cross section of $\ZZCombVal^{+\ZZCombTotUP}_{-\ZZCombTotDN}$~pb.

\end{abstract}

\maketitle


\section{\label{Section:Introduction}Introduction}
This Article reports a study of \wzlnull\ and \zznunull\ production 
in $p\bar{p}$ collisions at a center-of-mass energy of $\sqrt{s} = 1.96$ TeV. 
When not stated otherwise,
we denote as $Z$ the full $Z/\gamma^*$ propagator, with the requirement
of $60 < \mass < 120$~GeV on the dilepton invariant mass \mass\ for the
decay \zll.

The pair production of $W$ and $Z$ gauge bosons probes the electroweak component of the standard model (SM) as these cross sections 
are predicted to be significantly larger in the presence
of new resonances or anomalous triple-gauge-couplings~\cite{WZ_aTGCs,ZZ_aTGCs}.
Diboson production is a major source of background in many search channels for Higgs bosons.
For example, $ZZ$ decays correspond to some of the dominant backgrounds 
in searches for $ZH$ production at the Tevatron.
Understanding diboson production is therefore
crucial for demonstrating sensitivity to the presence of a SM Higgs boson at the Tevatron.

Production of \WZ\ pairs was first reported by the CDF Collaboration, in 1.1~\invfb\ of integrated luminosity
in the \lnull\ channel~\cite{cdf_wz_obs}.
Evidence for \wzlnull\ production was also presented by the D0 Collaboration in 1.0~\invfb\ of integrated luminosity~\cite{Dzero_WZ_IIa}.
The D0 Collaboration studied the same process with 4.1~\invfb~\cite{dzero_wz_4.1_plb},
measuring a production cross section of $3.90^{+1.06}_{-0.90}$~pb.
The production of \ZZ\ was first observed by D0
in the \llll\ final state with 2.7~\invfb~\cite{D0_ZZ_obs}.
Combination with an analysis of the \llnunu\ final state with 2.7~\invfb~\cite{dzero_zzllnunu_prd},
increased the significance from 5.3 to 5.7 standard deviations~\cite{D0_ZZ_obs}.
Evidence for \ZZ\ production was also presented by CDF in
1.9~\invfb~\cite{CDF_ZZ_2008} of integrated luminosity.
D0 has recently analyzed 6.4~\invfb\ of integrated luminosity in the \llll\ final state~\cite{dzero_zzllll}.
Combination with the earlier 2.7~\invfb\ analysis~\cite{dzero_zzllnunu_prd} of the 
\llnunu\ final state yielded a $ZZ$ 
production cross section of $1.40^{+0.43}_{-0.37}\mathrm{(stat)} \pm 0.14 \mathrm{(syst)}$~pb~\cite{Note_on_Dzero_ZZ}.
The CDF Collaboration recently measured a cross section
of $1.64^{+0.44}_{-0.38}$~pb for $ZZ$ production using 6~\invfb\
in the \llll\ and \llnunu\ final states~\cite{CDF_ZZ_2011}.
The ATLAS Collaboration has recently studied the \wzlnull\ and \zzllll\ 
processes in $pp$ collisions at $\sqrt{s}$~=~7~TeV 
using 1.1~fb$^{-1}$ of integrated luminosity~\cite{ATLAS_WZ_2011,ATLAS_ZZ_2011}.

Following the approach of the previous D0 analysis~\cite{dzero_zzllnunu_prd} of the \zzllnunu\ signal,
we measure the ratios of signal cross sections relative to the inclusive $Z$ cross section.
This has the advantage of cancelling the uncertainty on the luminosity,
while other systematic uncertainties
related to lepton identification and trigger efficiencies,
are also largely cancelled.

All selection requirements and analysis techniques are optimized
based on Monte Carlo (MC) simulation, or on data in signal-free control regions.
The data are examined in the region expected for signal
only after all selection criteria are defined through simulation.

\section{\label{Section:Apparatus} Apparatus}

The D0 detector~\cite{RunII_NIM,L1CAL,SMT} has a central-tracking system, consisting of a 
silicon microstrip tracker (SMT) and a central fiber tracker (CFT), 
both located within a 1.9~T superconducting solenoidal 
magnet, with designs optimized for tracking and 
vertexing at pseudorapidities~\cite{coordinates}
$|\eta|<3$ and $|\eta|<2.5$, respectively. 
A liquid-argon and uranium calorimeter has a 
central section (CC) covering $|\eta|$ up to 
$\approx 1.1$ and two end calorimeters (EC) that extend coverage 
to $|\eta|\approx 4.2$, with all three housed in separate 
cryostats. 
The inter-cryostat (IC) region (1.1 $<$ $|\eta|$ $<$ 1.5) 
has little electromagnetic (EM) calorimetry,
and reduced hadronic coverage.
Additional scintillating tiles covering the region 1.1 $<$ $|\eta|$ $<$ 1.4
provide improved energy resolution for hadronic jets.
An outer muon system, covering $|\eta|<2$, 
consists of a layer of wire chambers and scintillation trigger 
counters in front of 1.8~T toroidal magnets, followed by two similar layers 
after the toroids.

\section{\label{Section:DataSet} Data and initial event selection}

The data used in this analysis were collected with the D0 detector at the Fermilab
Tevatron $p\bar{p}$ collider at a center-of-mass energy of $\sqrt{s} = 1.96$ TeV.
An integrated luminosity of 8.6 fb$^{-1}$ is available
for analysis, following application of data quality requirements.

Unlike the previous D0 analyses of these channels, 
we do not restrict the offline event selection to events satisfying specific trigger conditions.  
Rather, we analyse all recorded data in order to maximise the event yields.
The analyzed events are mostly recorded by triggers that require one or two 
electrons or muons with high transverse momentum, $p_T$.

Since both signal processes involve the decay \zll,
a natural starting point is to select an inclusive sample of dilepton events.
In addition to the \diem\ and \dimu\ dilepton channels, 
the \ZZ\ analysis makes extensive use of the \emmu\ channel
for verifying modelling of dominant backgrounds (mostly \wwlnulnu).
In all channels we require that there are two oppositely charged leptons
with an invariant mass \mass\ between 60 and 120~GeV.
The regions $40 < \mass < 60$~GeV and $\mass > 120$~GeV are used as control regions.
The two leptons are required to originate from a common vertex that
is  located within $\pm$~80~cm of the detector center along the $z$ axis, defined by the
beam direction.

We define three different qualities for electrons and muons
and refer to electrons or muons satisfying the corresponding selection criteria,
discussed below, as loose, medium, and tight, respectively.
Electrons are treated differently when they are reconstructed in the
CC, EC, and IC regions of the calorimeter.
Loose CC/EC electron candidates are reconstructed using EM energy clusters in the calorimeter that satisfy
minimal shower shape and isolation requirements and that are spatially matched to central tracks.
A boosted decision tree (BDT)~\cite{TMVA_manual} is trained to separate electrons from jets,
based on tracking, isolation, and shower shape information.
Medium and tight CC/EC electrons must satisfy increasingly stringent requirements on the output from this BDT.

In the IC region, there is no dedicated reconstruction of electrons.
However, electrons traversing this region are identified using 
an algorithm that searches for hadronic decays of tau leptons.
A neural network is trained to use
shower shape, isolation, and tracking information to recover electrons from 
reconstructed taus while rejecting hadronic jets.
Electrons found in the IC region must be matched to a central track with at least one hit in the SMT and ten hits in the CFT.
Loose, medium, and tight IC electrons must satisfy increasingly stringent requirements
on the neural network output. 
In addition, medium(tight) IC electrons must satisfy $\isohcfour/p_T$ $<$ 0.2(0.1), 
where \isohcfour\ is the scalar $p_T$ sum of all tracks within an annulus of radius $0.05 < \DR < 0.4$~\cite{DR_ref}
around the candidate electron.
IC electrons are placed into two sub-categories:
type-2(1) IC electrons do (not) contain a cluster of energy in the EM layers of the calorimeter.
For type-2 IC electrons the momentum is determined from the calorimeter energy,
whereas for type-1 IC electrons, we rely on the central track, which provides
a relatively poor momentum resolution.

Loose muons are required to have a track segment that has wire and scintillator hits 
in at least one layer of the muon spectrometer and a spatially matched track in the 
central detector.
Loose muons must also satisfy a calorimeter isolation requirement of $\icalmuon/p_T$ $<$ 0.4,
where \icalmuon\ is the scalar sum of transverse energies of all calorimeter cells 
within an annulus of radius $0.1 < \DR < 0.4$
around the candidate muon.
A track isolation requirement, $\itrkmuon/p_T$ $<$ 0.4, is also imposed on loose muons,
where \itrkmuon\ is the scalar $p_T$ sum of all tracks within a cone of radius $\DR < 0.5$ relative to the candidate muon.
Medium(tight) muons must satisfy $\icalmuon/p_T$ $<$ 0.2(0.1) and $\itrkmuon/p_T$ $<$ 0.2(0.1).

The number of events that pass an inclusive dilepton selection is used as the denominator
for the purposes of 
measuring the ratio of \WZ\ and \ZZ\ cross sections to the $Z$ cross section.
The \WZ\ dilepton selection requires two opposite charge medium quality leptons
of the same flavor.
Hard and soft electron(muon) $p_T$ requirements are defined as $p_T$~$>$~20(15)~GeV and $p_T$~$>$~15(10)~GeV,
respectively.
IC electrons are considered only if they satisfy $p_T$ $>$ 20 GeV.
The \diem\ and \dimu\ channels require that both leptons satisfy the appropriate soft $p_T$ requirement,
and that at least one lepton satisfies the appropriate hard $p_T$ requirement.

\begin{figure*}[htbp]\centering
\includegraphics[width=0.32\linewidth]{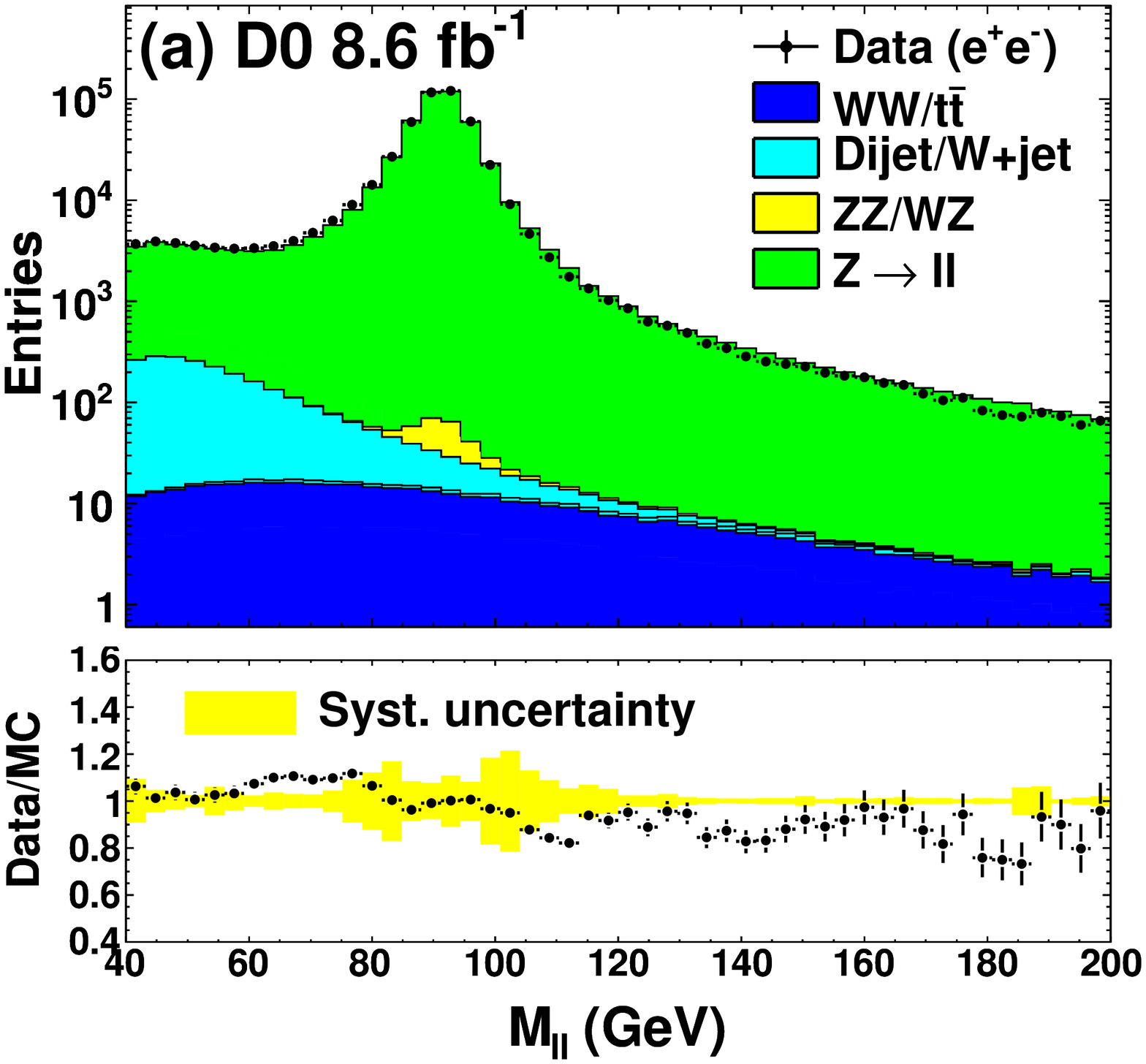}
\includegraphics[width=0.32\linewidth]{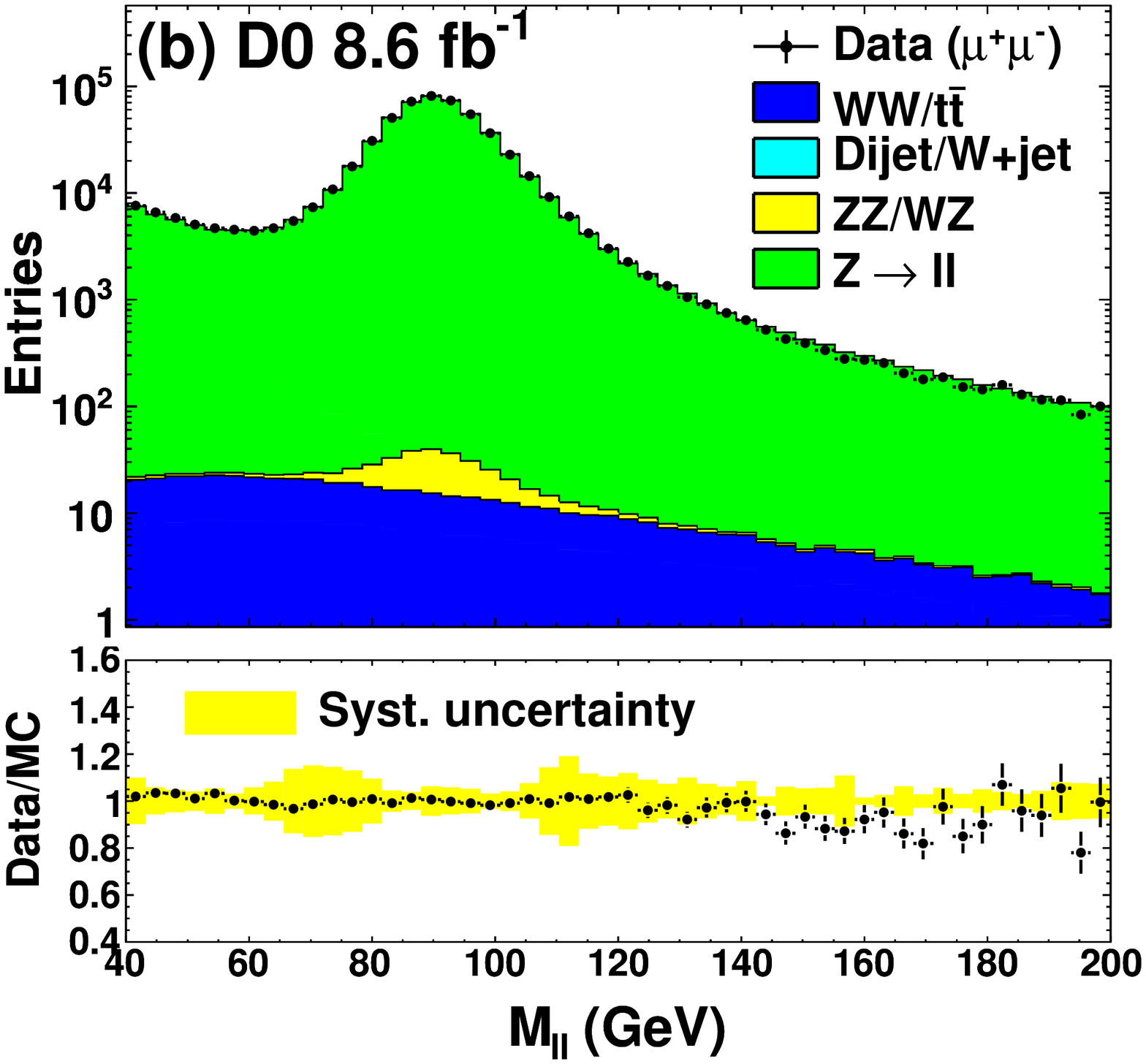}
\includegraphics[width=0.32\linewidth]{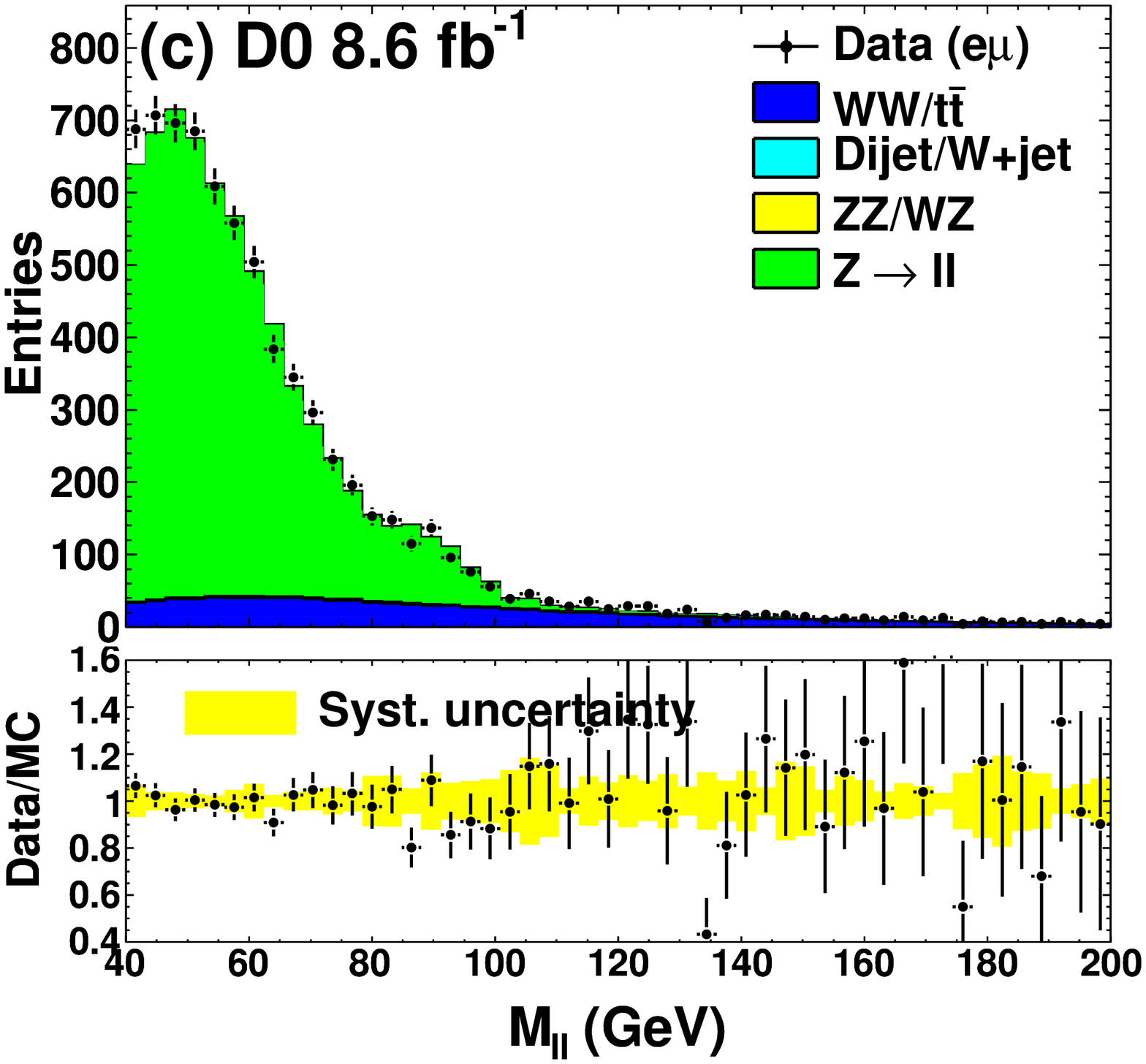}
\caption{Comparison of data and simulation in the \mass\ distribution after selecting an oppositely charged pair of tight quality leptons
in the (a) \diem, (b) \dimu, and (c) \emmu\ channels.
The lower halves of the plots show the ratio of data to simulation, with the 
yellow band representing the systematic uncertainty (see Section~\ref{Section:Systematics}) on the 
simulation.}
\label{Figure:ZZnorm_plots}
\end{figure*}

The \ZZ\ dilepton selection requires tight rather than medium leptons,
and also includes the \emmu\ control channel.
The lepton $p_T$ requirements are the same as in the \WZ\ analysis.
Since a reliable energy/momentum measurement is needed to suppress
background from mis-reconstructed \zll\ events,
type-1 IC electrons are excluded. 
In the \emmu\ channel, electrons are considered only in the CC and EC regions.
Figure~\ref{Figure:ZZnorm_plots} compares the \mass\ distribution in data and 
simulation (described in Section~\ref{Section:bgd_and_signal})
after the \ZZ\ dilepton selection, prior to any additional requirements.
The data are well described by the simulation 
apart from the region of large \mass\ in the dielectron channel
where the simulation yields more events than the data due to a mis-modelling of the
resolution tails.
This is shown not to have a significant effect on the analysis.

The \zll\ selections used as denominators in the cross section ratio measurements
include some further requirements that are detailed in 
Sections~\ref{Section:WZ} and~\ref{Section:ZZ}.

\section{\label{Section:bgd_and_signal} Prediction for background and signal}

Background yields are estimated using a combination of control data samples and MC simulation.
The signal processes and certain backgrounds (\wwlnulnu, \zzllll, \zll, \ttbar, \zgam\ and \wgam)
are estimated based on simulations using the {\sc pythia}~\cite{PYTHIA} event
generator and leading order CTEQ6L1~\cite{CTEQ6L1} parton distribution functions (PDF).
Events are passed through a {\sc geant}~\cite{GEANT} based simulation of the D0 detector response.
In addition, randomly triggered bunch crossings from data
are added to the simulated events to model the effect of additional $p\bar{p}$ collisions.
The {\sc geant} based simulation does not include efficiency of the trigger.
Instead, the efficiencies of certain triggers (single-electron and single-muon) 
are measured using \zll\ candidate events from data.
A parameterization of these efficiencies is applied to the simulated events,
with a correction determined from data for the additional
efficiency that is gained from the remaining triggers
(mostly dilepton and lepton-plus-jet triggers).
The MC simulation does not accurately describe the 
dilepton $p_T$ distribution in $Z$ production.
Weights are therefore assigned to the simulated events as a function
of their generated dilepton $p_T$, based on a comparison with more accurate predictions from {\sc resbos}~\cite{ResBos}.
The diboson events are similarly corrected as a function
of diboson $p_T$ to match predictions from the next-to-leading order (NLO) event generator {\sc powheg}~\cite{powheg,powheg2}.
The simulation of $WZ$ production by {\sc pythia} does not include diagrams with $\gamma^* \rightarrow \ell^+\ell^-$.
Weights are assigned to the generated events based on comparison of the \zll\ invariant mass distribution
with {\sc powheg}, which does include these diagrams.
The simulated events are further corrected for inaccuracies in reconstruction efficiency and
energy/momentum resolution for leptons.
The MC predictions are normalized to match the observed event 
yield after the inclusive dilepton selection,
such that knowledge of the integrated luminosity is not required.

Instrumental backgrounds arise from the mis-reconstruction of hadronic jets 
as isolated electron and muon candidates.
These backgrounds are estimated from data using the so-called ``Matrix Method''~\cite{MatrixMethod_ref}, 
since their rates cannot be modelled sufficiently well by the MC.
We select a sample of events in which a high $p_T$ jet 
satisfies the trigger conditions for the event
and is back-to-back in $\phi$ with an 
electron or muon that satisfies the loose requirements.
We measure the efficiency ($\epsilon_{\rm jet}$) for a jet that is mis-reconstructed as a loose lepton to
also satisfy the medium or tight lepton requirements.
The equivalent efficiency for genuine electrons and muons ($\epsilon_{\rm sig}$)
is measured in a sample of \zll\ candidate events.
The $Z$+jets background in the \wzlnull\ sample is estimated by
selecting a sample of events in which the lepton assigned to the $W\rightarrow \ell\nu$ decay
is of loose rather than tight quality.
Given the estimates of $\epsilon_{\rm sig}$ and $\epsilon_{\rm jet}$,
we solve a set of simultaneous equations to estimate the amount of background in the tight sample.
A sample of $Z$+jets events generated with {\sc pythia} is 
normalized to the estimate from data.
The $W$+jets background in the \zzllnunu\ sample is estimated in a similar way,
with a sample of MC events normalized to an estimate from data.
In the inclusive dilepton sample, 
there is a small background from 
multijet events in the \diem\ channel.
This background is estimated by fitting the observed \mass\ distribution
with the sum of simulated contributions and a multijet template that is obtained by inverting 
the electron quality cuts in real data.

\section{\label{Section:WZ} Selection of WZ candidates}

Four channels are considered for $WZ$ decay: 
$e^+e^-e^{\pm}$, 
$e^+e^-\mu^{\pm}$, 
$\mu^+\mu^-e^{\pm}$, 
$\mu^+\mu^-\mu^{\pm}$.
Events must contain two oppositely charged medium quality leptons
satisfying the $p_T$ requirements described earlier and with \mass\ between 60 and 120 GeV.
The selection of $WZ$ candidates further requires an additional electron (CC or EC) or muon with $p_T$ $>$ 15~GeV, and tight quality.
This lepton must originate from a common vertex with the lepton pair
that is assigned to the \zll\ decay.
If there are three like flavor leptons, there
are two possible combinations of opposite charge leptons.
In this case, the pair with smallest $|\mass-m_{Z}|$, where $m_Z$ is the $Z$ boson mass~\cite{PDG},
is assigned to the \zll\ decay.
Simulation shows that this assignment is correct in 93\% of cases in the three electron channel,
and 87\% of cases in the three muon channel.
In order to suppress the contribution from \zzllll\ decays, no additional 
reconstructed EM clusters are allowed for the $\ell^+\ell^-e^\pm$ selection, 
and no additional reconstructed muons for the $\ell^+\ell^-\mu^\pm$ selection. 
The additional EM clusters must satisfy $E_T$~$>$~5 GeV.
Clusters that are not matched to a central track must satisfy loose shower shape requirements.
There are no explicit $p_T$ or isolation requirements imposed on the additional muons.
Events that satisfy these requirements, excluding the requirement of a third lepton, 
are considered as 
$Z$ candidates, to be used in the denominator when measuring the 
ratio of $WZ$ and $Z$ cross sections.
We choose to include a veto on more than two leptons in the $Z$ selection,
such that uncertainties in the veto efficiency are mostly cancelled in the ratio
of $WZ$ to $Z$ cross sections.

The \WZ\ events are characterised by large missing transverse energy,
\met, defined as the 
magnitude of a vector sum of the
transverse energies of cells in the calorimeter.
The \met\ is corrected for muons, 
which only deposit a small fraction of their energy in the calorimeter,
and for the energy loss of electrons.
The variable \metconstr\ is defined by recalculating the
\met\ after rescaling the transverse momenta of the leptons that are assigned to be $Z$ daughters
within 3 standard deviations of their resolution $\sigma(p_T^{(i)})$, through a fit that minimises the $\chi^2$ function:
\begin{equation}
\chi^2 = \left( \frac{\mass-m_Z}{\Gamma_Z} \right)^2 + \left(\frac{\delta p_T^{(1)}}{\sigma(p_T^{(1)})}\right)^2 + \left(\frac{\delta p_T^{(2)}}{\sigma(p_T^{(2)})}\right)^2, 
\end{equation}
where $\Gamma_Z$ is the total width of the $Z$ boson~\cite{PDG},
and $\delta p_T^{(i)}$ is the amount by which the $p_T$ of lepton $i$ 
is shifted.
This kinematic constraint is only used for the purposes of calculating the variable \metconstr.
The requirement of \metconstr\ $> 20$~GeV is imposed
in order to reject $Z$+jets and $Z\gamma$ backgrounds.
A background to the sub-channels with a $W\rightarrow e\nu$ decay
is the radiation of a high $p_T$ photon from a lepton in
a \zll\ decay. 
We therefore require that $|M_{\ell\ell\ell}-91.2|$ $>$ $|\mass-91.2|$, where $M_{\ell\ell\ell}$ is the invariant
mass of the three leptons.
In addition, at least one of the leptons from the \zll\ decay is required to have $p_T$ $>$ 25 GeV.

\begin{figure*}[htbp]\centering
\includegraphics[width=0.32\linewidth]{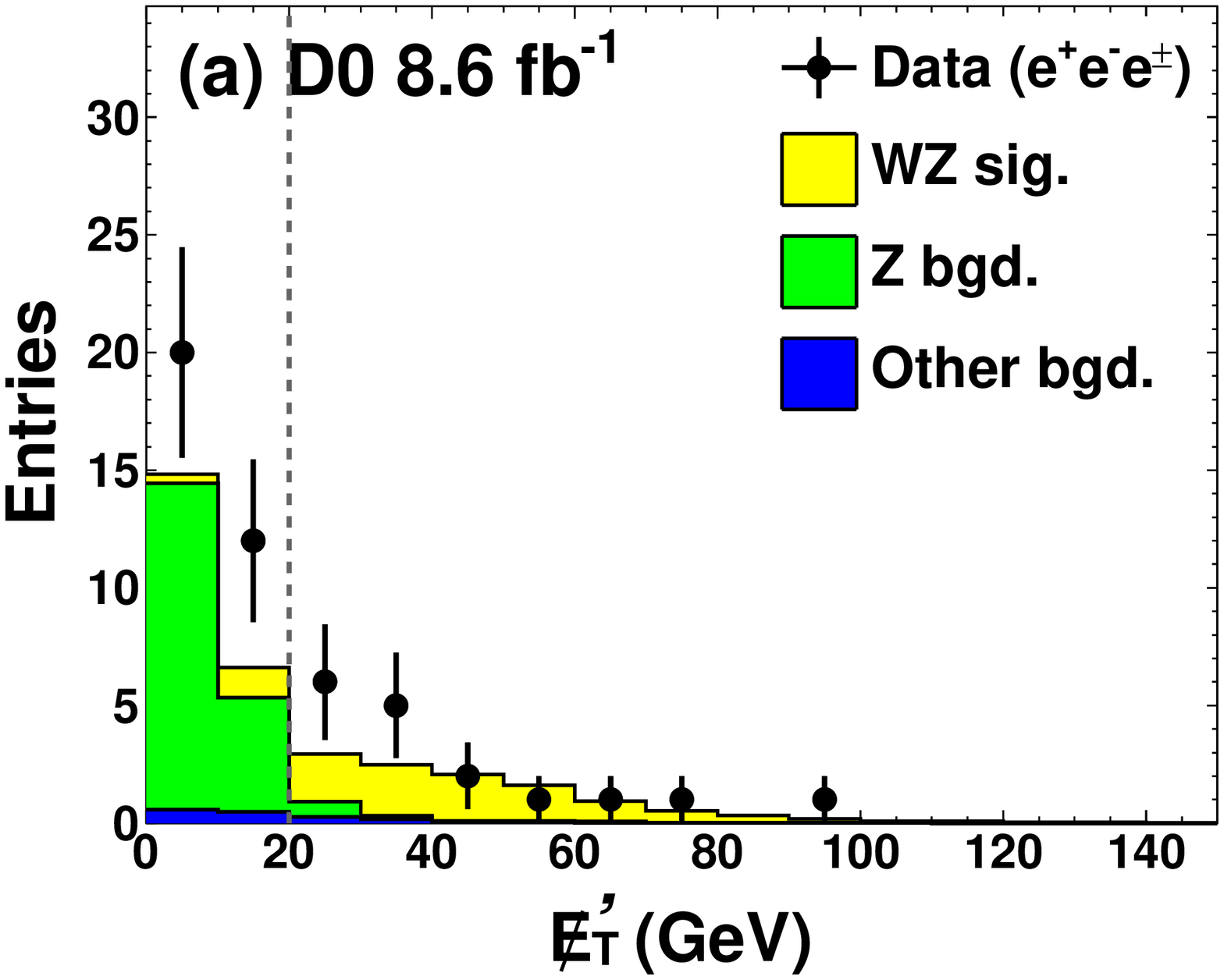}
\includegraphics[width=0.32\linewidth]{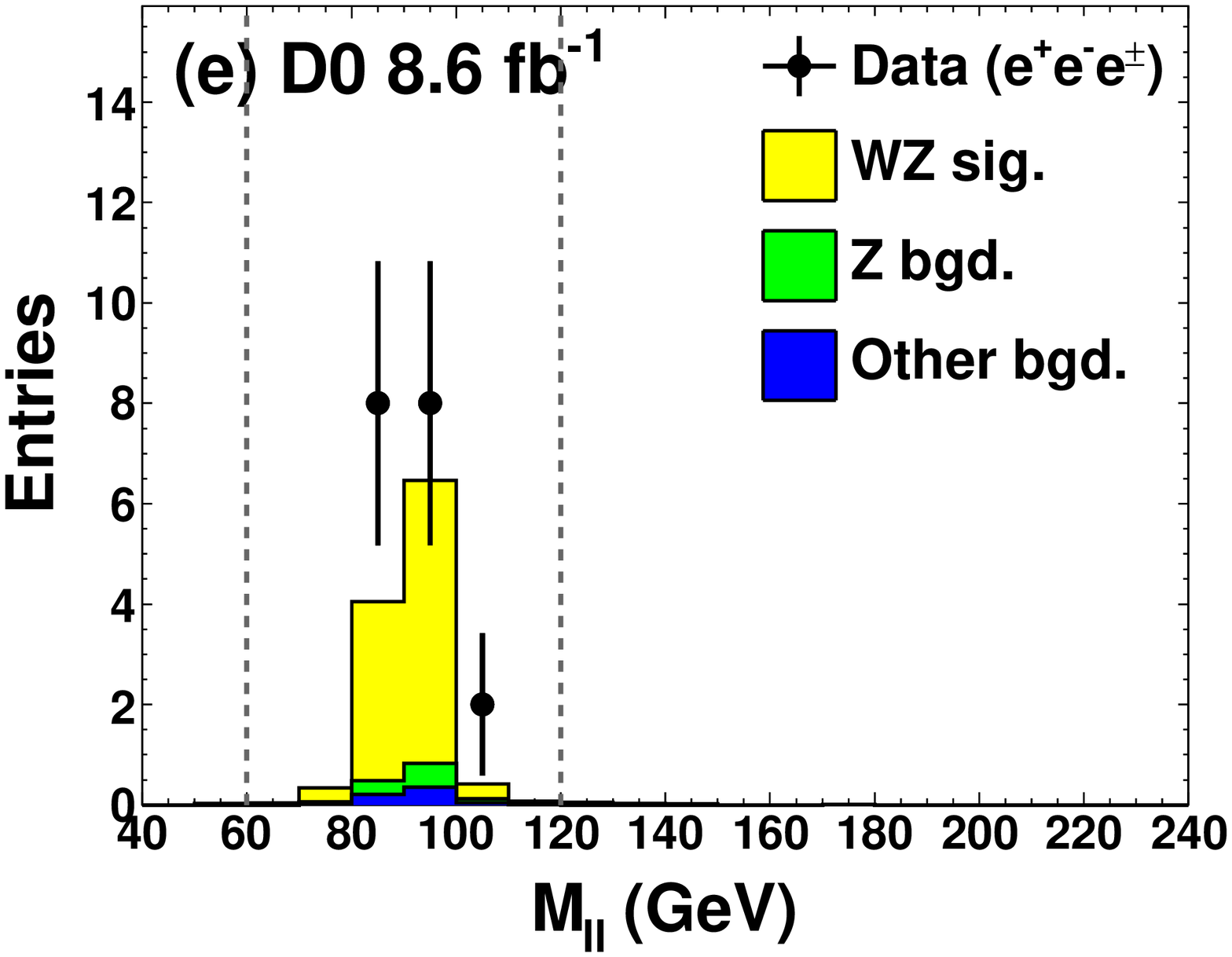}
\includegraphics[width=0.32\linewidth]{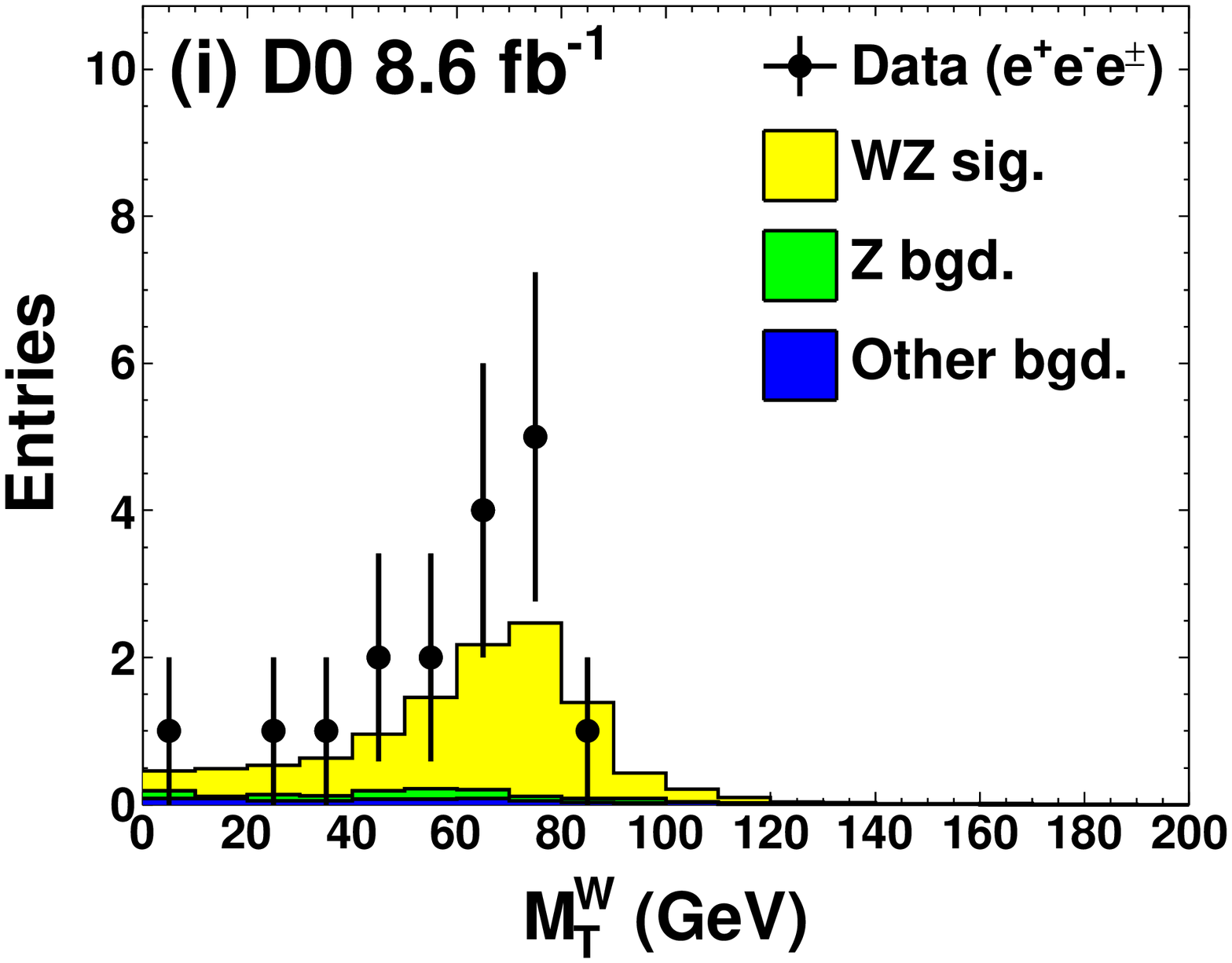}
\includegraphics[width=0.32\linewidth]{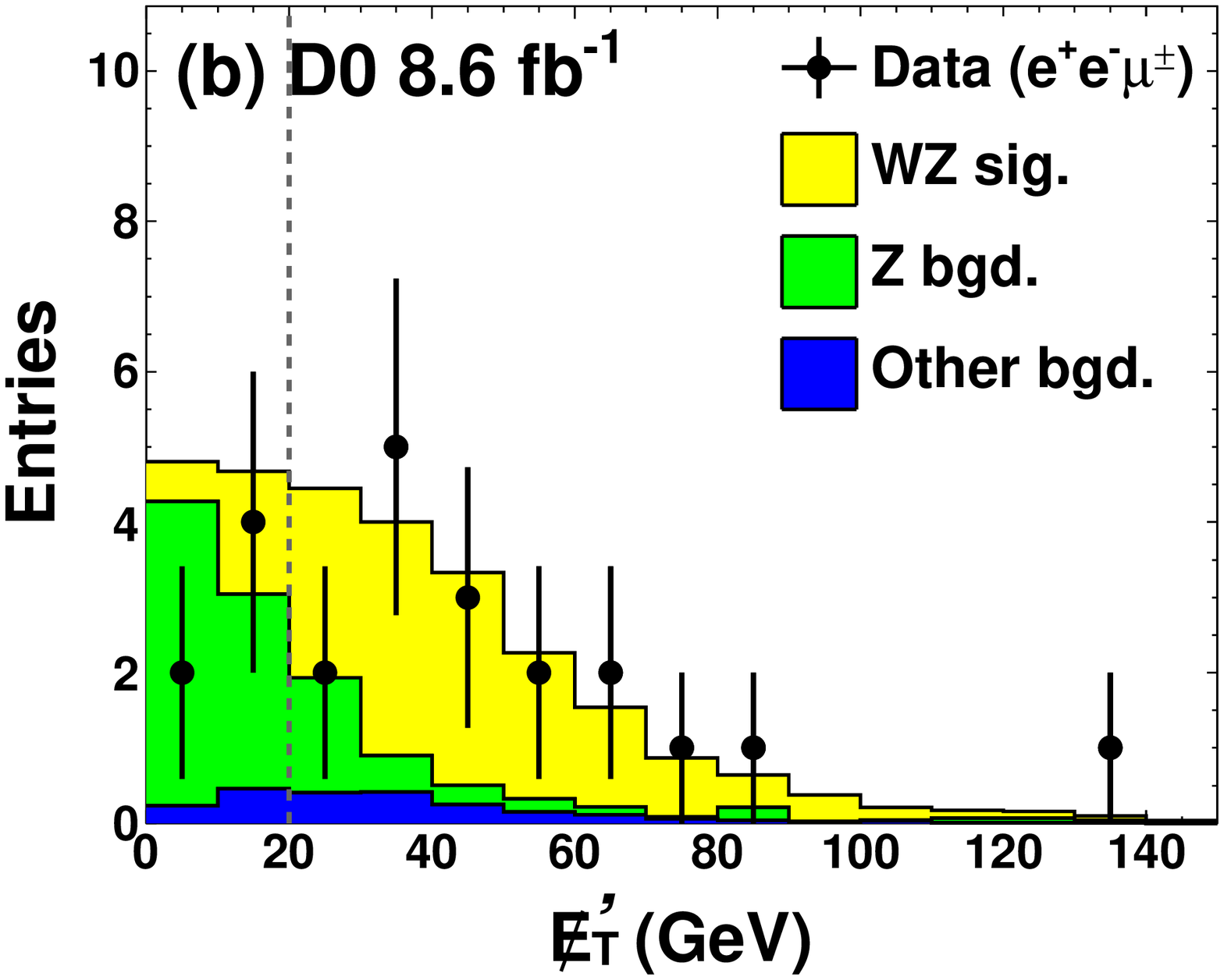}
\includegraphics[width=0.32\linewidth]{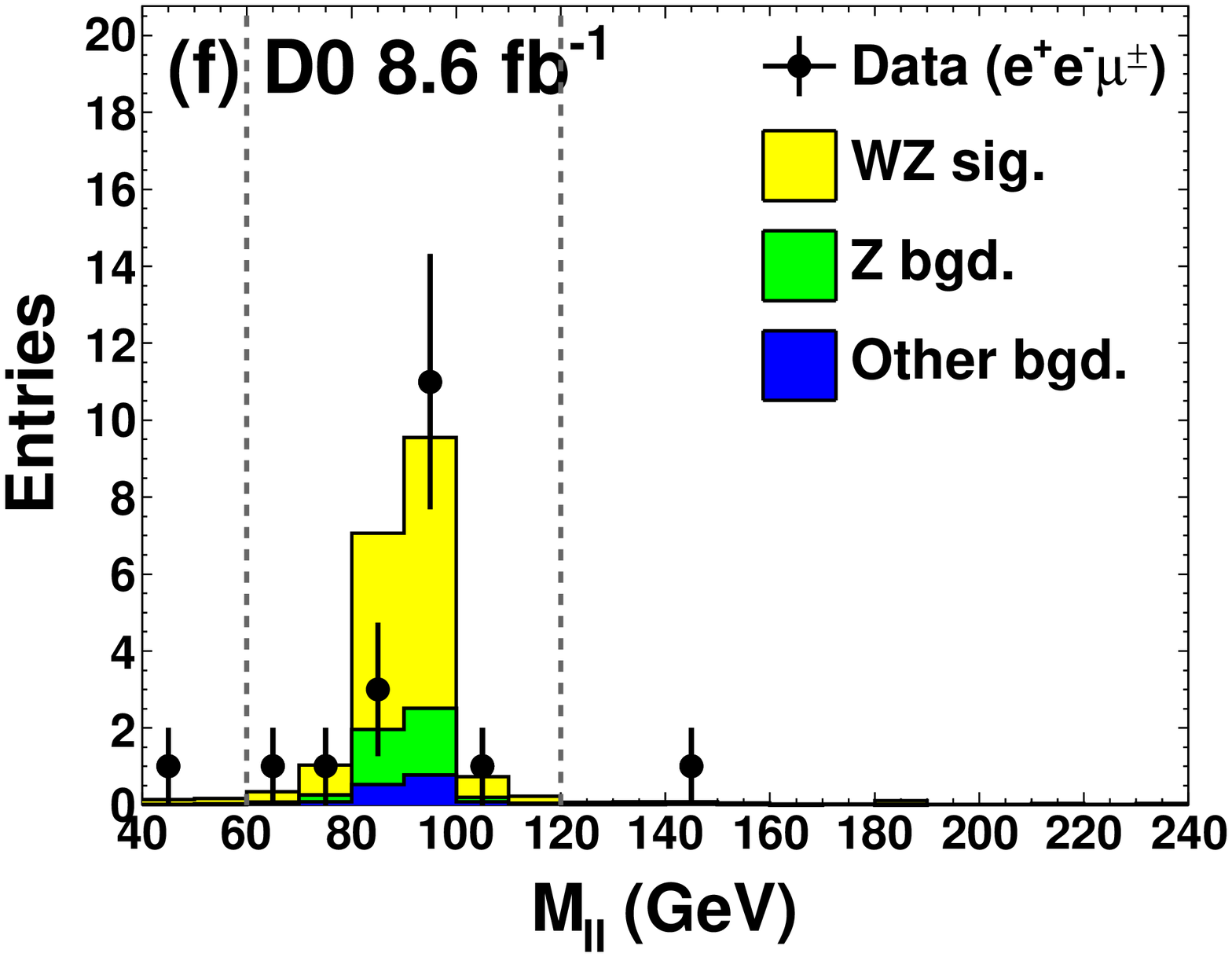}
\includegraphics[width=0.32\linewidth]{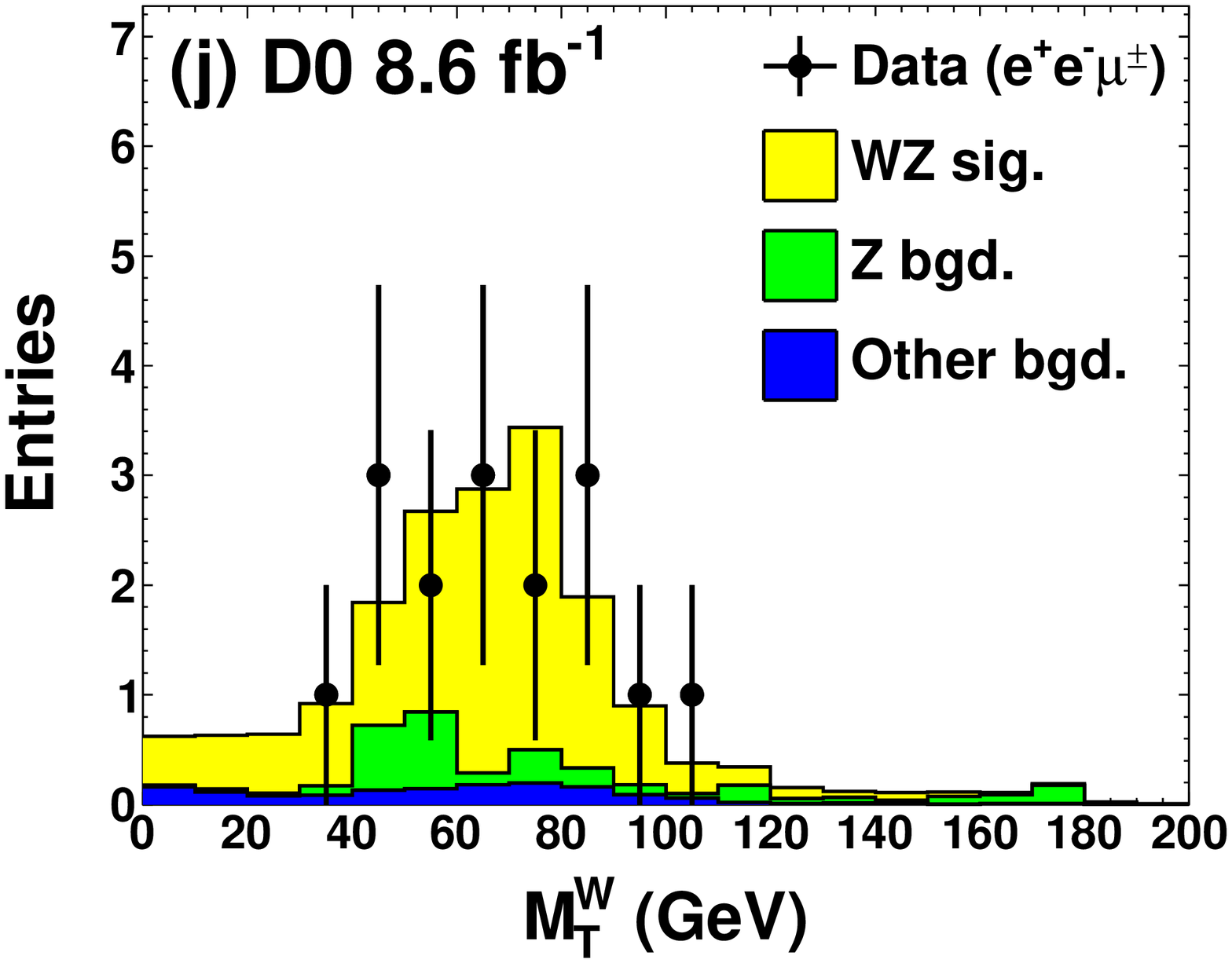}
\includegraphics[width=0.32\linewidth]{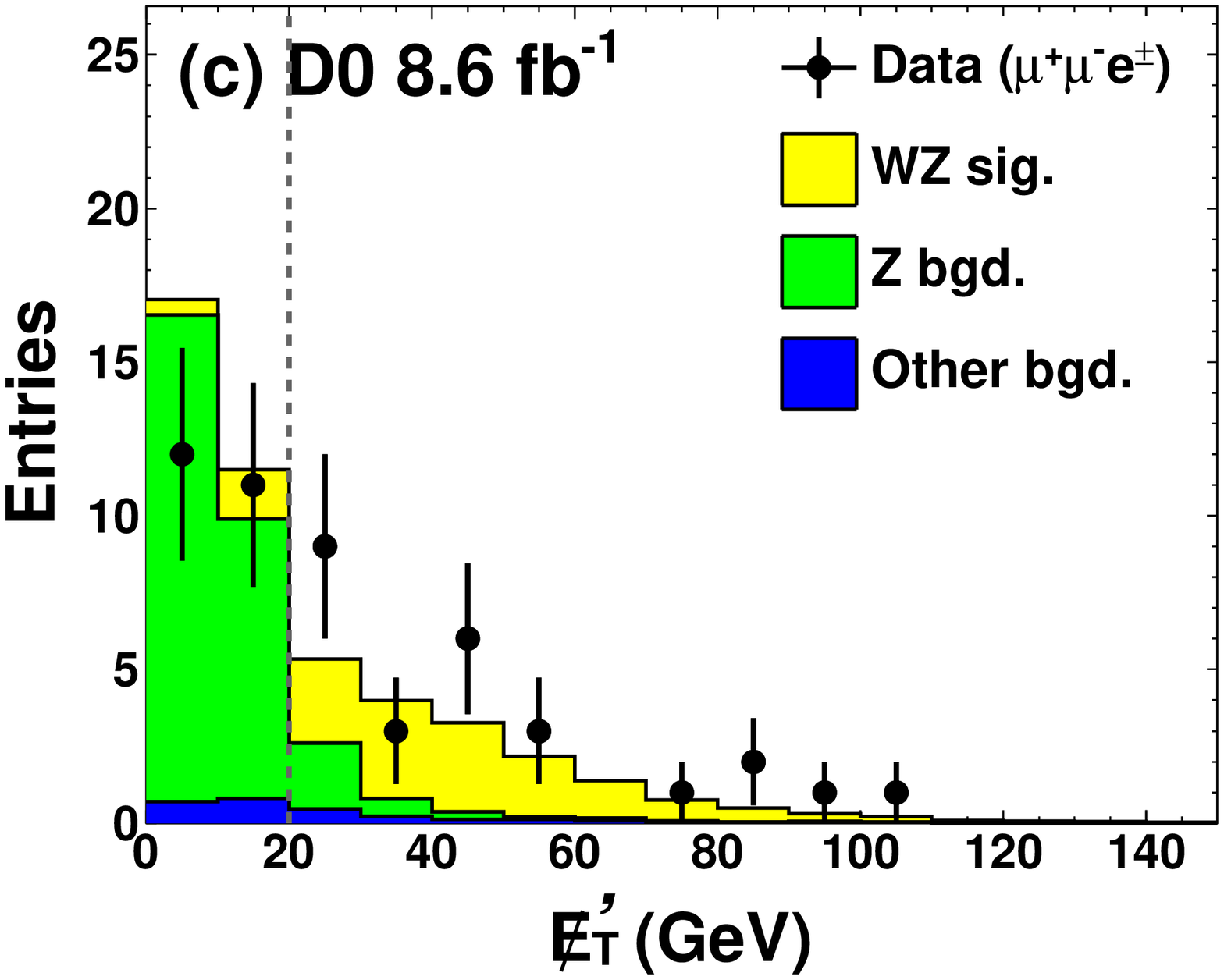}
\includegraphics[width=0.32\linewidth]{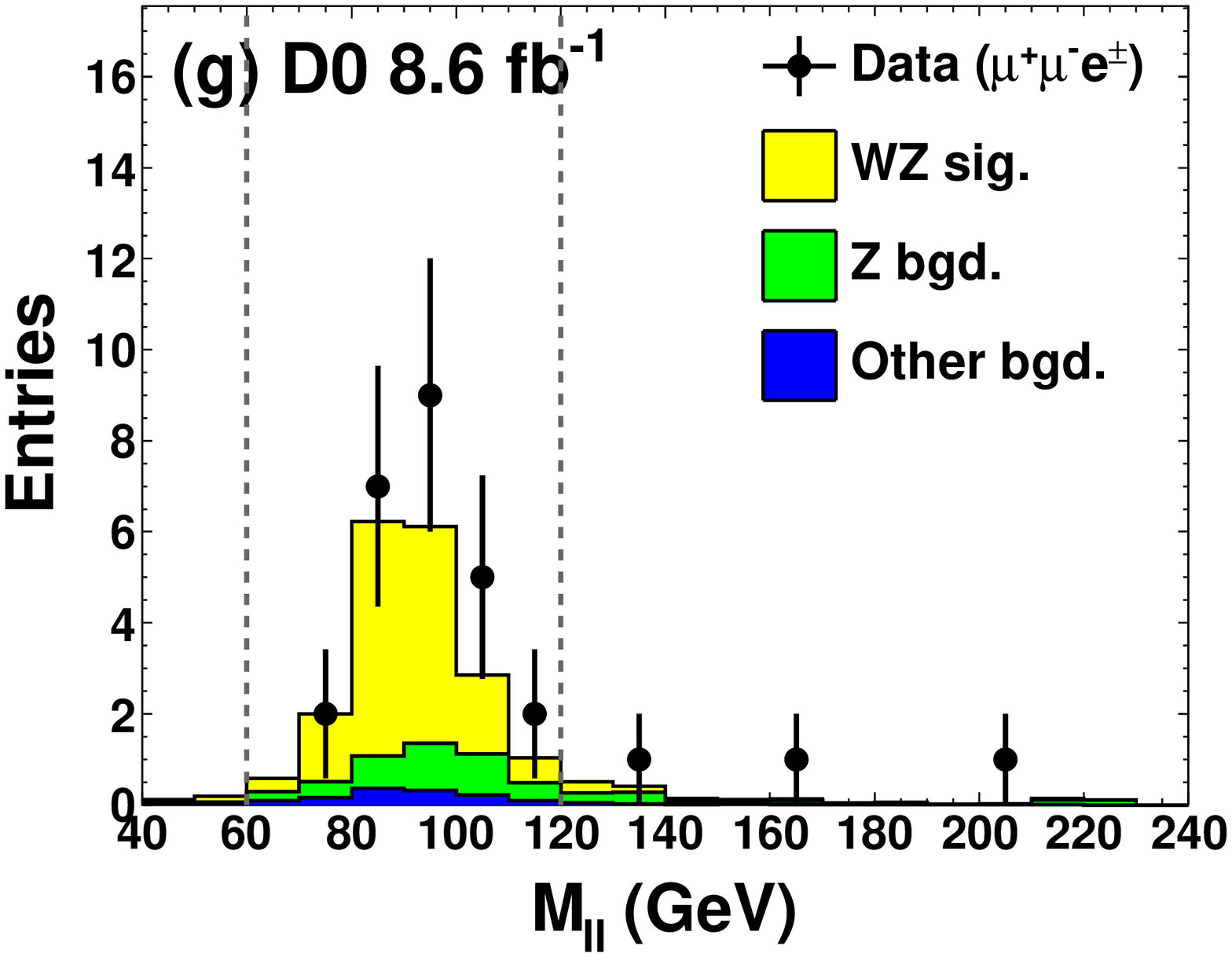}
\includegraphics[width=0.32\linewidth]{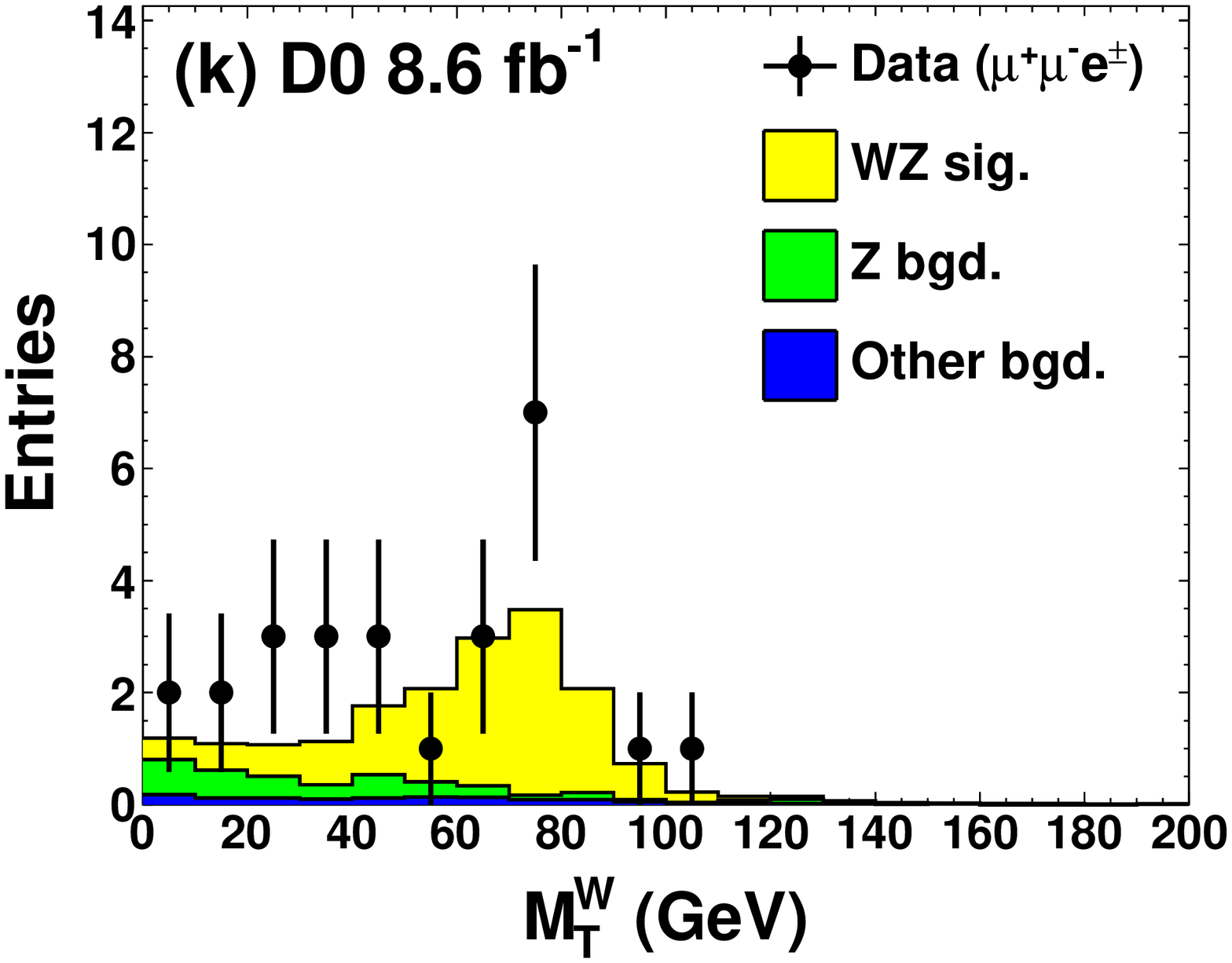}
\includegraphics[width=0.32\linewidth]{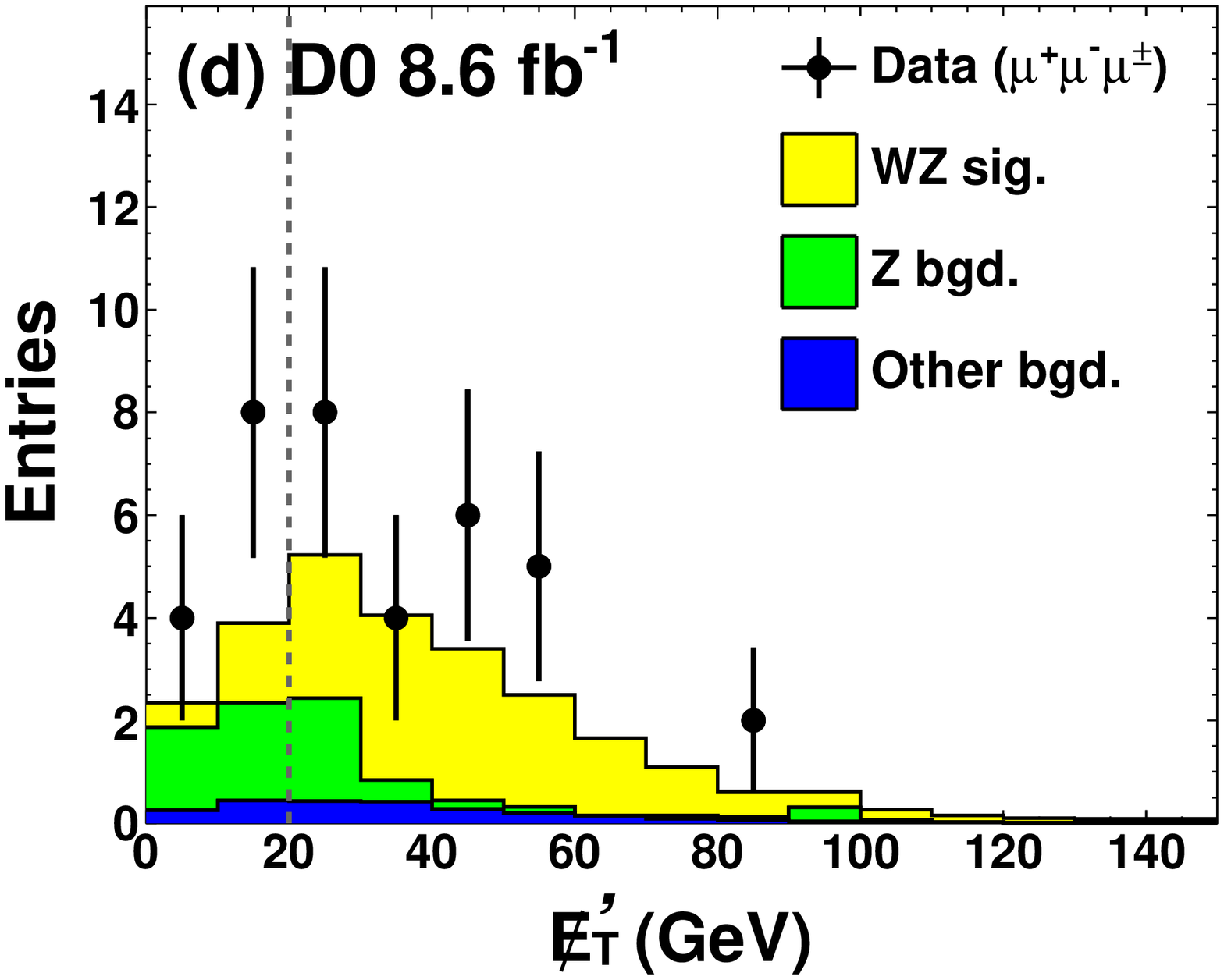}
\includegraphics[width=0.32\linewidth]{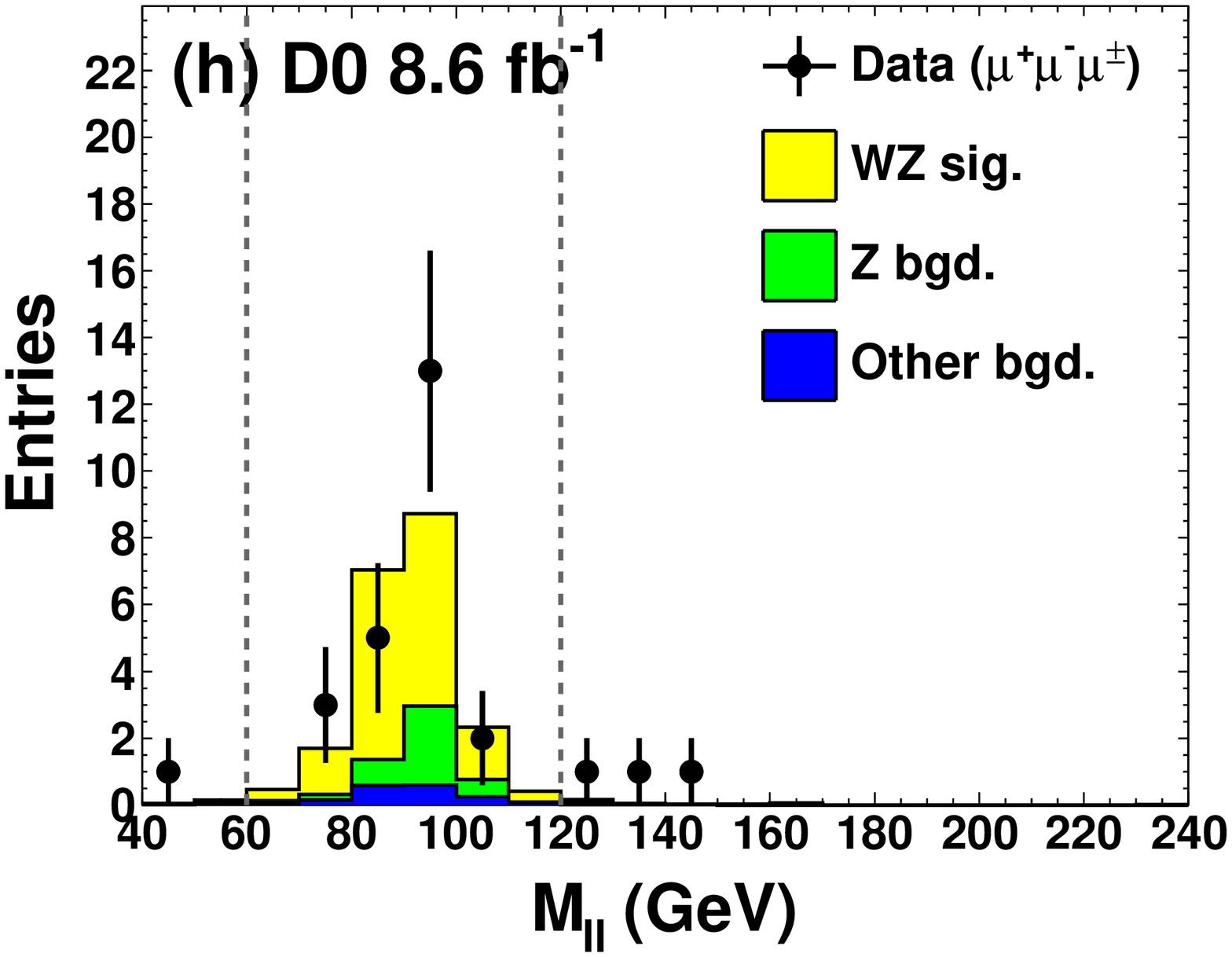}
\includegraphics[width=0.32\linewidth]{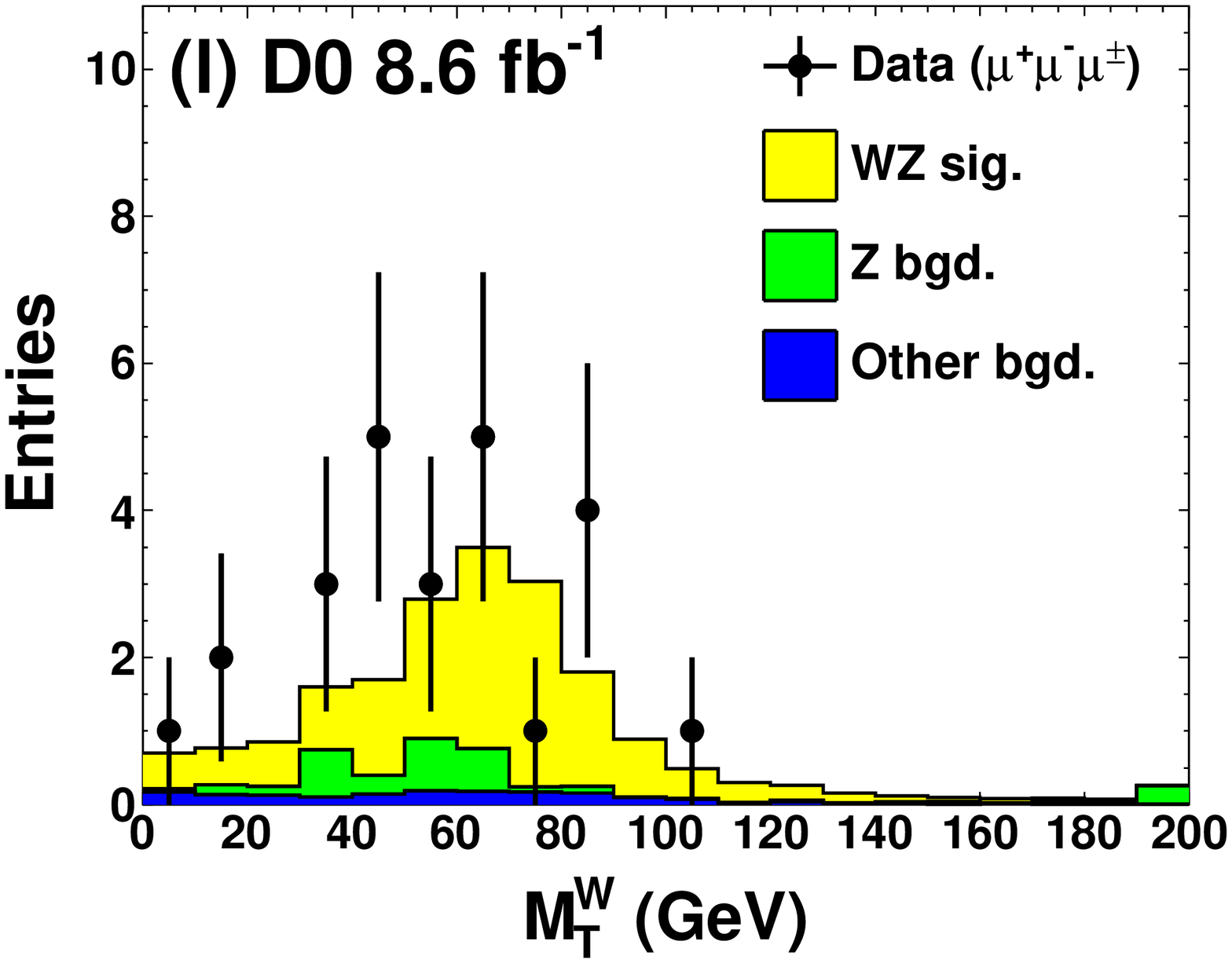}
\caption{
The distribution of (a-d) \metconstr, (e-h) \mass\ and (i-l) the $W$ transverse mass of the \WZ\ candidate events.
The \metconstr\ requirement is not imposed for (a-d), and the \mass\ requirement is not imposed for (e-h).
The rows correspond to different sub-channels as indicated on the figures.
The vertical dashed lines indicate the requirements on \metconstr\ and \mass.
The signal normalization is as described in Section~\ref{Section:bgd_and_signal}.
}
\label{Figure:wz_selection1}
\end{figure*}

\begin{figure*}[htbp]\centering
\includegraphics[width=0.32\linewidth]{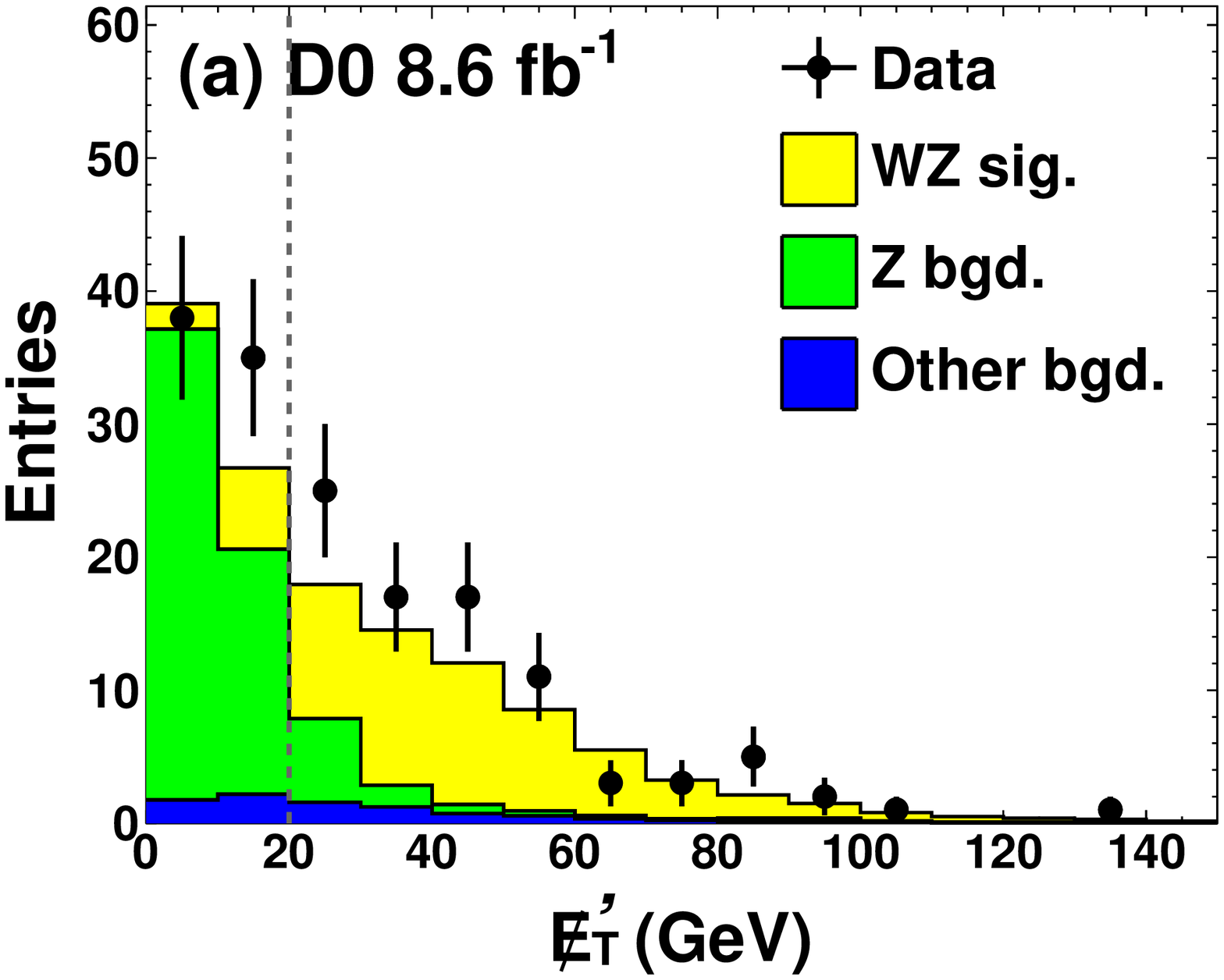}
\includegraphics[width=0.32\linewidth]{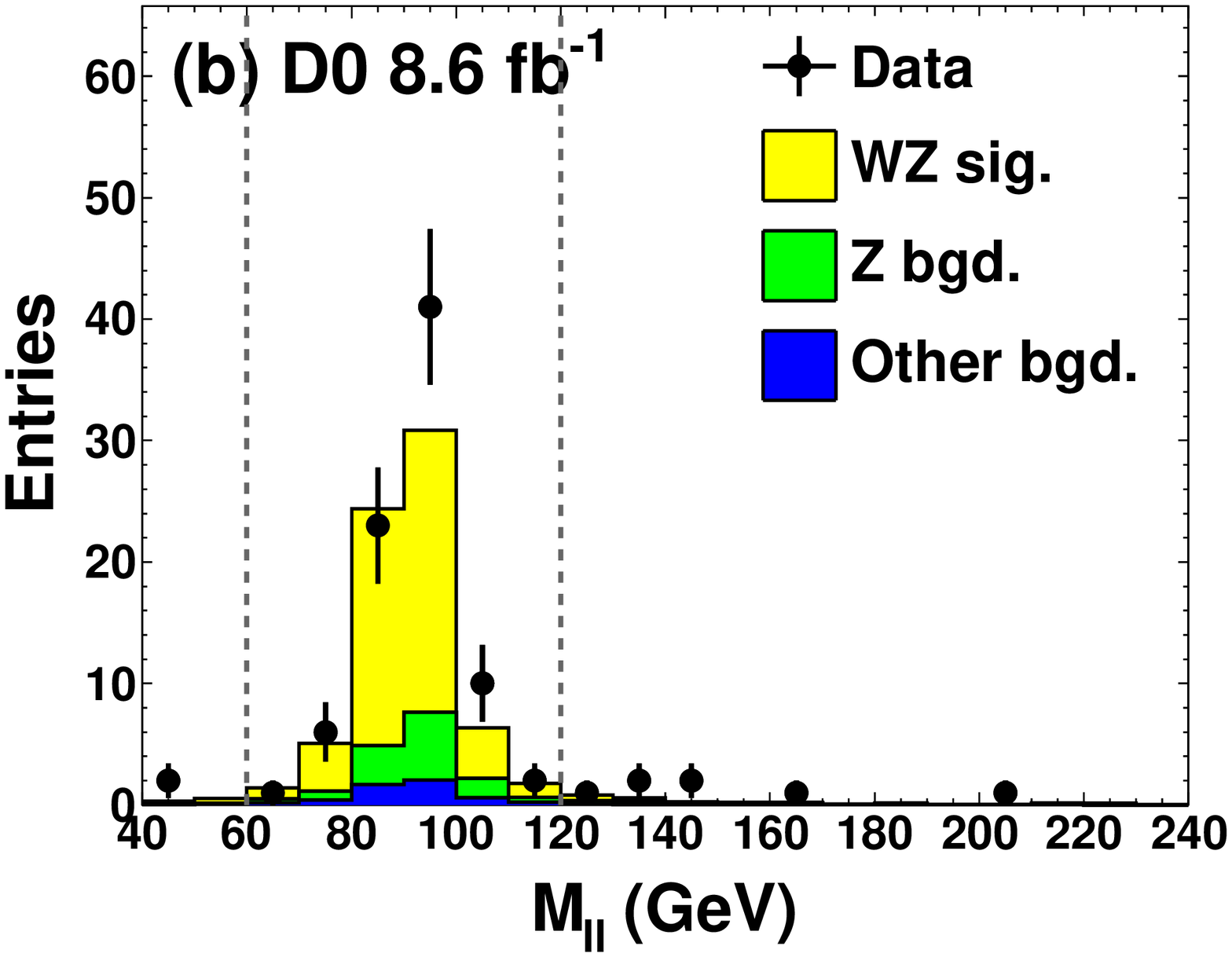}
\includegraphics[width=0.32\linewidth]{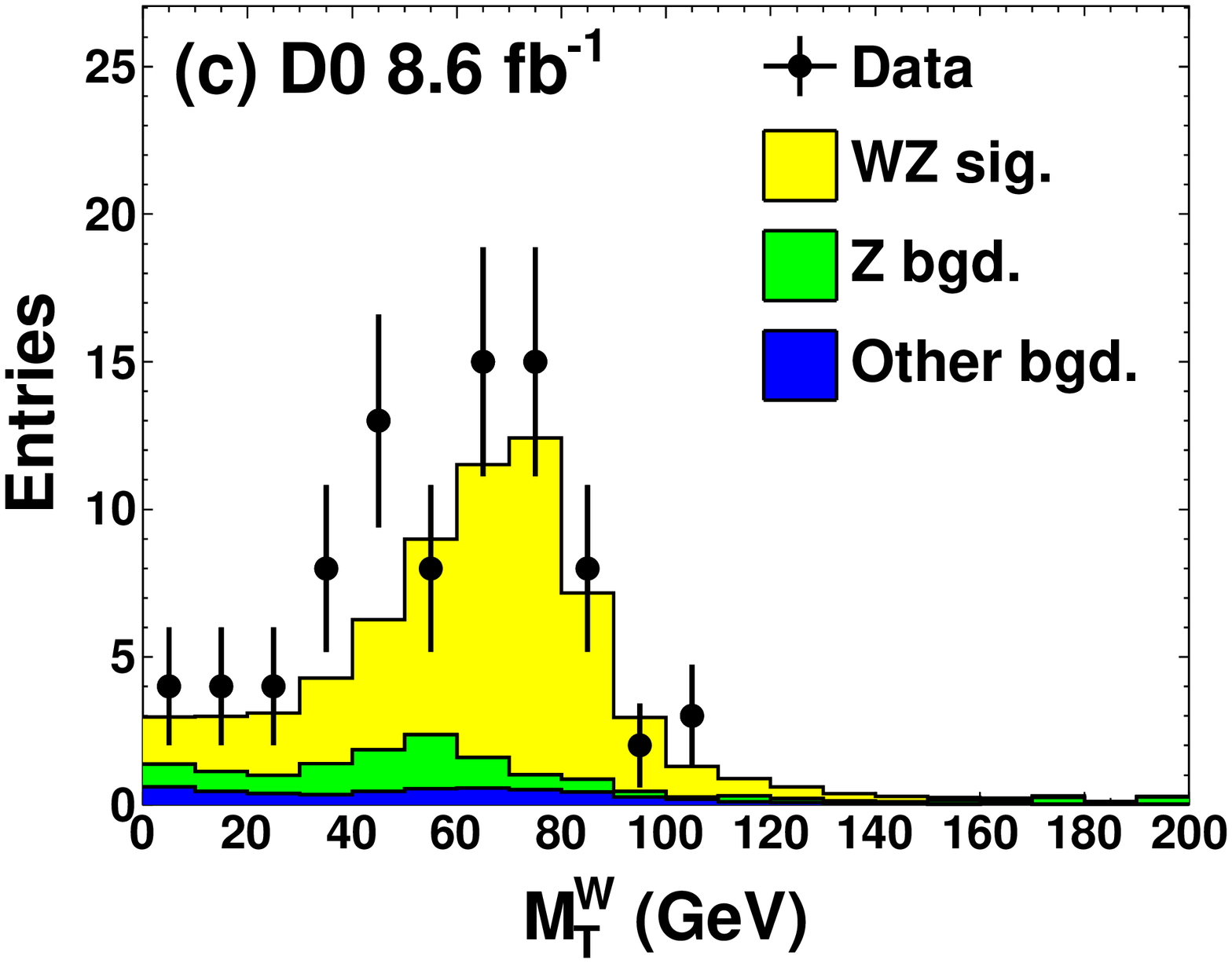}
\includegraphics[width=0.32\linewidth]{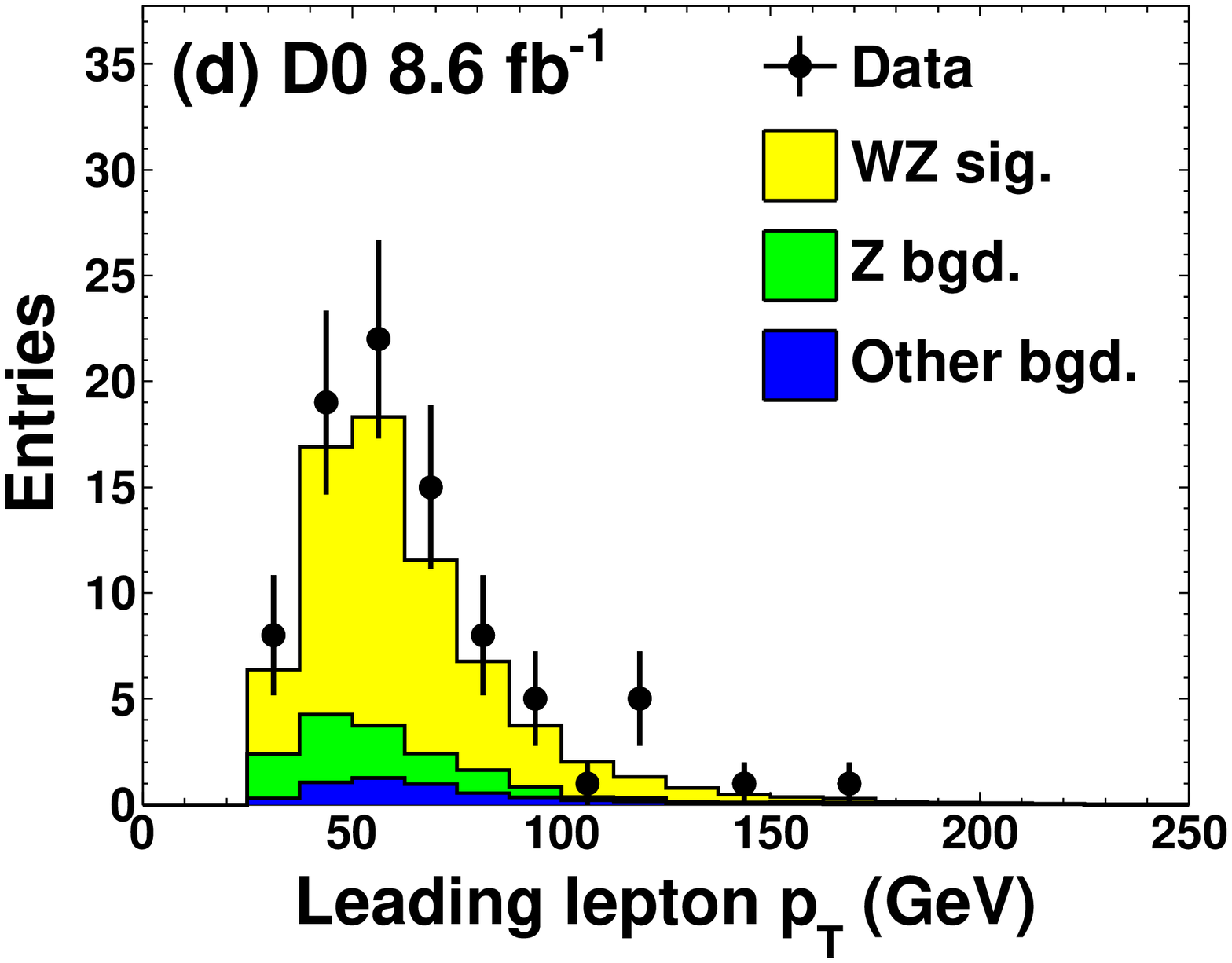}
\includegraphics[width=0.32\linewidth]{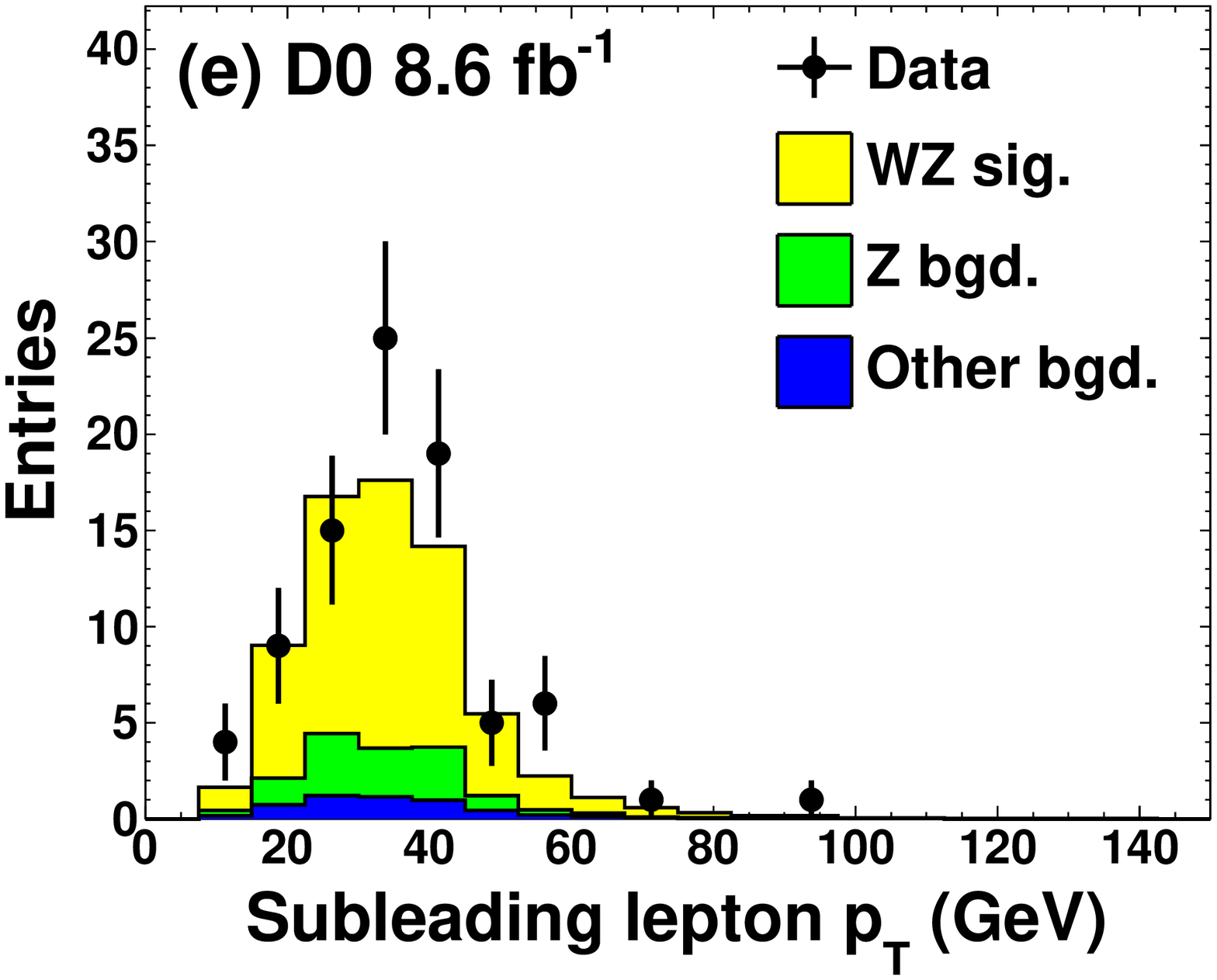}
\includegraphics[width=0.32\linewidth]{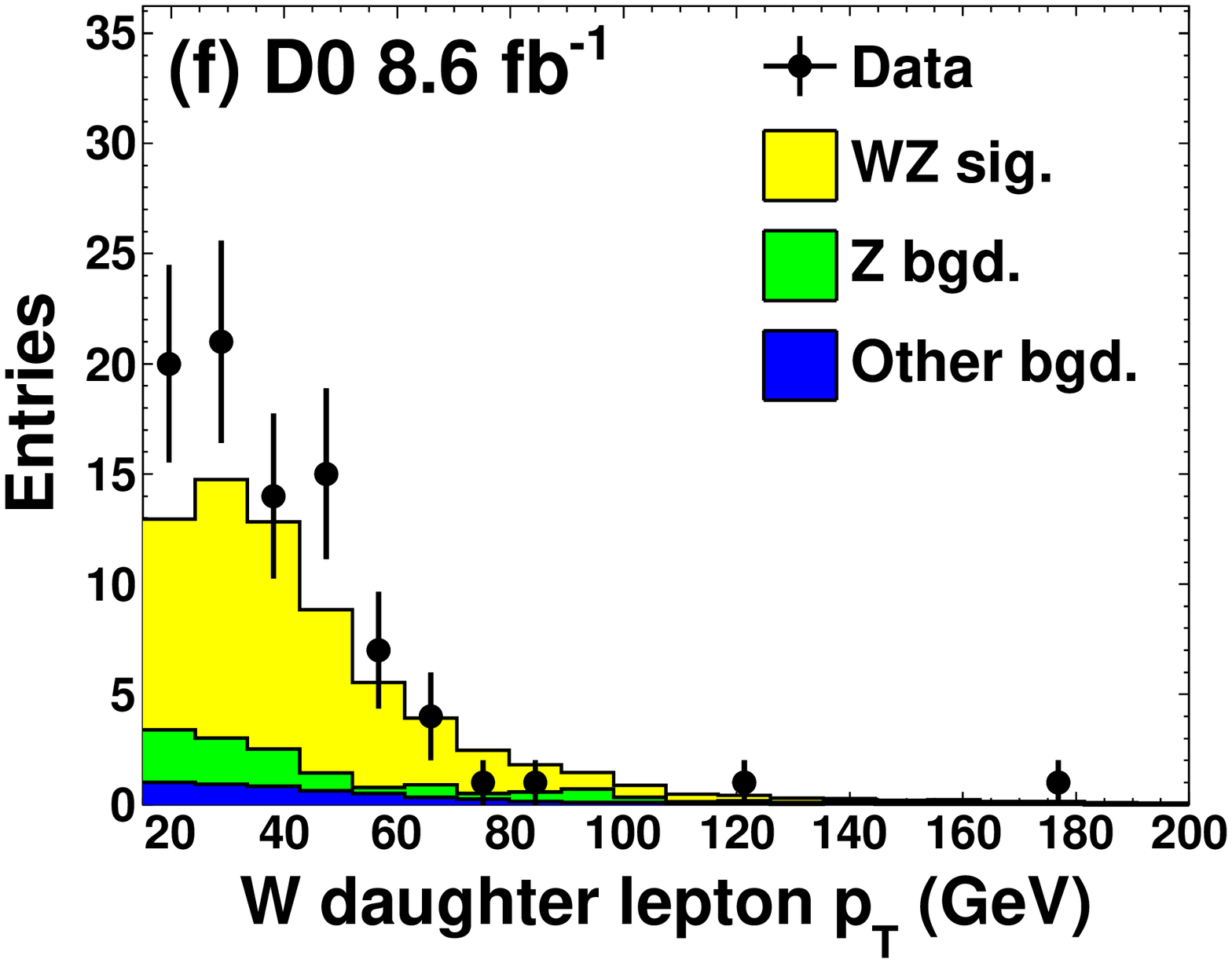}
\includegraphics[width=0.32\linewidth]{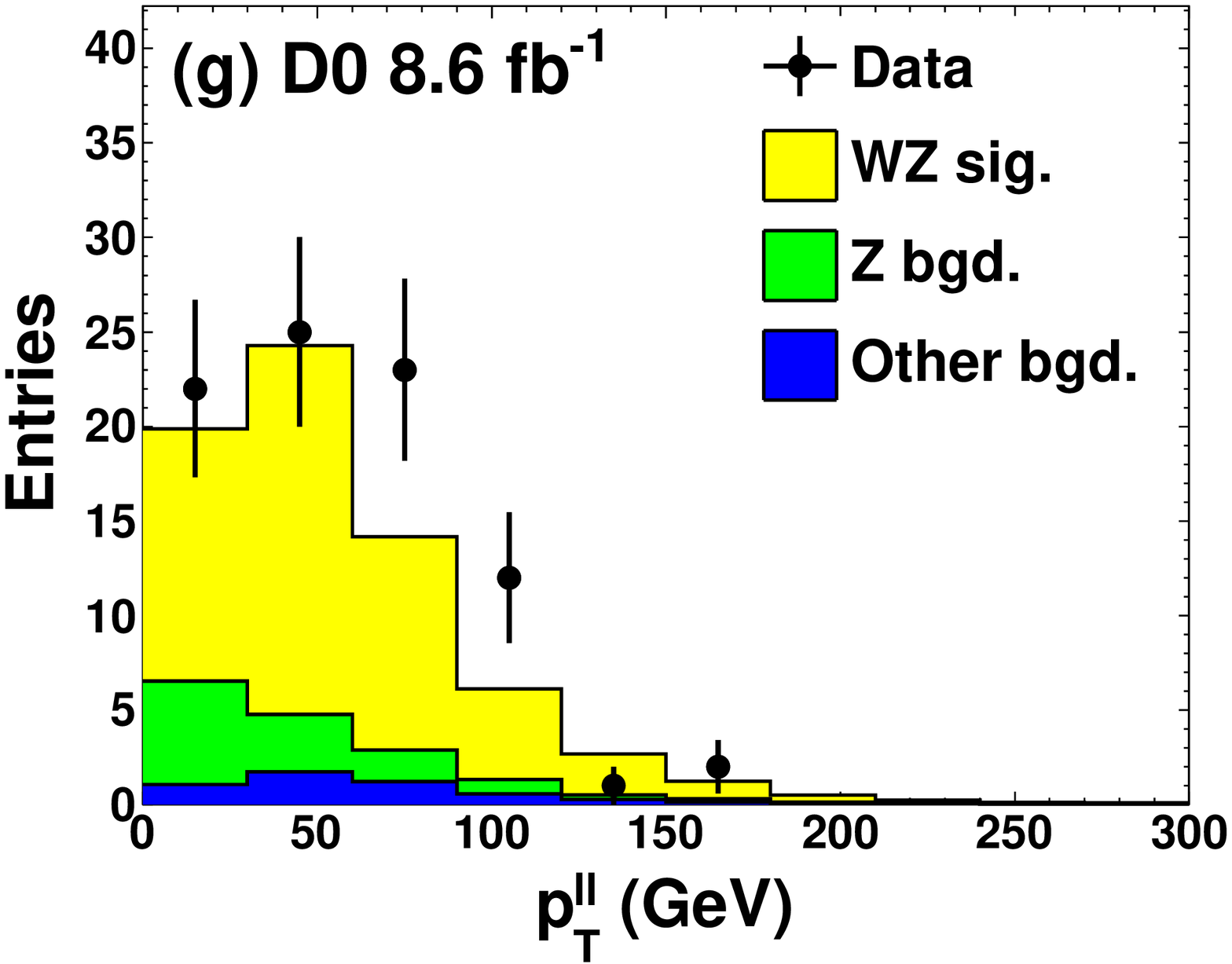}
\includegraphics[width=0.32\linewidth]{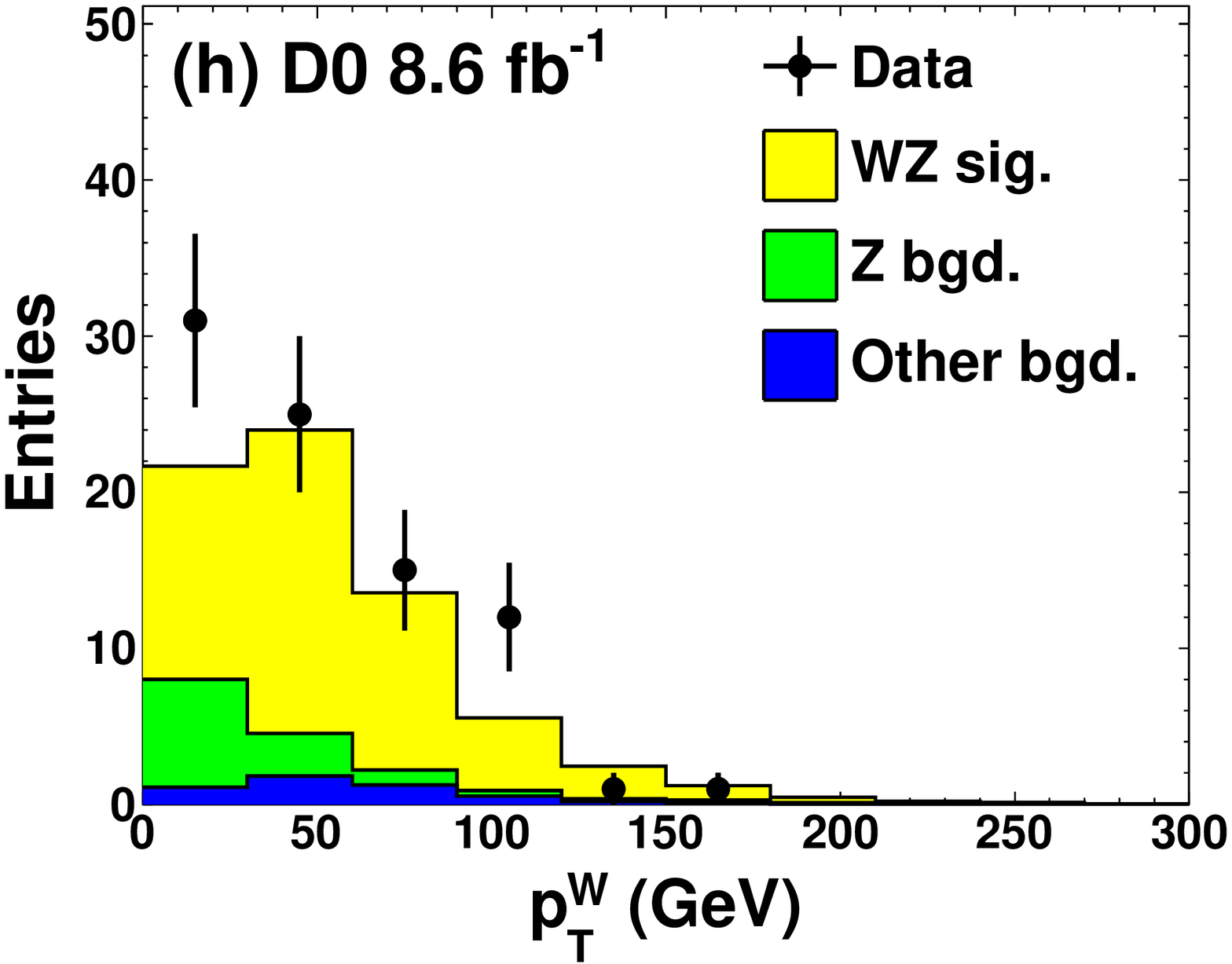}
\caption{Kinematic distributions for the \wzlnull\ signal candidates after combining
the different sub-channels.
The following variables are shown:
(a) the \metconstr;
(b) the invariant mass of the \zll\ decay;
(c) the $W$ transverse mass;
the transverse momenta of the (d) leading and (e) subleading leptons from the \zll\ decay
and (f) the charged lepton from the $W$ decay;
the transverse momenta of the reconstructed (g) \zll\ and (h) $W\rightarrow l\nu$ decays.
The vertical dashed lines indicate the requirements on \metconstr\ and \mass.
The signal normalization is as described in Section~\ref{Section:bgd_and_signal}.
}
\label{Figure:wz_candidates}
\end{figure*}

Tables~\ref{Table:wz_composition_diem} and~\ref{Table:wz_composition_dimu} list 
the observed and predicted event yields after all \WZ\ selection requirements.
The yields are also listed for the samples that exclusively fail the \met\ or \mass\ requirements,
but satisfy all other requirements.
Figure~\ref{Figure:wz_selection1} shows the \met, \mass, and $M_T^W$ distributions for \wzlnull\ candidate events, 
before imposing the requirement on each variable.
The transverse mass is defined as 
$M_T^W = \sqrt{2p_T\met(1-\cos\Delta\phi)}$, with $p_T$ being the
transverse momentum of the charged lepton that is assigned as the $W$ daughter,
and $\Delta\phi$ being the azimuthal angle between this lepton and the
\met\ vector.
Figure~\ref{Figure:wz_candidates} shows the distributions of various kinematic 
quantities  after combining the four sub-channels.

\begin{table*}
\centering
\caption{\wzlnull\ selection: Predicted and observed yields in the \zee\ sub-channels.
The systematic uncertainties are provided for the predictions. }
\input{\plotdir/Table1.tex}
\label{Table:wz_composition_diem}
\end{table*}

\begin{table*}
\centering
\caption{\wzlnull\ selection: Predicted and observed yields in the \zmm\ sub-channels.
The systematic uncertainties are provided for the predictions. }
\input{\plotdir/Table2.tex}
\label{Table:wz_composition_dimu}
\end{table*}

\section{\label{Section:metprime}Missing transverse momentum estimators}

\input{metprime.tex}

\section{\label{Section:ZZ} Selection of ZZ candidates}

Decays of $ZZ$ into the following final states are considered:
$e^+e^-\nu\bar{\nu}$,
$\mu^+\mu^-\nu\bar{\nu}$.
Events must contain two oppositely charged tight quality leptons
satisfying the $p_T$ requirements described earlier and with \mass\ between 60 and 120 GeV.
To reject events in which the missing transverse momentum estimators defined in Section~\ref{Section:metprime}
are poorly reconstructed, we require that there are no more than two jets with $p_T > 15$~GeV 
and separated by at least $\DR = 0.3$ from the leptons.
In order to reject \wzlnull\ and \zzllll\ events, there must be no additional 
EM clusters or muons according to the criteria in Section~\ref{Section:WZ}.
In addition, there must be no isolated tracks or hadronic taus 
with $p_T$~$>$~5~GeV.
This requirement is not necessary in the \WZ\ analysis, for which
the \zzllll\ background is less significant.
These four types of objects are only considered if they are separated by at least $\DR = 0.3$ from the leptons.
The jet and additional lepton vetoes are also effective in suppressing the background from \ttbar\ decays.
The number of events that satisfy these requirements is used as the denominator
in the measurement of the $ZZ/Z$ cross section ratio.
Including the jet and lepton vetoes in the $Z$ selection helps to reduce the impact
of uncertainties in the corresponding efficiencies.

\begin{figure*}[htbp]\centering
\includegraphics[width=0.32\linewidth]{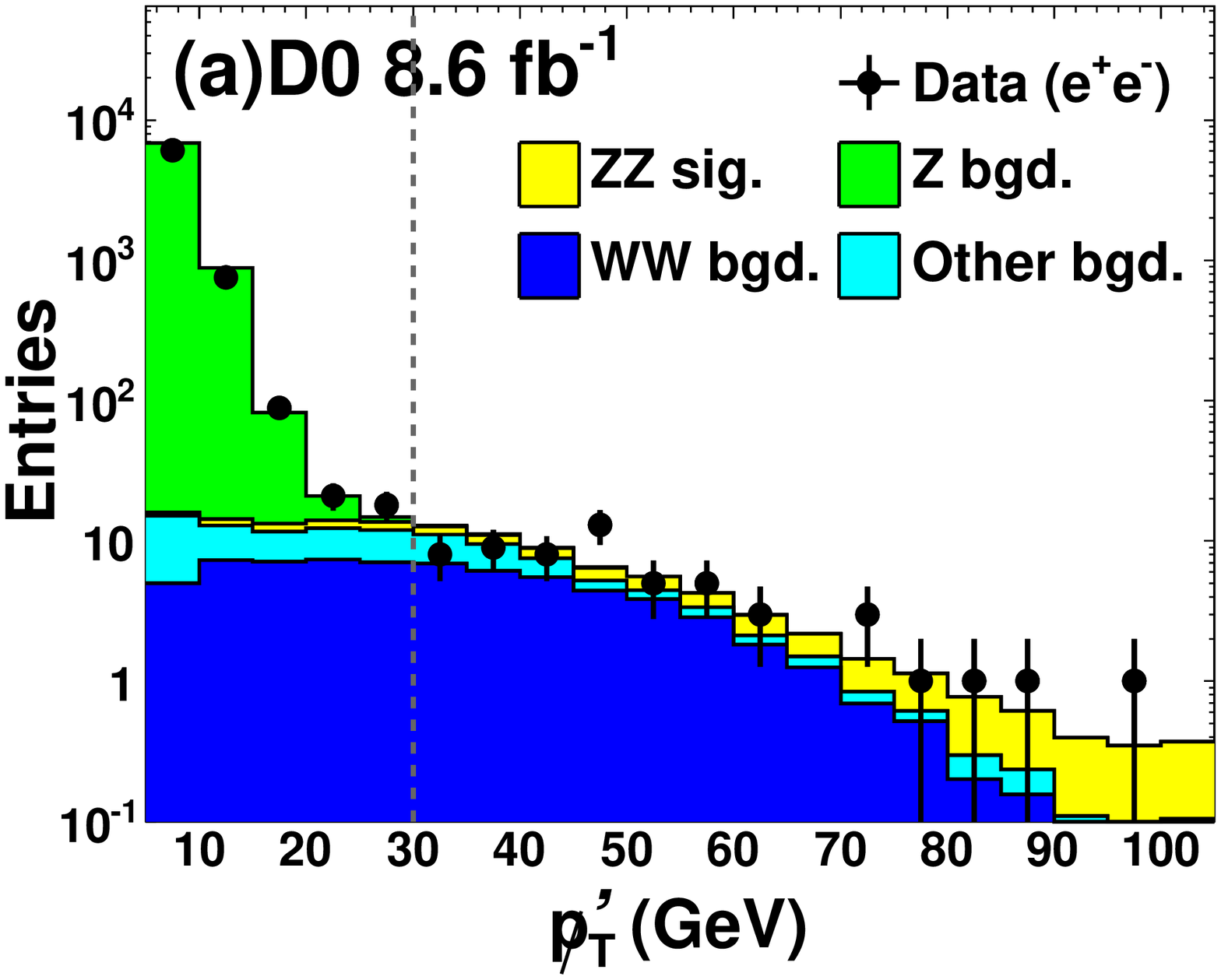}
\includegraphics[width=0.32\linewidth]{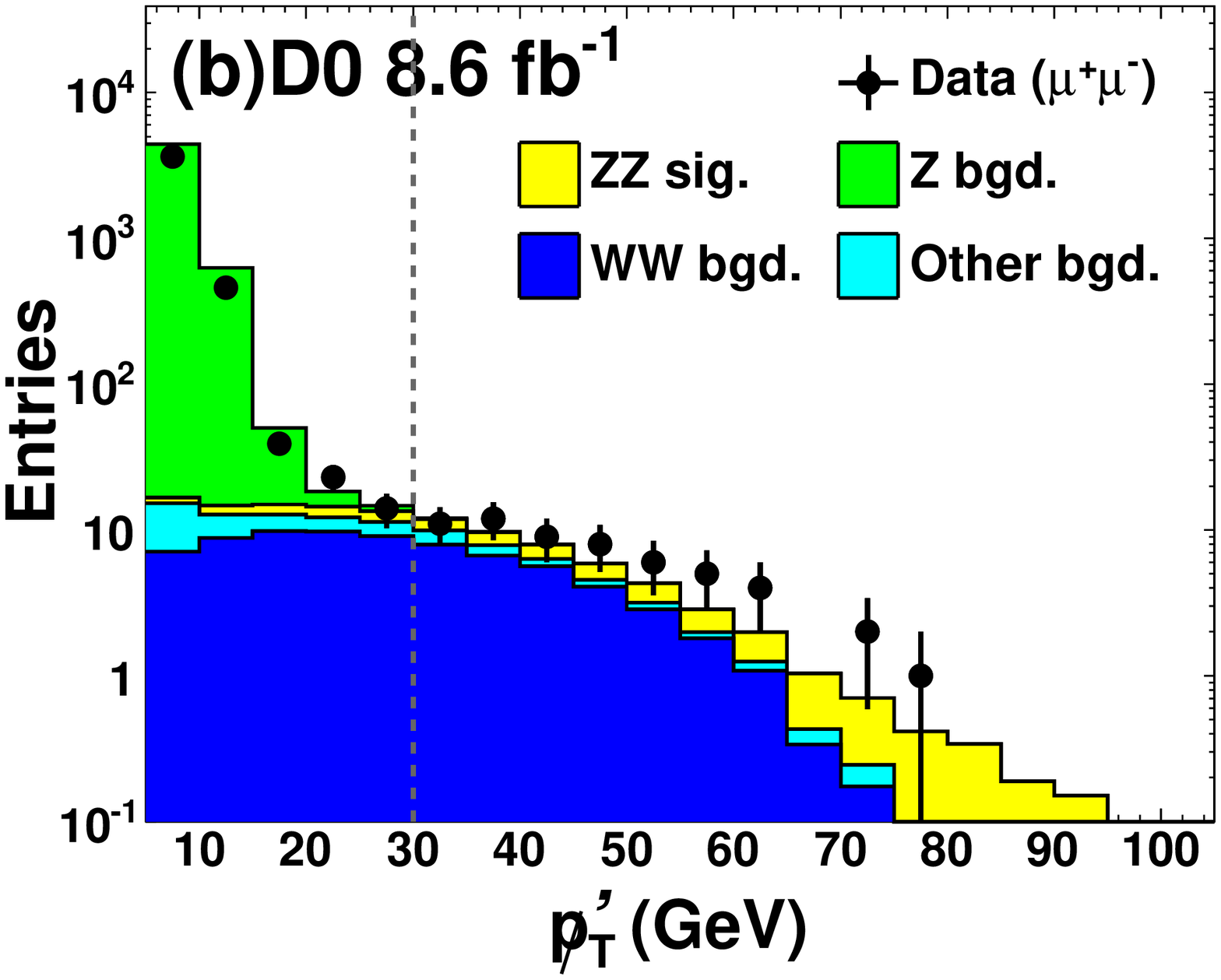}
\includegraphics[width=0.32\linewidth]{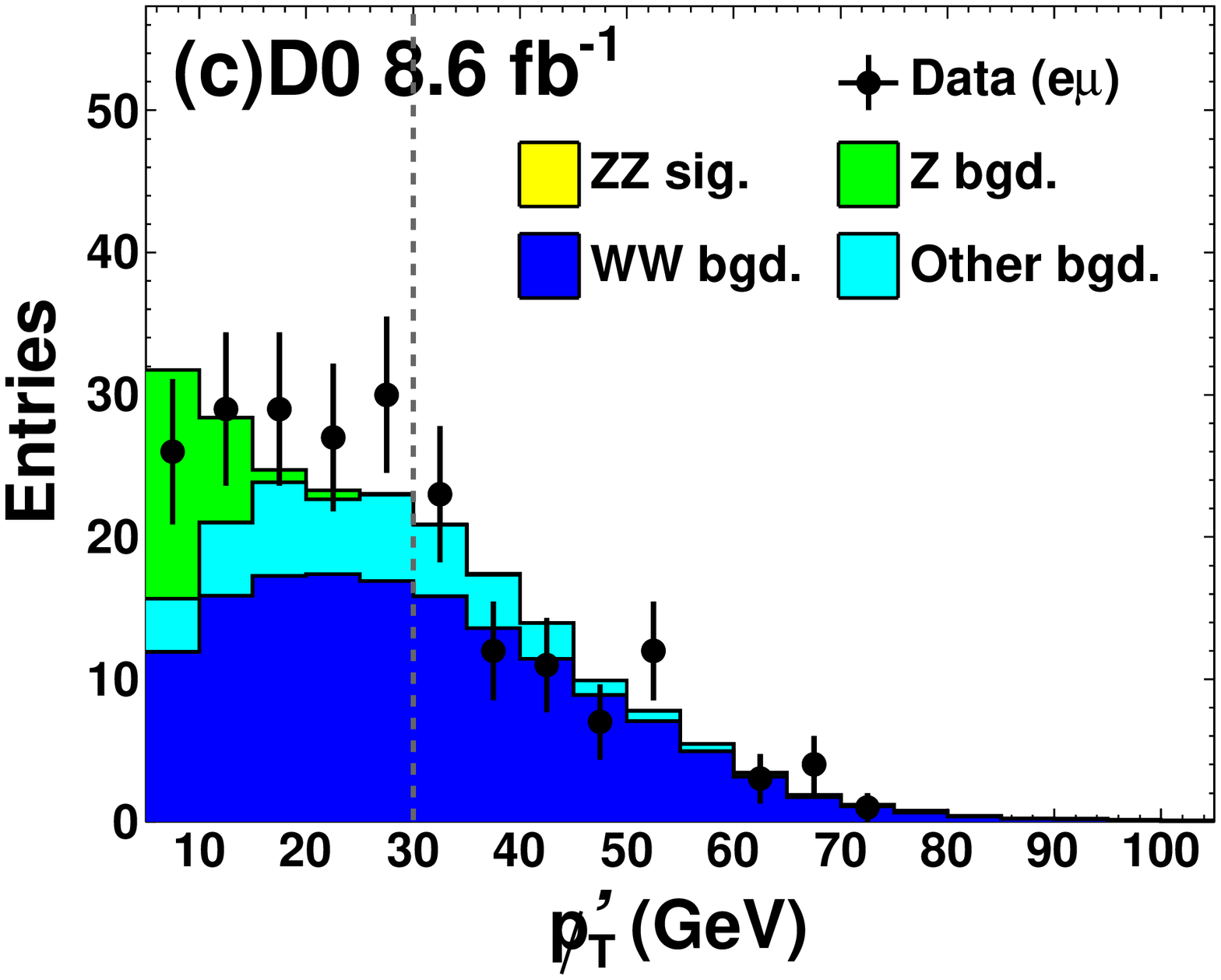}
\includegraphics[width=0.32\linewidth]{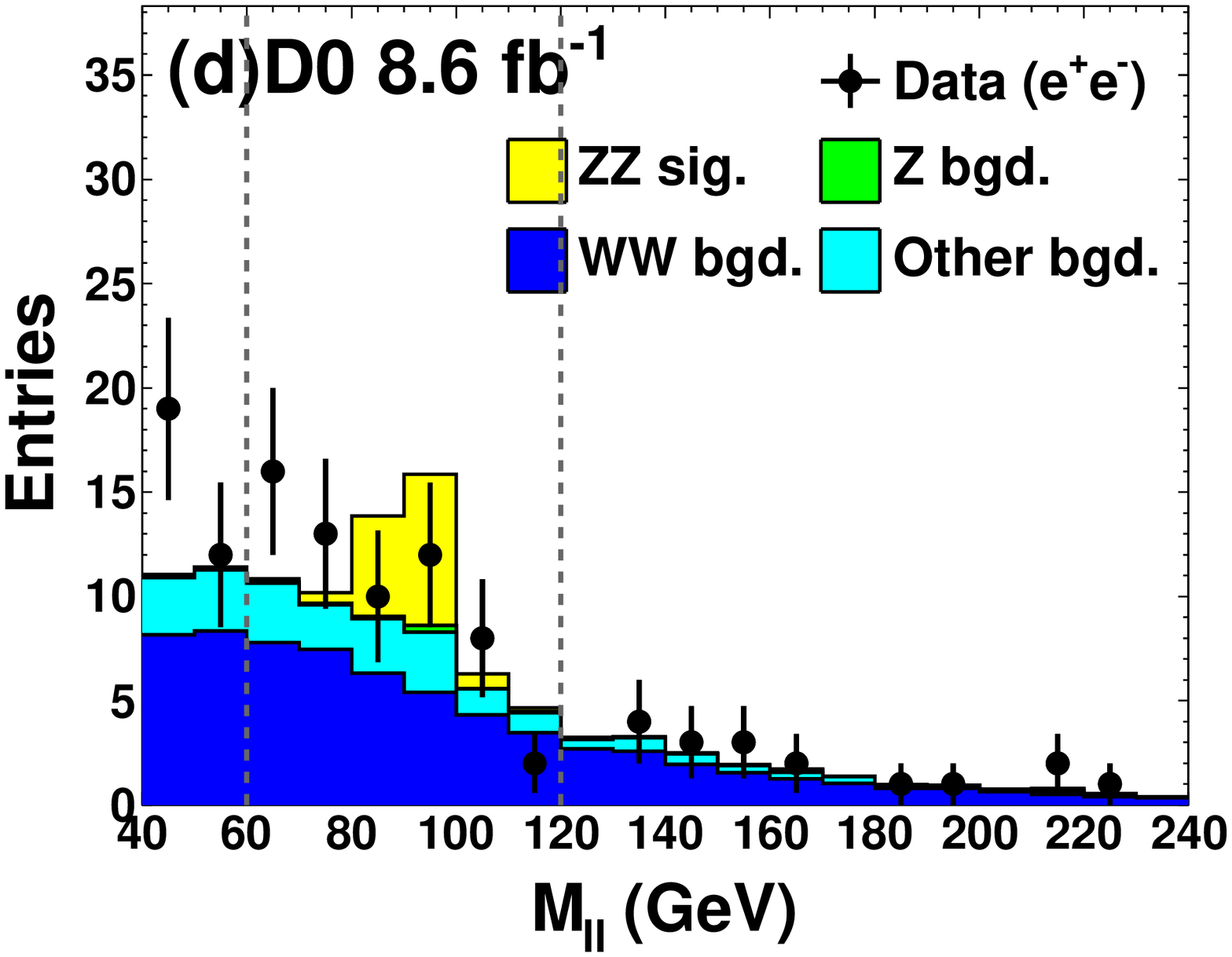}
\includegraphics[width=0.32\linewidth]{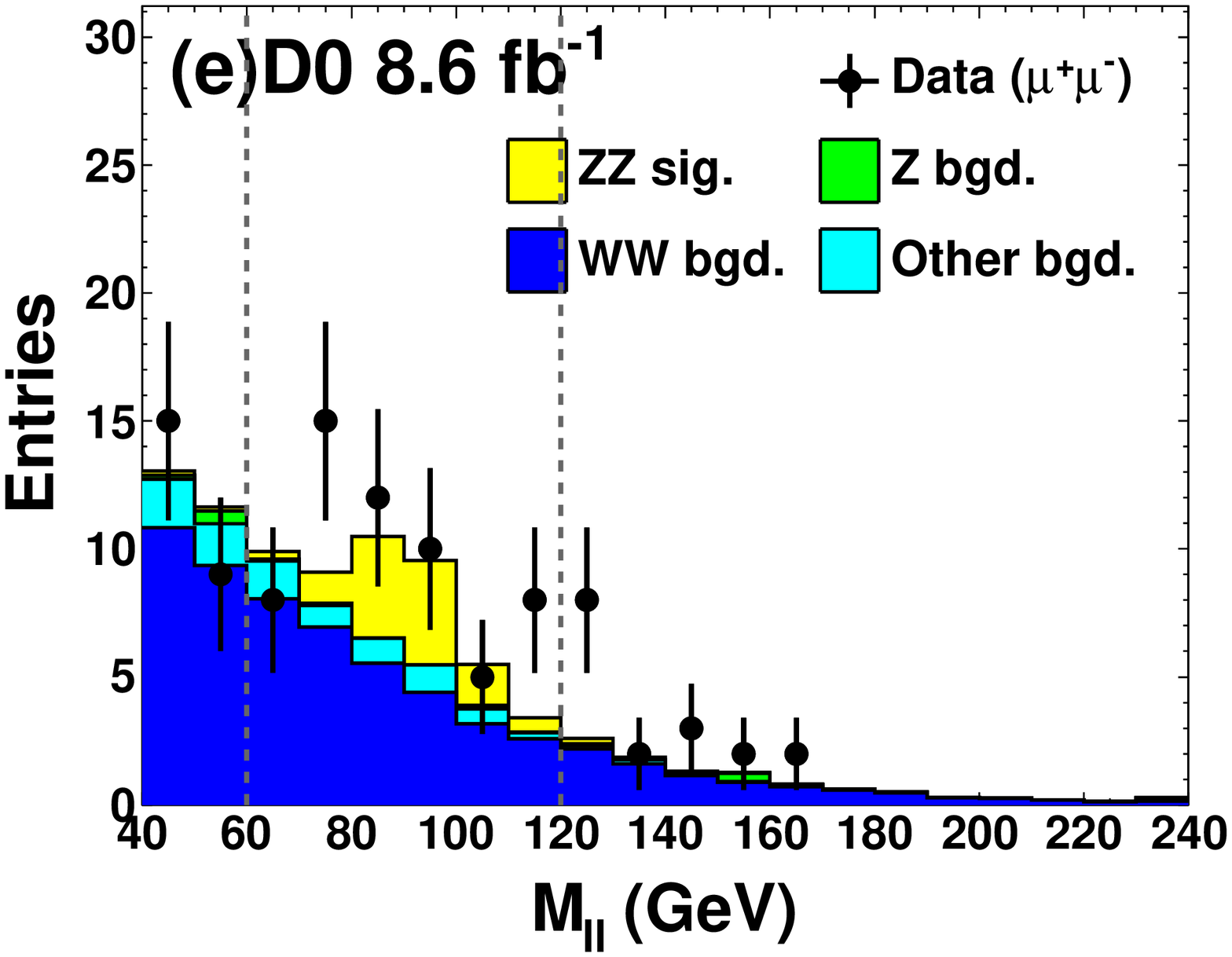}
\includegraphics[width=0.32\linewidth]{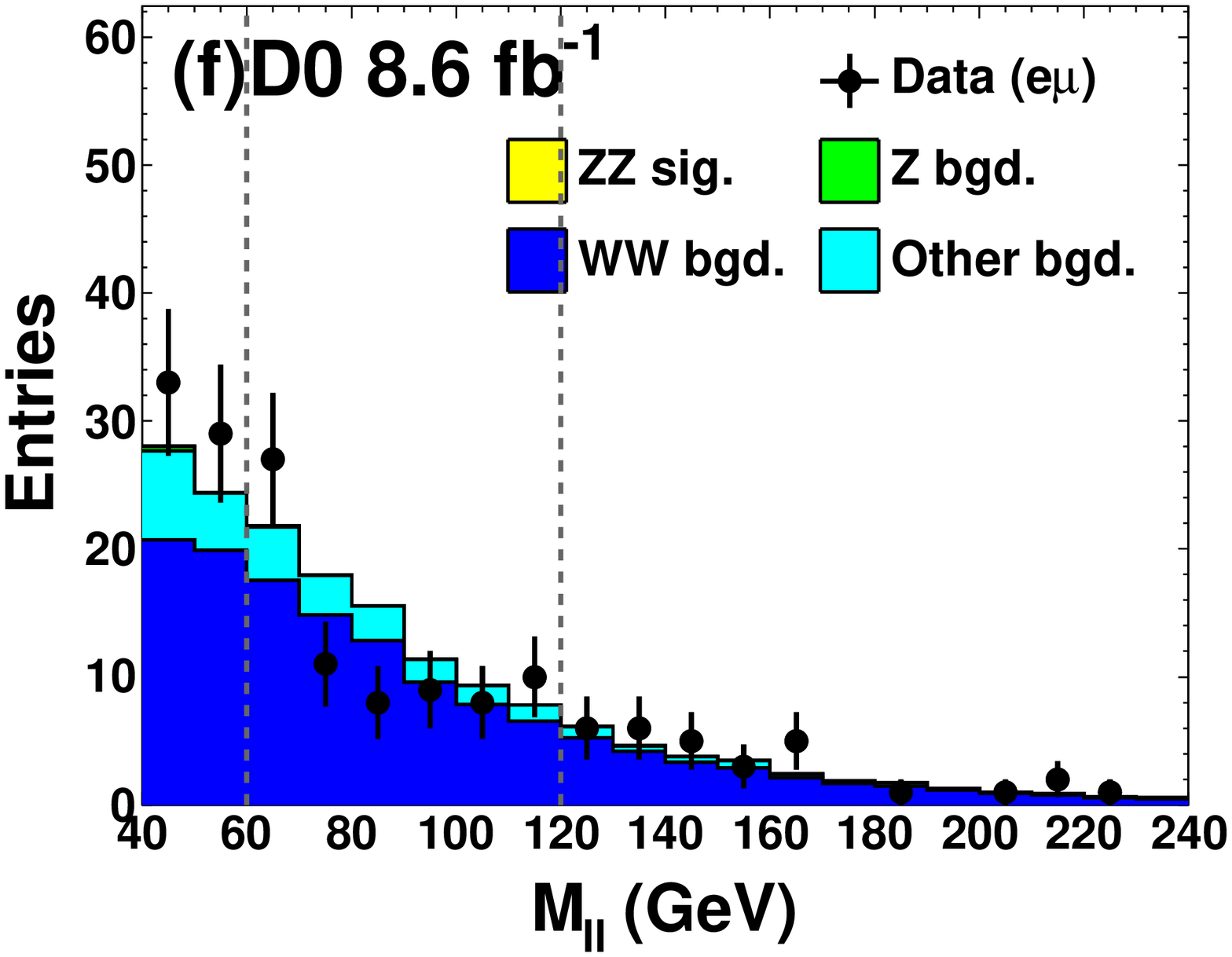}
\includegraphics[width=0.32\linewidth]{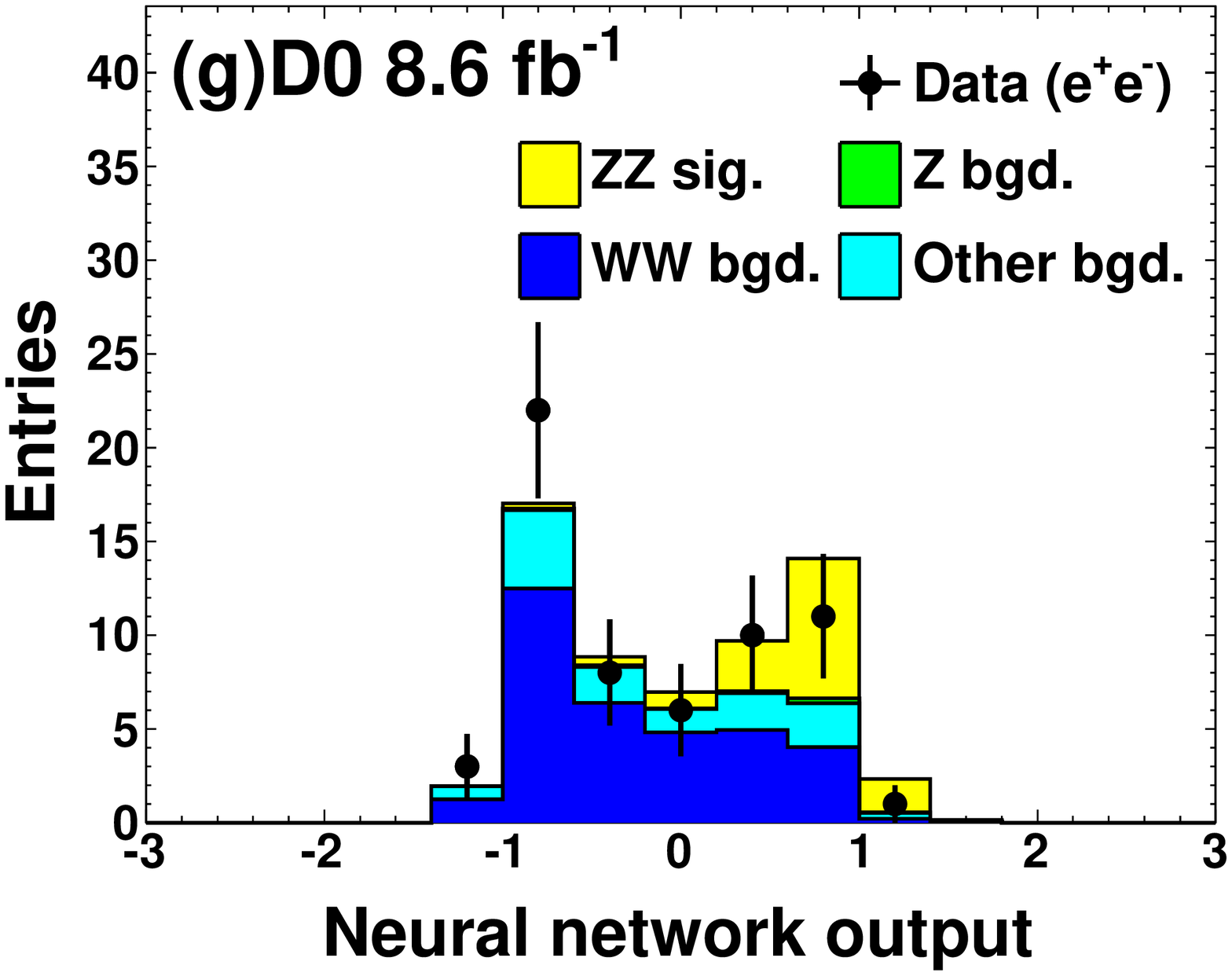}
\includegraphics[width=0.32\linewidth]{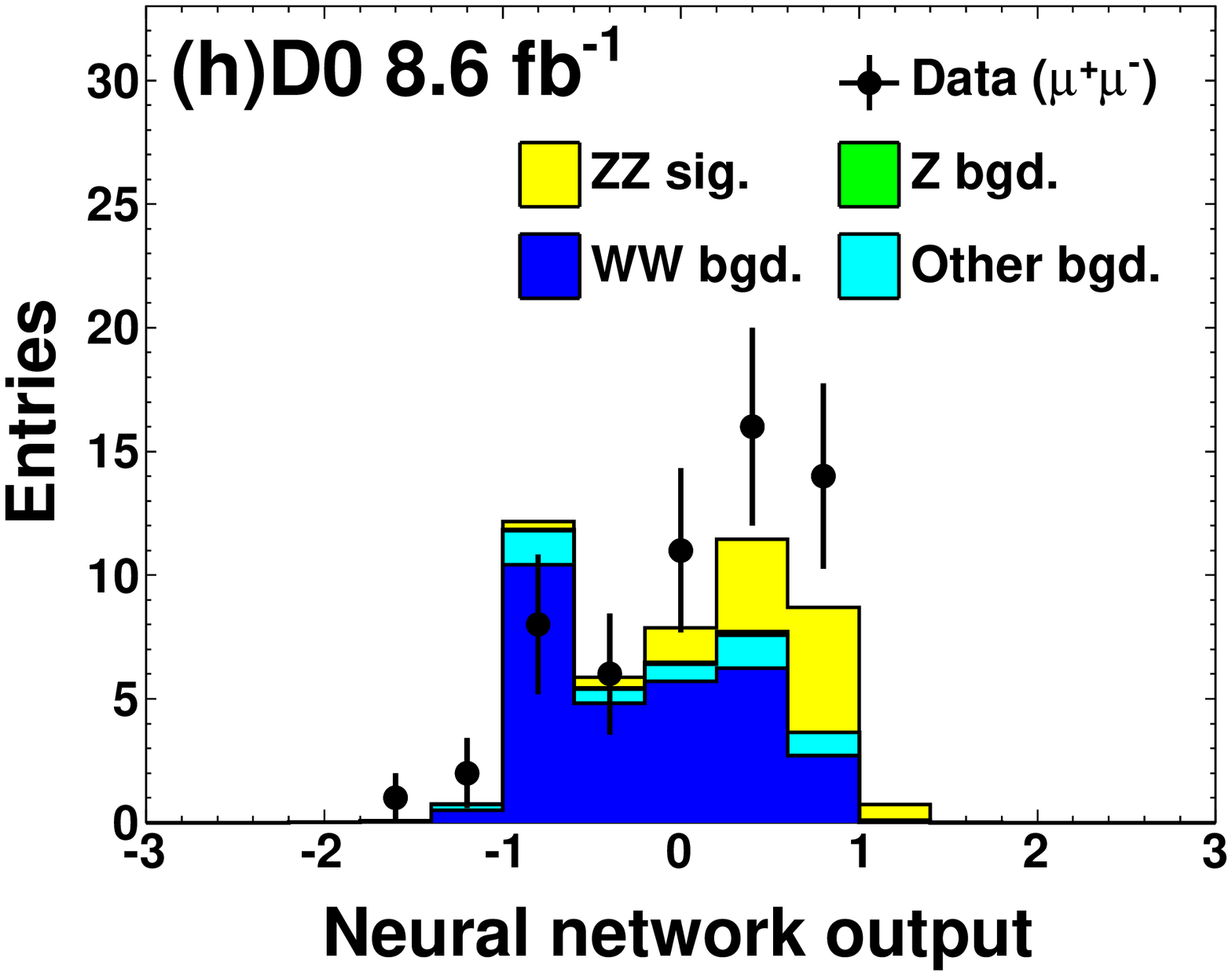}
\includegraphics[width=0.32\linewidth]{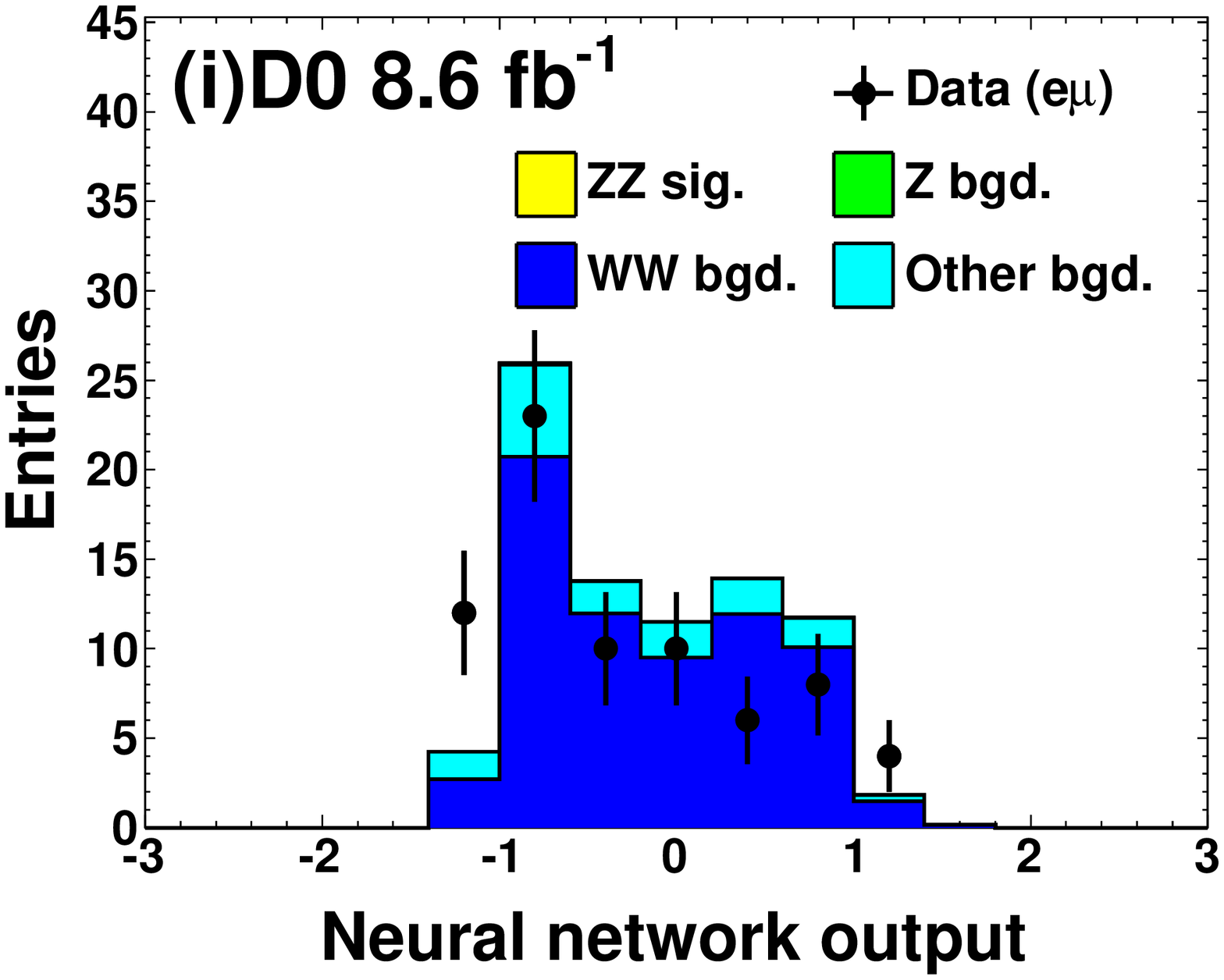}
\caption{
(a-c) The \pTmissPrime\ distribution of the \zzllnunu\ candidate events before imposing the \pTmissPrime\ requirement.
(d-f) The \mass\ distribution of the \zzllnunu\ candidate events before imposing the \mass\ requirement.
(g-i) The neural network output distribution of the accepted \zzllnunu\ candidate events.
For the \emmu\ channel, the neural network trained in the \diem\ channel is shown.
The vertical dashed lines indicate the requirements on \pTmissPrime\ and \mass.
The signal normalization is as described in Section~\ref{Section:bgd_and_signal}.}
\label{Figure:ZZ_selection_cuts}
\end{figure*}

\begin{figure*}
\includegraphics[width=0.32\linewidth]{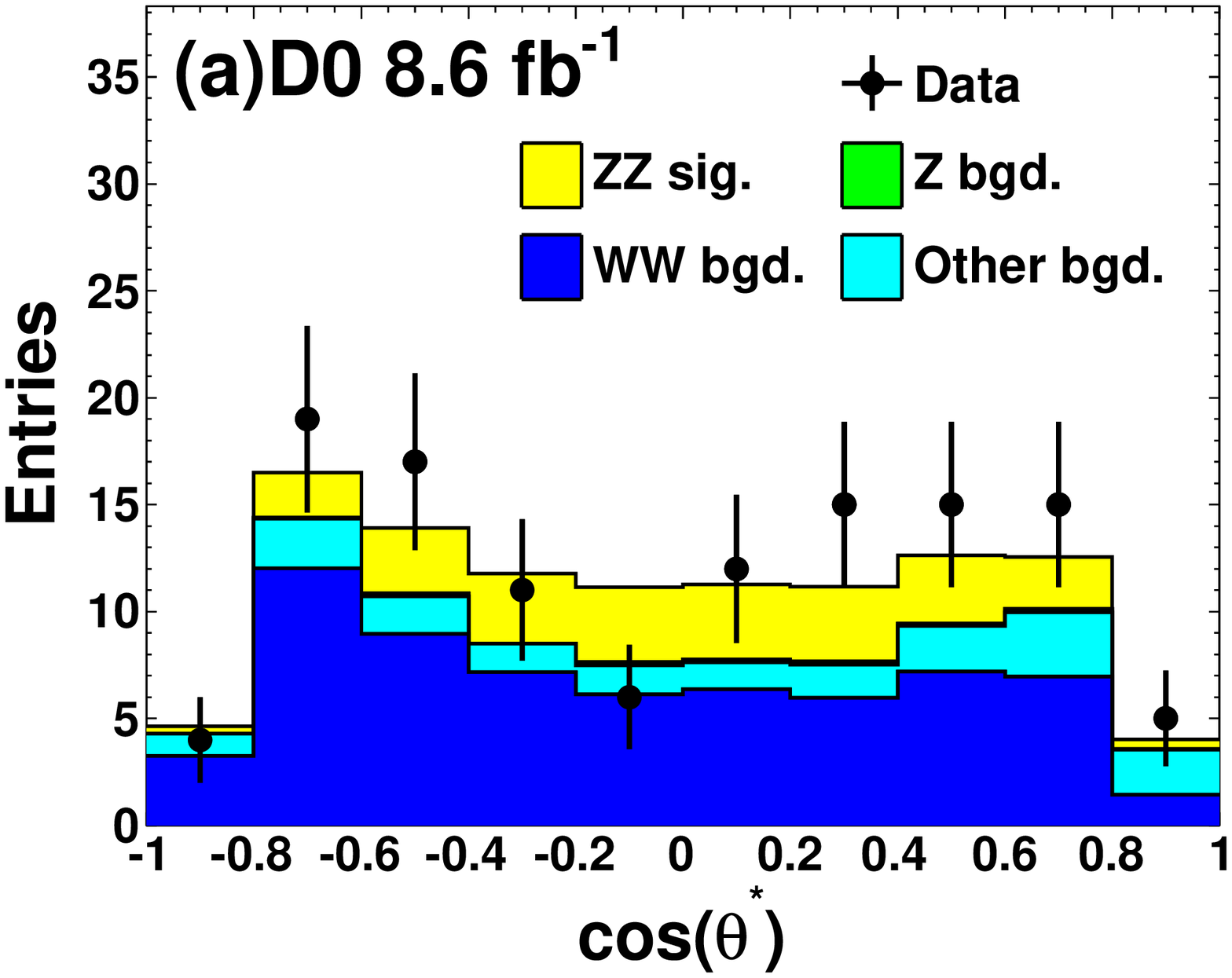}
\includegraphics[width=0.32\linewidth]{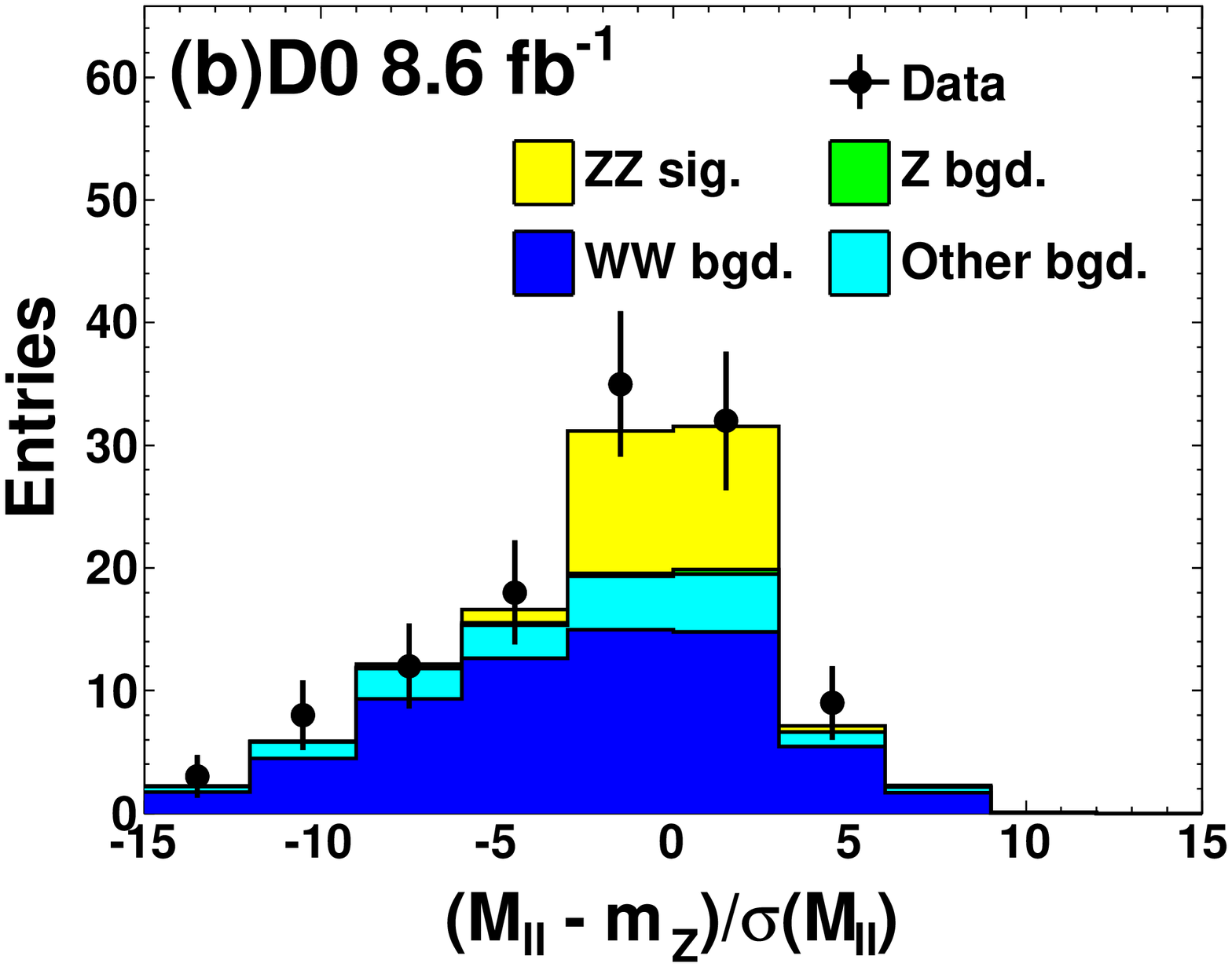}
\includegraphics[width=0.32\linewidth]{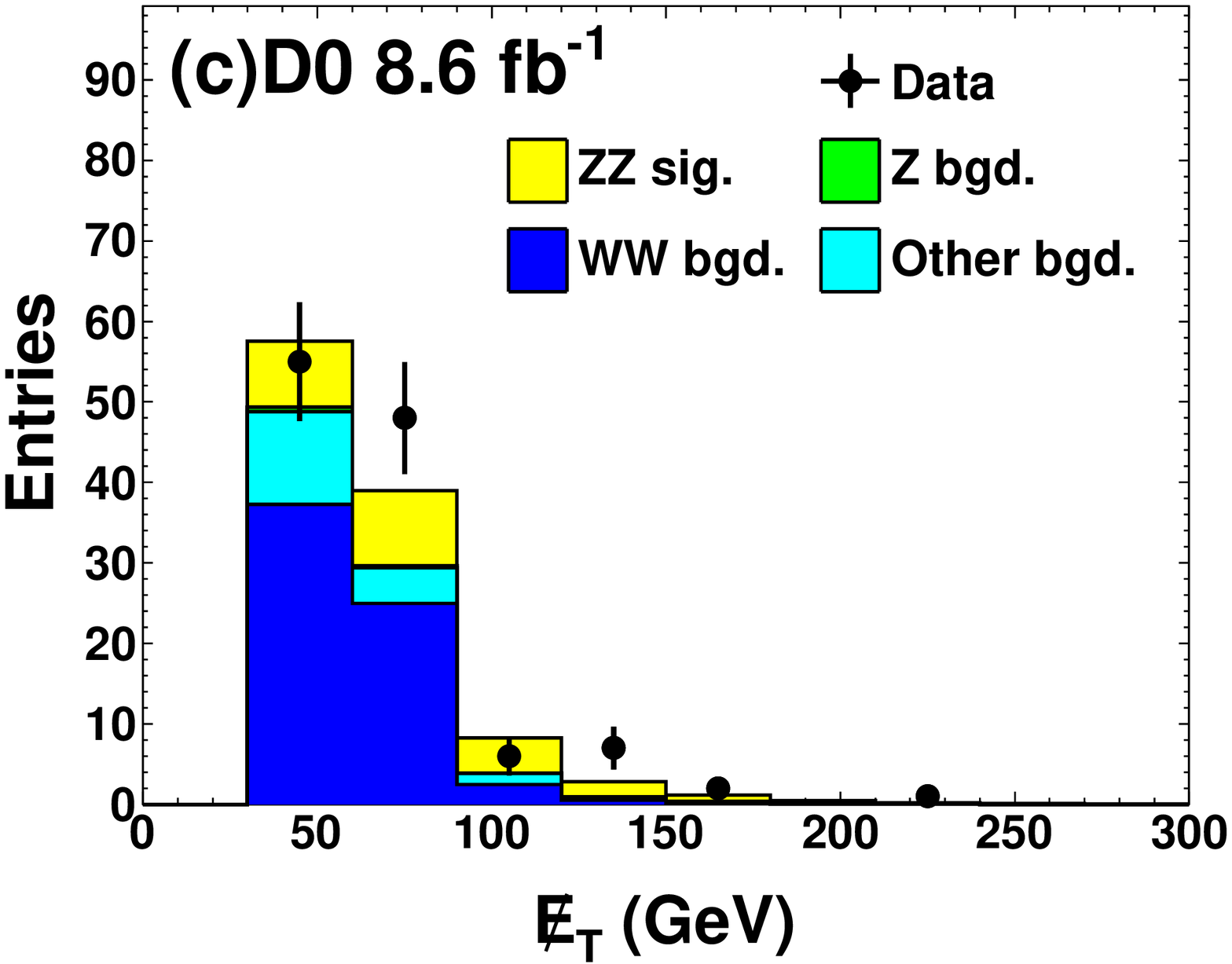}
\includegraphics[width=0.32\linewidth]{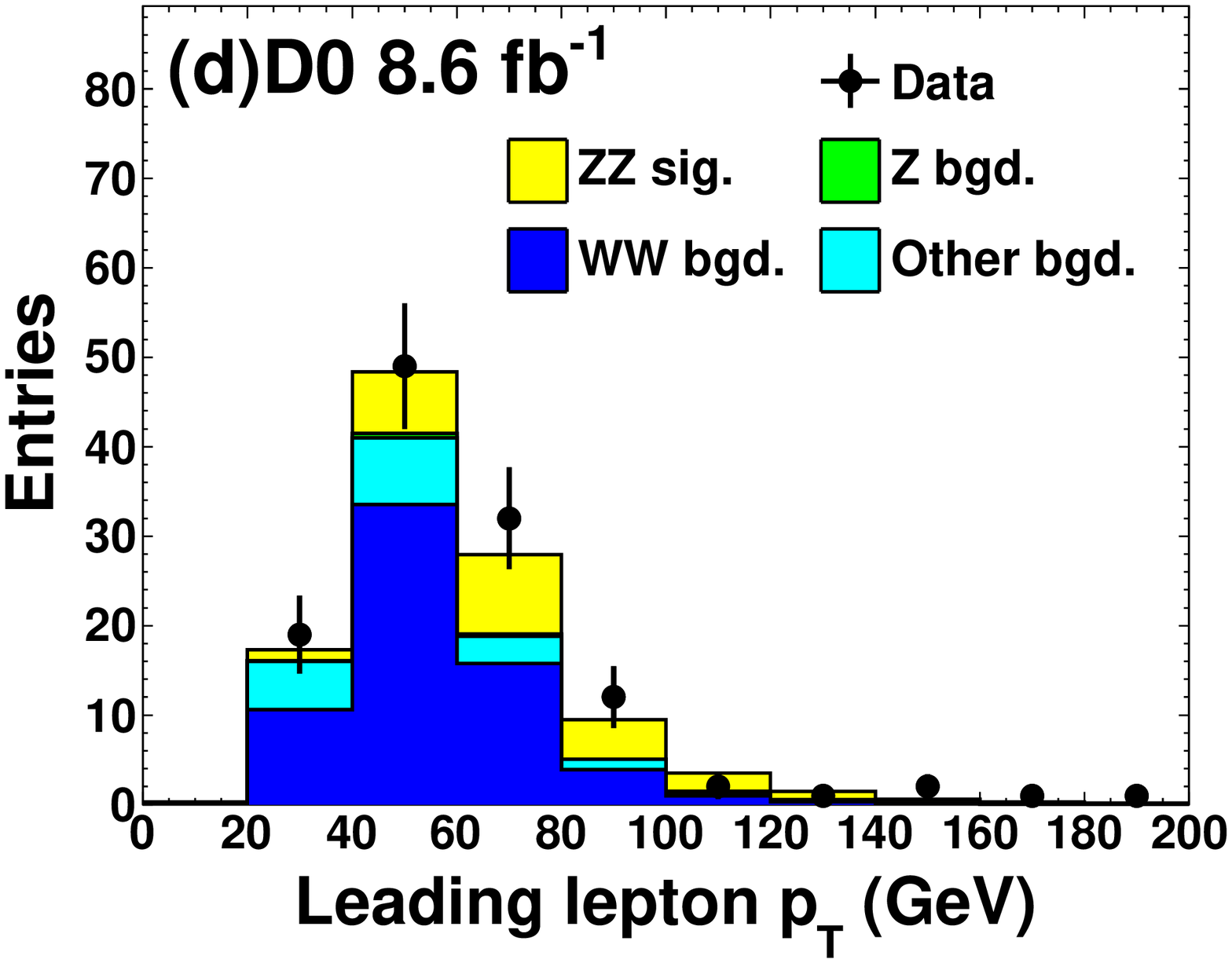}
\includegraphics[width=0.32\linewidth]{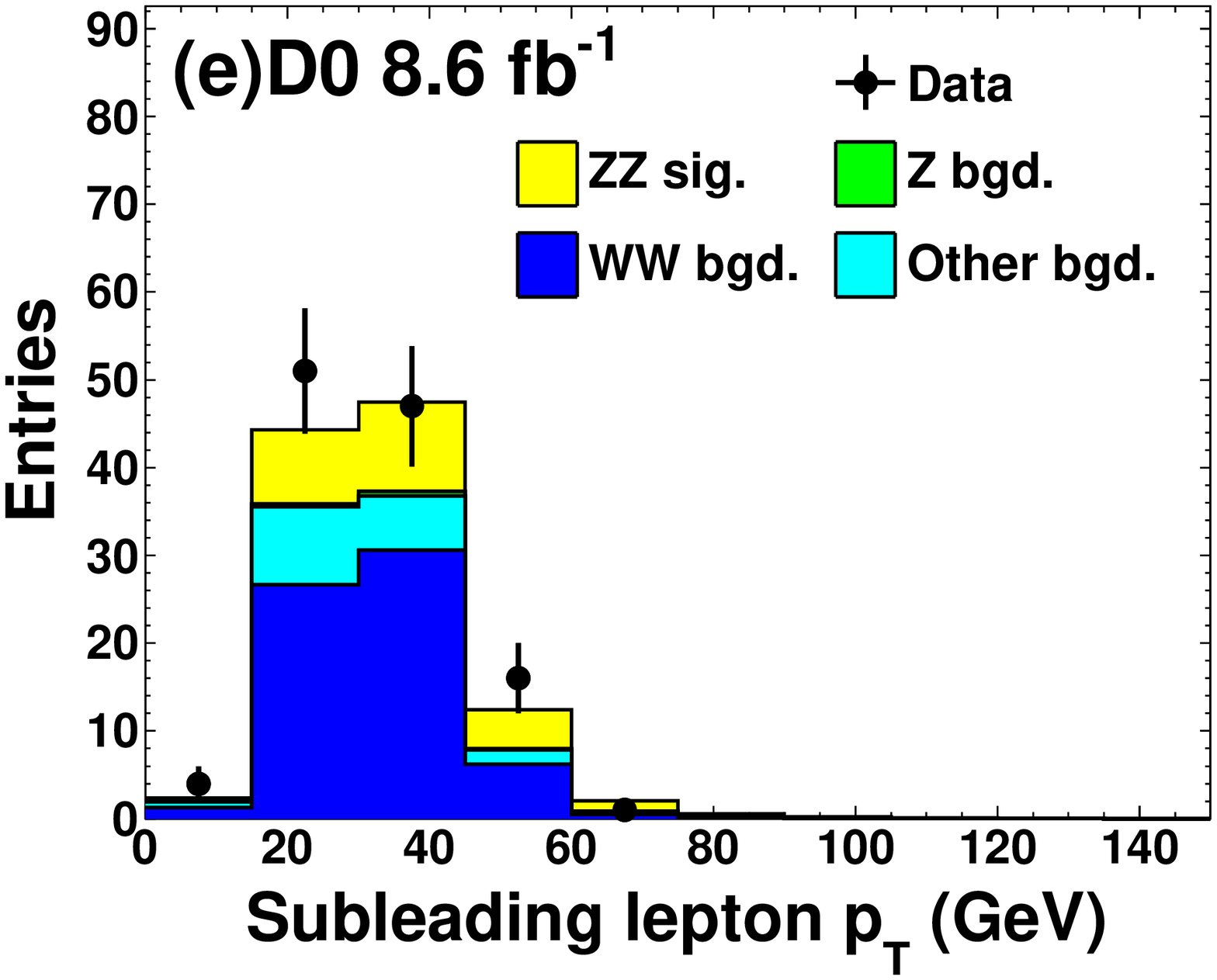}
\includegraphics[width=0.32\linewidth]{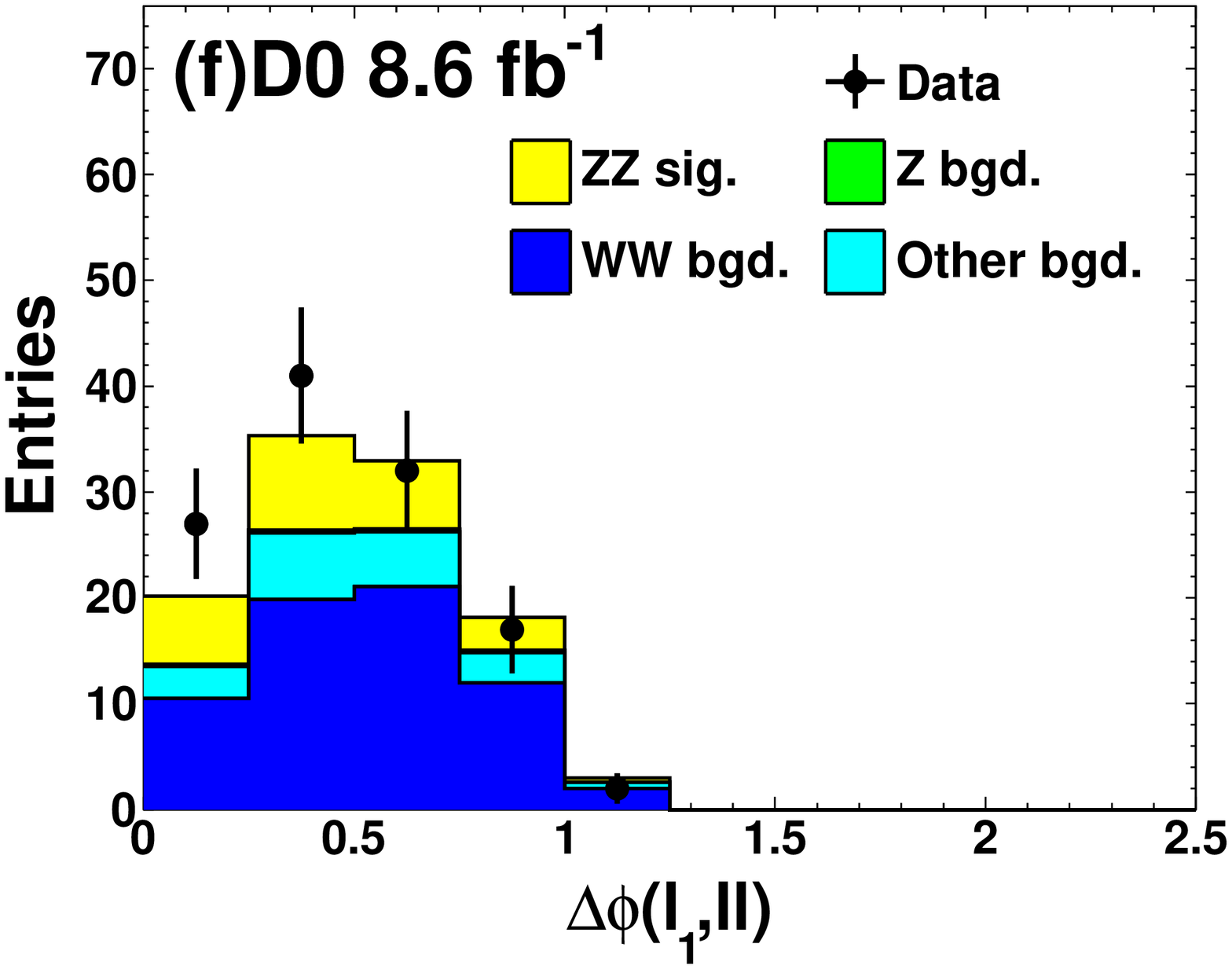}
\caption{Distributions of 
(a) $\cos(\theta^*)$, 
(b) $(\mass-m_{Z})/\sigma(\mass)$, 
(c) \met, 
transverse momenta of the (d) leading and (e) subleading lepton,
(f) the azimuthal angle between the leading lepton and the dilepton system $\Delta\phi(\ell_{1},\ell\ell)$
for the combination of \zzeenunu\ and \zzmmnunu\ candidates.
The signal normalization is as described in Section~\ref{Section:bgd_and_signal}.}
\label{Figure:ZZ_candidates}
\end{figure*}

Events are considered as \zzllnunu\ candidates if they further satisfy
\aTmissPrime\ $>$ 5~GeV (to reject \ztt) and \pTmissPrime\ $>$ 30~GeV
(to reject \zee\ and \zmm).
Tables~\ref{Table:zz_composition_diem},~\ref{Table:zz_composition_dimu} and~\ref{Table:zz_composition_emmu} show, for the three
sub-channels, the predicted yields for each process.
The yields are also presented for events that fail each requirement exclusively.
Figure~\ref{Figure:ZZ_selection_cuts} shows the \pTmissPrime\ and \mass\ distributions
before imposing their respective requirements.
A neural network (NN) is trained to discriminate \zzllnunu\ from the dominant background in the
final event sample (\wwlnulnu).
The following input variables are used: 
the $p_T$ of each lepton, the \met, the center of mass scattering angle $\cos\theta^*_{\eta}$~\cite{phistar_EPJ},
the azimuthal angle between the leading lepton and the dilepton system $\Delta\phi(\ell_{1},\ell\ell)$,
and $(\mass-m_Z)/\sigma(\mass)$ where $\sigma(\mass)$ is the estimated uncertainty on the measured
dilepton invariant mass.
Figure~\ref{Figure:ZZ_selection_cuts} also shows the
NN output distribution of the selected signal candidate events.
Separate NNs are trained for the \diem\ and \dimu\ channels,
and the \diem\ version is shown for the \emmu\ channel.
Figure~\ref{Figure:ZZ_candidates} shows a number of kinematic distributions
for the combination of \zzeenunu\ and \zzmmnunu\ candidate events.

\begin{figure*}[htbp]\centering
\includegraphics[width=0.45\linewidth]{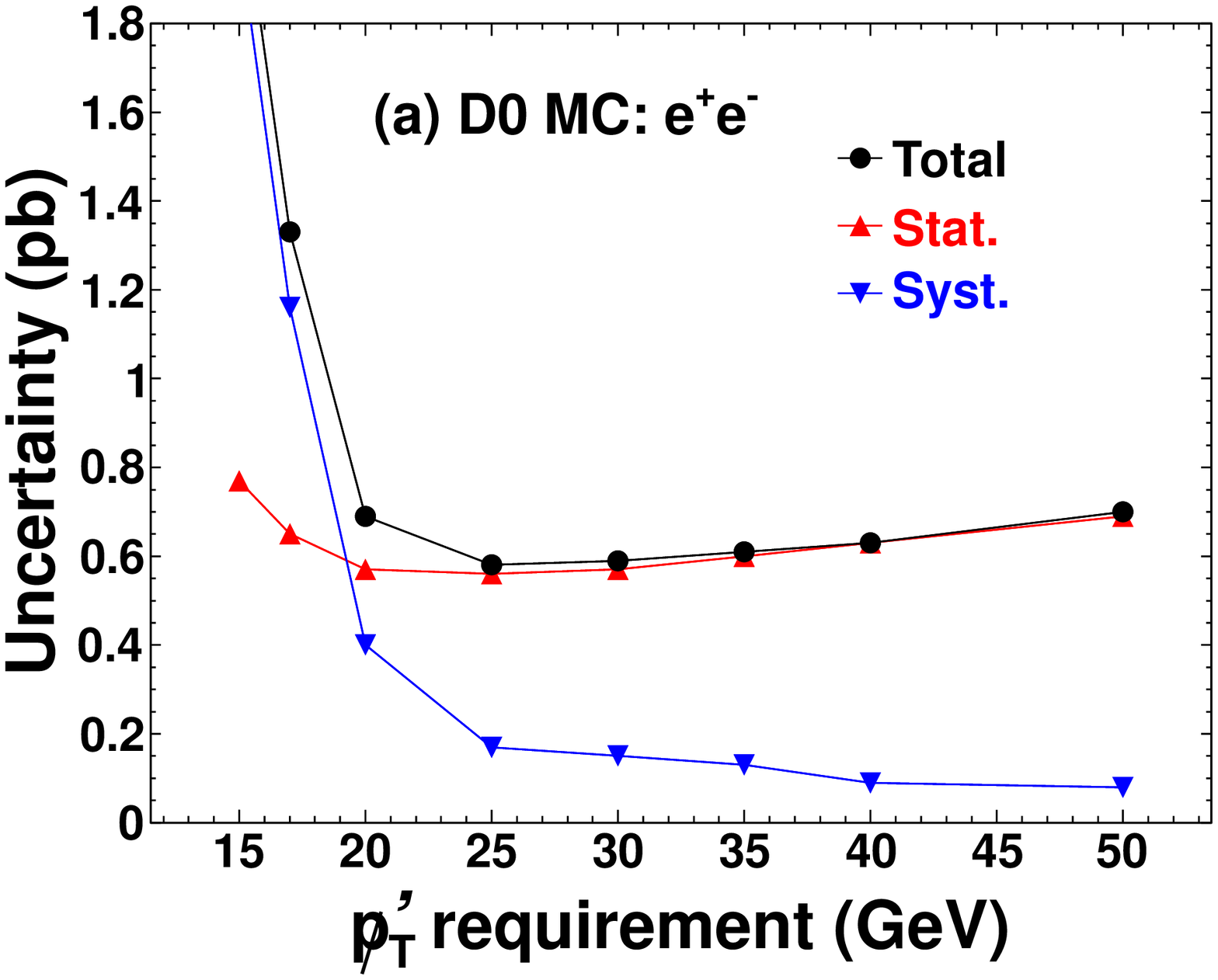}
\includegraphics[width=0.45\linewidth]{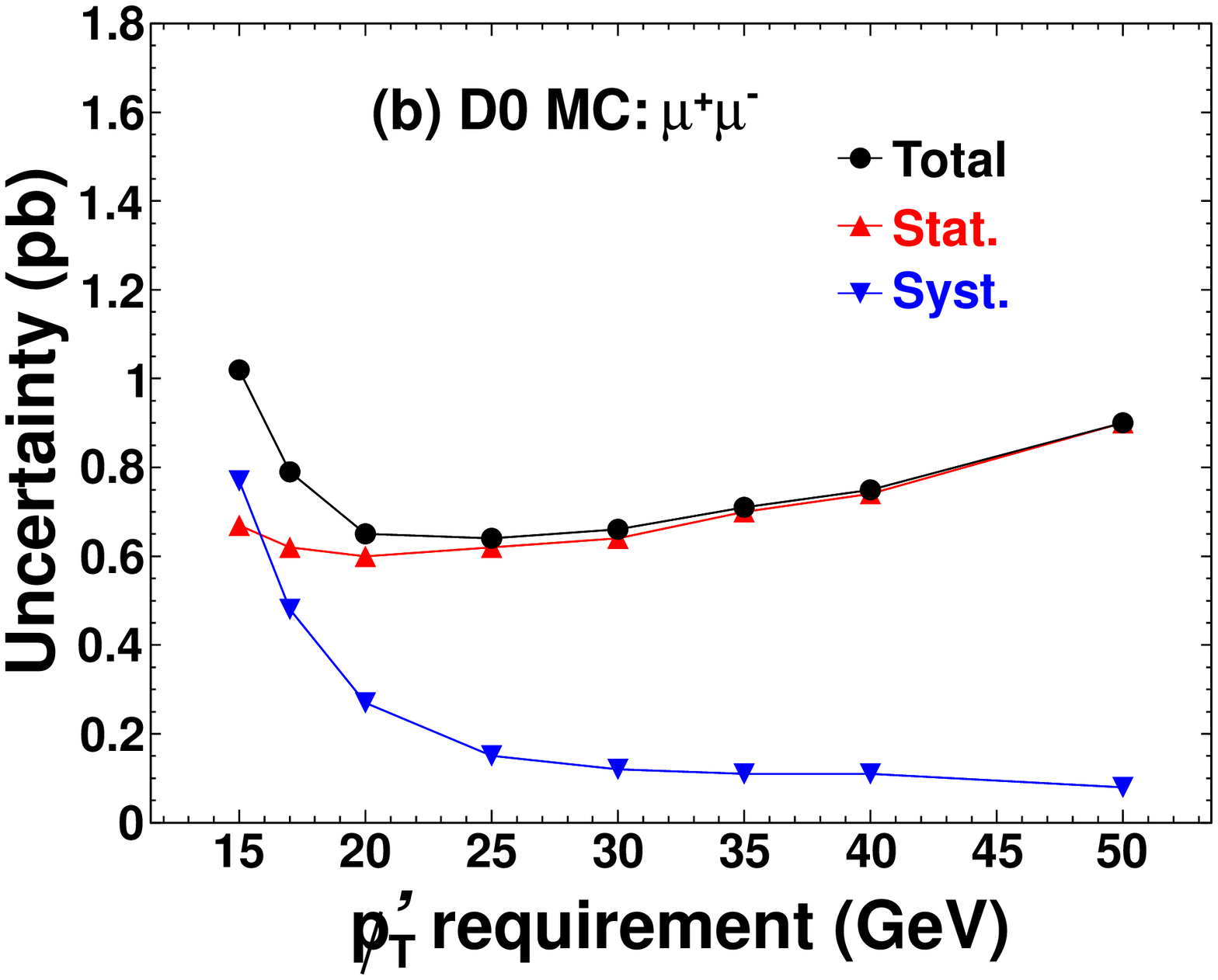}
\caption{Variation of the predicted uncertainties on the measured \ZZ\ cross section with the choice of \pTmissPrime\ requirement 
in the (a) \diem\ and (b) \dimu\ channels.}
\label{Figure:optimise_metprime_cuts}
\end{figure*}

Figure~\ref{Figure:optimise_metprime_cuts} shows how the 
predicted \ZZ\ cross section measurement uncertainty varies as 
a function of the \pTmissPrime\ requirement.
The expected systematic uncertainty rises rapidly below 25 GeV as the \zll\ background 
starts to contaminate the sample.
The requirement \pTmissPrime\ $> 30$~GeV is close to the minimum 
and is in a region where the systematic uncertainty is small.

\begin{table*}\centering
\caption{Table of predicted signal and background yields for the \zzeenunu\ signal and control regions.
The systematic uncertainties are provided for the predictions. }
\input{\plotdir/Table3.tex}
\label{Table:zz_composition_diem}
\end{table*}

\begin{table*}\centering
\caption{Table of predicted signal and background yields for the \zzmmnunu\ signal and control regions.
The systematic uncertainties are provided for the predictions. }
\input{\plotdir/Table4.tex}
\label{Table:zz_composition_dimu}
\end{table*}

\begin{table*}\centering
\caption{Table of predicted yields in the \emmu\ channel.
The systematic uncertainties are provided for the predictions. }
\input{\plotdir/Table5.tex}
\label{Table:zz_composition_emmu}
\end{table*}

\section{\label{Section:Systematics} Systematic uncertainties}

We measure the ratios of \WZ\ and \ZZ\ cross sections 
relative to the inclusive $Z$ cross section.
Lepton reconstruction, identification, and trigger efficiency uncertainties are largely cancelled
in the ratio, as are those arising from the vetoes on additional lepton candidates or other activity.
The \WZ\ analysis is sensitive to the lepton identification
efficiencies, since the signal and \zll\ samples differ by
the requirement of an additional tight quality reconstructed electron or muon.
The \ZZ\ analysis is sensitive to the modelling of the diboson $p_T$,
since requirements on the missing $p_T$ estimators are less efficient in signal events
with a large hadronic recoil.
Tables~\ref{Table:syst_wz} and~\ref{Table:syst_zzllnunu} list the sources of systematic
uncertainty on the \WZ\ and \ZZ\ cross section measurements, respectively.
We list the fractional variations in the number of predicted background events $N_{\rm bgd}$,
the acceptances (multiplied by efficiencies) for signal ($A_{\rm sig}$) and \zll\ ($A_{\ell\ell}$),
and the measured signal cross section $\sigma_{\rm sig}$.
The following sources of systematic uncertainty are considered.
\begin{itemize}
\item {\bf Beam conditions}\\
The differential distributions of the instantaneous luminosity and vertex $z$ position are varied
to cover any disagreement with the data.
\item {\bf Physics modelling}\\
The value of the $g_2$ parameter in \resbos\ is varied when determining the 
corrections that are applied to the simulated \zll\ events.
This is a model parameter that describes the intrinsic transverse momentum
of the partons within the colliding hadrons.
As a test of sensitivity to the diboson $p_T$ modelling, 
the reweighting in this variable is switched off.
\item {\bf Jet reconstruction}\\
The jet energy scale, resolution, and reconstruction efficiencies
are varied within their uncertainties.
The simulation requires additional corrections to the energy response
for jets in the IC region. 
An additional systematic uncertainty is assigned to these corrections.
The track jet reconstruction efficiency is also varied to cover an 
observed disagreement with the data.
\item {\bf Lepton momentum scale and resolution}\\
The lepton momentum scales and resolutions are varied within their uncertainties,
as are the reconstruction and identification efficiencies.
Non-Gaussian tails in the lepton momentum resolution are also considered.
\item{\bf Instrumental backgrounds}\\
The $W$+jets and $Z$+jets background normalizations are varied within
the uncertainties of the estimate from data.
All other variations on the simulation (e.g., lepton momentum scales and resolutions)
are allowed to vary the shape of these backgrounds.
Since \pythia\ does not include the matrix element for wide angle photon emission
in $W\gamma$ production, the normalization of this process is varied by a factor of two,
which is considered to be an overestimate but introduces no significant uncertainty on the
\ZZ\ cross section measurement.
\item {\bf Trigger efficiencies}\\
The trigger efficiencies are estimated to introduce a negligible uncertainty
into the cross section measurements.
\end{itemize}

\begin{table}\centering
\caption{Table of uncertainty sources in the \WZ\ cross section measurement after combining the four sub-channels.
All values are given in percent.}
\input{\plotdir/Table6.tex}
\label{Table:syst_wz}
\end{table}

\begin{table}\centering
\caption{Table of uncertainty sources in the \ZZ\ cross section measurement after combining the \diem\ and \dimu\ channels.
All values are given in percent.}
\input{\plotdir/Table7.tex}
\label{Table:syst_zzllnunu}
\end{table}

\section{\label{Section:CrossSection} Measurement of cross sections}

The ratios of the signal ($WZ$ or $ZZ$) cross sections to the inclusive $Z$ cross section
are determined as follows:
\[ \mathcal{R} = \frac{N_{\rm sig}^{\rm obs}/(A_{\rm sig} \times B_{\rm sig} \times \mathcal{L})}{N_{\ell\ell}^{\rm obs}/(A_{\ell\ell} \times B_{\ell\ell} \times \mathcal{L})}, \]

\noindent where $\mathcal{L}$ is the integrated luminosity; 
$B_{\ell\ell}$ and $B_{\rm sig}$ are the known branching fractions for \zll\
and the signal decay, respectively~\cite{PDG}.
We choose an acceptance window of $60 < \mass <  120$~GeV.

The number of observed signal events, $N_{\rm sig}^{\rm obs}$ is determined by allowing the predicted signal
yield to float such that the following likelihood function is maximized:
\begin{equation}
 L = \prod_{i = 0}^{\rm bins} \mathcal{P}(N_{i}^{\rm obs};N_{i}^{\rm pred}), 
\end{equation}
where $\mathcal{P}$ is the Poisson probability to 
observe $N_{i}^{\rm obs}$ events in the $i$th bin, given a prediction of $N_{i}^{\rm pred}$.
In the \WZ\ analysis, the $M_T$ distribution is used,
while the neural network output distribution is used in the \ZZ\ analysis.
The 68\% C.L.~interval on the signal yield is defined by $\delta (\ln L) = 0.5$,
with respect to the maximum of $\ln L$.

Table~\ref{Table:acceptances} lists, for the six different sub-channels, the 
ratios of inclusive $Z$ and signal acceptances that are estimated from the 
simulation.
Table~\ref{Table:R_table} lists the measured $\mathcal{R}$ values.
The $p$-values for consistency of the different sub-channels 
are 54\% and 11\% for the \WZ\ and \ZZ\ analyses, respectively,
evaluated using a $\chi^{2}$ test.
For the combination of respective sub-channels, we measure:

\[ \mathcal{R}(WZ) = \WZRatVal \pm \WZRatStat \mathrm{(stat)} \pm \WZRatSyst \mathrm{(syst)} (\times 10^{-3}), \]
\[ \mathcal{R}(ZZ) = \ZZRatVal \pm \ZZRatStat \mathrm{(stat)} \pm \ZZRatSyst \mathrm{(syst)} (\times 10^{-3}). \]

A theoretical calculation of the $Z$ cross section can be used 
to translate these into signal cross section measurements.
The product of the cross section and branching fraction for 
\zll\ (one lepton flavor) is calculated 
using a modified version of the next-to-NLO (NNLO) code of Ref.~\cite{Hamberg_1991}
with the MRST2004 NNLO PDFs~\cite{MRST_2004}.
Since this code excludes the $\gamma^*$ and $Z/\gamma^*$ interference,
a correction factor is determined using 
\pythia\ and the NLO event generator \mcatnlo~\cite{mcatnlo}.
For $60 < \mass < 120$ GeV, the result is,

\[ \sigma(p\bar{p} \rightarrow Z/\gamma^*) \times B_{\ell\ell} = \sigmaZ\ ^{+\sigmaZErrUP}_{-\sigmaZErrDN}~\mathrm{pb}, \]

\noindent  where the uncertainties arise from variations in the PDFs and the renormalization and factorization scales,
and with $B_{\ell\ell} =  3.3658 \pm 0.0023$~\%~\cite{PDG}.
The measured $WZ$ cross section with $60 < \mass < 120$ GeV is

\[ \sigma(p\bar{p} \rightarrow WZ)= \WZValue \pm \WZStat \mathrm{(stat)} ^{+\WZSystUP}_{-\WZSystDN} \mathrm{(syst)}~\mathrm{pb}. \]

\noindent This result is slightly higher than, but still consistent with a prediction of $\WZtheoryVAL \pm \WZtheoryERR$~pb 
from the NLO program {\sc MCFM}~\cite{MCFM} with the
MSTW2008 NLO PDFs~\cite{MSTW_2008}, and setting the renormalization and factorization scales equal to $m_W+m_Z$.
The measured $ZZ$ cross section with $60 < \mass < 120$ GeV is 

\[ \sigma(p\bar{p} \rightarrow ZZ)= \ZZValue \pm \ZZStat \mathrm{(stat)}  ^{+\ZZSystUP}_{-\ZZSystDN} \mathrm{(syst)}~\mathrm{pb}. \]

This can be compared to a prediction of $\ZZtheoryVAL \pm \ZZtheoryERR$~pb from MCFM setting the renormalization and factorization scales equal to $2m_Z$.
For comparing to and combining with  previous measurements
it is more convenient to correct the cross sections
for the presence of diagrams involving $\gamma^*$.
A correction of +3.4\% is obtained by comparing $ZZ$ cross sections
with and without $\gamma^*$ and $Z/\gamma^*$ interference from \mcfm~\cite{MCFM}.
Combining this corrected $ZZ$ cross section with a previous D0 measurement~\cite{dzero_zzllll} in the \zzllll\ channel yields
\[ \sigma(p\bar{p}~\rightarrow~ZZ)~=~\ZZCombVal^{+\ZZCombStatUP}_{-\ZZCombStatDN}~\mathrm{(stat)}~^{+\ZZCombSystUP}_{-\ZZCombSystDN}~\mathrm{(syst)}~\mathrm{pb}.\]

\begin{table}
\centering
\input{\plotdir/Table8.tex}
\caption{
Table of acceptance ratios for the different sub-channels,
where the quoted uncertainties are systematic.
Also shown are the numbers of observed events at the dilepton selection
stage, $N_{\ell\ell}^{\rm obs}$.
}
\label{Table:acceptances}
\end{table}

\begin{table}[hbt]
\input{\plotdir/Table9.tex}
\caption{Table of $\mathcal{R}$ values measured for each of 
the sub-channels, where the uncertainties correspond to statistical
and systematic components added in quadrature.
}
\label{Table:R_table}
\end{table}

\section{\label{Section:Conclusions} Conclusions}

\begin{figure}
\centering
\includegraphics[width=0.95\linewidth]{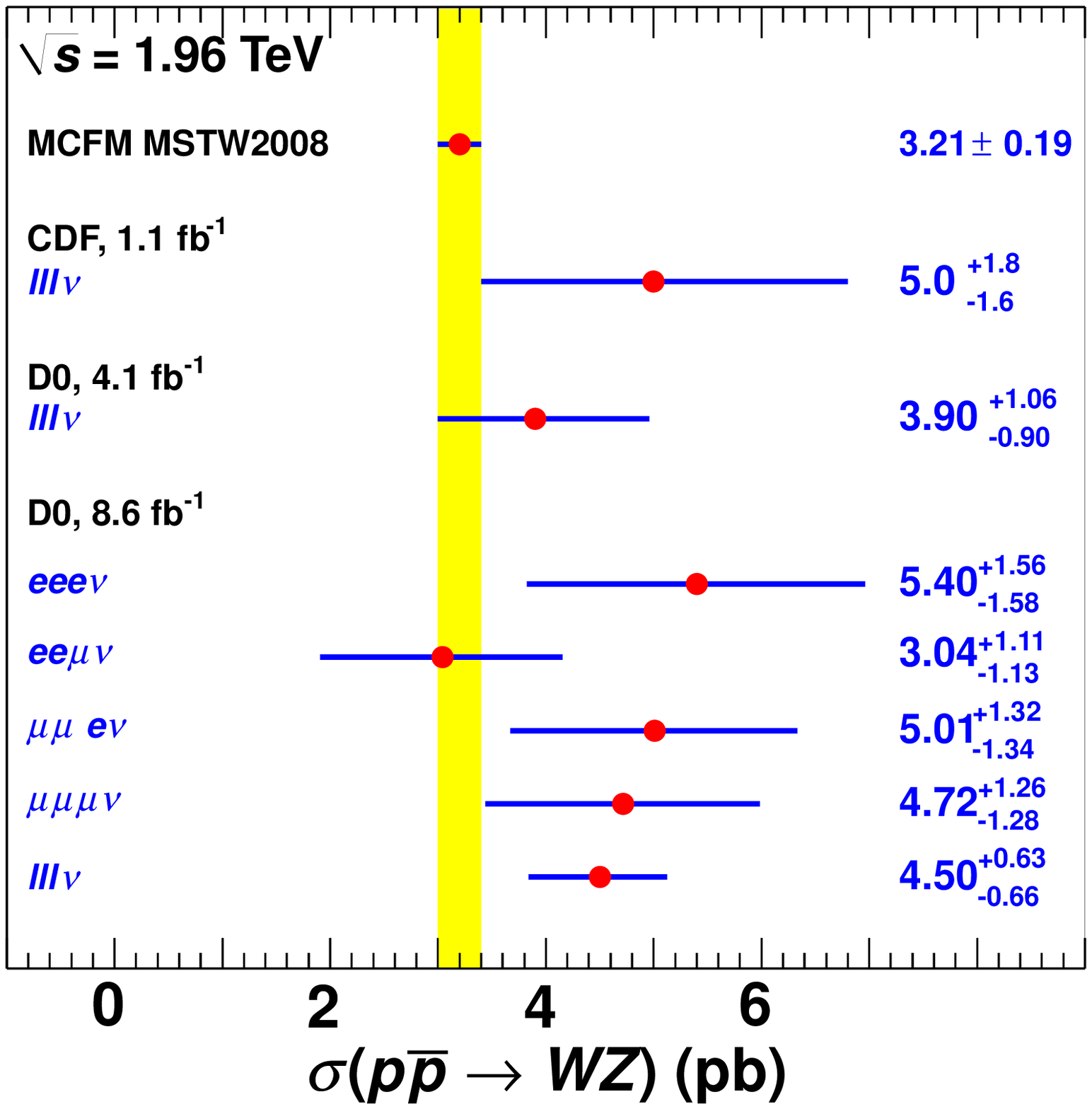}
\includegraphics[width=0.95\linewidth]{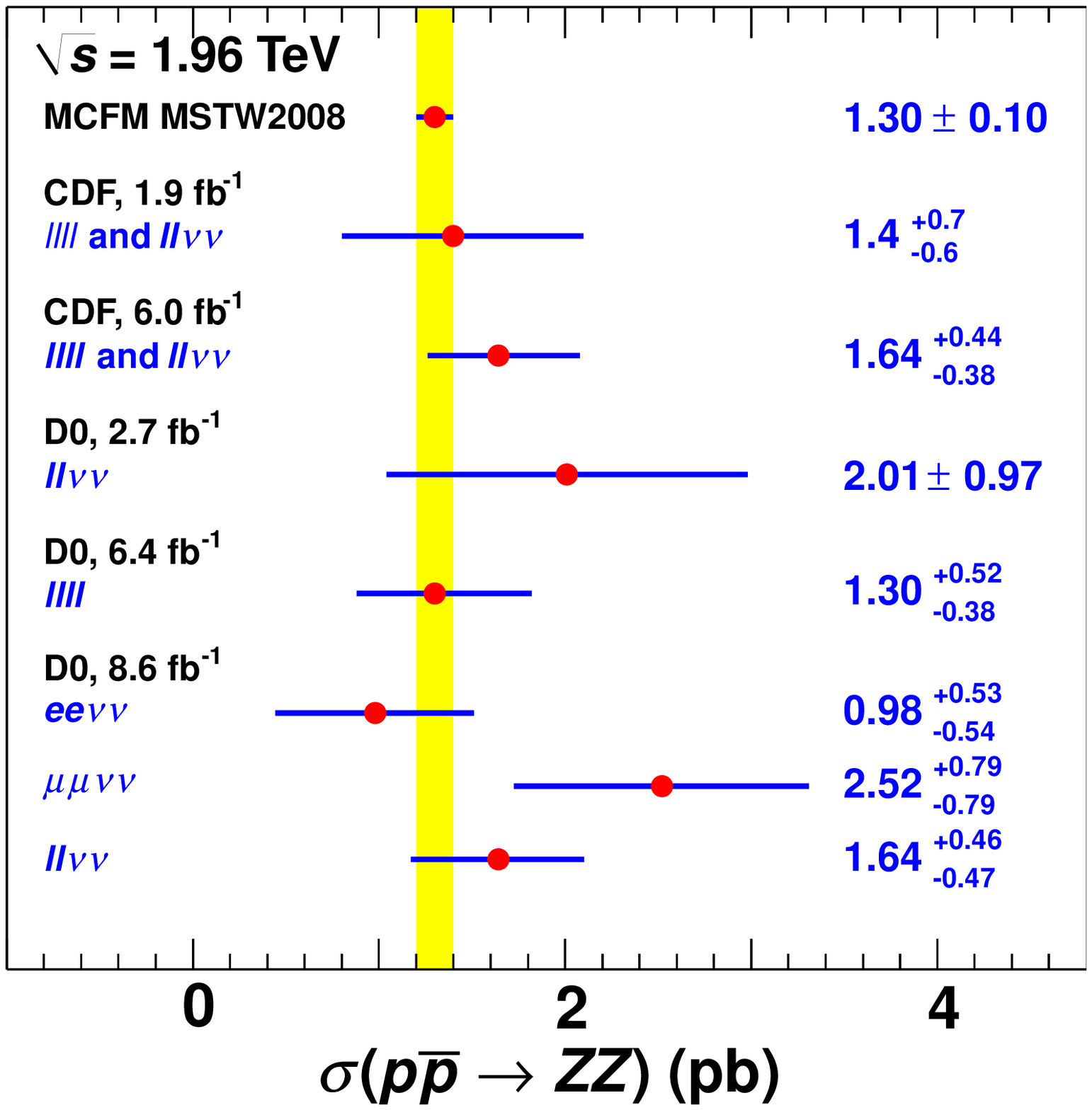}
\caption{
Comparison of the measured \ZZ\ and \WZ\ cross sections 
with SM predictions, and with previous measurements
in leptonic final states.
The $ZZ$ cross section measured by D0 in the \zzllll\ channel
has been corrected to the same dilepton invariant mass range
as considered in this analysis.
}
\label{Figure:xsec_comp_unblind}
\end{figure}

We measure the production cross sections
for the processes $p\bar{p} \rightarrow WZ \rightarrow l\nu l^+l^-$ 
and $p\bar{p} \rightarrow ZZ \rightarrow \nu\bar{\nu} l^+l^-$,
using 8.6 \invfb\ of integrated luminosity collected by the D0 experiment at the Fermilab Tevatron collider.
For decay channels involving electrons and muons,
we observe agreement between the different sub-channels
as can be seen in Fig.~\ref{Figure:xsec_comp_unblind}.
Combining the sub-channels yields a $WZ$ cross section of
\WZValue~$^{+\WZTotUP}_{-\WZTotDN}$~pb,
which is slightly above, but still consistent with a
standard model prediction of $\WZtheoryVAL \pm \WZtheoryERR$~pb.
The $ZZ$ cross section is measured to be
\ZZValue~$\pm$~\ZZTot~pb,
which is also in 
agreement with a standard model prediction of $\ZZtheoryVAL \pm \ZZtheoryERR$~pb.
These are the most precise measurements to date of the 
\WZ\ and \ZZ\ cross sections in $p\bar{p}$ collisions at 
$\sqrt{s} = 1.96$~TeV.
Correcting for the contribution from $\gamma^*$ and $Z/\gamma^*$ interference
and combining with a previous measurement in the \llll\ channel yields
a $ZZ$ cross section of 
$\ZZCombVal^{+\ZZCombTotUP}_{-\ZZCombTotDN}$~pb.

\input acknowledgement.tex   

\bibliographystyle{h-physrev3.bst}
\bibliography{bib}

\end{document}

%% file: Commands.tex
\def\mcfm{{\sc mcfm}}
\def\mcatnlo{{\sc mc@nlo}}
\def\pythia{{\sc pythia}}

\def\resbos{{\sc resbos}}

\def\diem{$e^+e^-$}
\def\dimu{$\mu^+\mu^-$}
\def\emmu{$e^\pm\mu^\mp$}

\newcommand{\DR}{\mbox{$\Delta\mathcal{R}$}}
\newcommand{\mass}{\mbox{$M_{\ell \ell}$}}

\newcommand{\isohcfour}{\mbox{$\mathcal{I}_{\rm trk}^{\rm elec}$}}
\newcommand{\itrkmuon}{\mbox{$\mathcal{I}_{\rm trk}^{\rm muon}$}}
\newcommand{\icalmuon}{\mbox{$\mathcal{I}_{\rm cal}^{\rm muon}$}}

\newcommand{\llnunu}{\mbox{$\ell^+\ell^-\nu\bar{\nu}$}}

\newcommand{\llll}{\mbox{$\ell^+\ell^-\ell^+\ell^-$}}
\newcommand{\zzllll}{\mbox{$ZZ$ $\rightarrow$ $\ell^+\ell^-\ell^+\ell^-$}}
\newcommand{\zzeenunu}{\mbox{$ZZ$ $\rightarrow$ $e^+e^-\nu\bar{\nu}$}}
\newcommand{\zzmmnunu}{\mbox{$ZZ$ $\rightarrow$ $\mu^+\mu^-\nu\bar{\nu}$}}
\newcommand{\zzllnunu}{\mbox{$ZZ$ $\rightarrow$ $\ell^+\ell^-\nu\bar{\nu}$}}
\newcommand{\zznunull}{\mbox{$ZZ$ $\rightarrow$ $\ell^+\ell^-\nu\bar{\nu}$}}
\newcommand{\wwlnulnu}{\mbox{$WW$ $\rightarrow$ $\ell^+\nu \ell^-\bar{\nu}$}}
\newcommand{\WZ}{\mbox{$WZ$}}
\newcommand{\ZZ}{\mbox{$ZZ$}}

\newcommand{\wzlnull}{\mbox{$WZ$ $\rightarrow$ $\ell \nu \ell^+\ell^-$}}
\newcommand{\lnull}{\mbox{$\ell\nu \ell^+\ell^-$}}
\newcommand{\wzenuee}{\mbox{$WZ$ $\rightarrow$ $e\nu e^+e^-$}}
\newcommand{\wzenumm}{\mbox{$WZ$ $\rightarrow$ $e\nu \mu^+\mu^-$}}
\newcommand{\wzmnuee}{\mbox{$WZ$ $\rightarrow$ $\mu\nu e^+e^-$}}
\newcommand{\wzmnumm}{\mbox{$WZ$ $\rightarrow$ $\mu\nu \mu^+\mu^-$}}
\newcommand{\zgam}{\mbox{$Z\gamma$ $\rightarrow$ $\ell^+\ell^-\gamma$}}
\newcommand{\wgam}{\mbox{$W\gamma$ $\rightarrow$ $\ell\nu\gamma$}}

\newcommand{\ttbar}{\mbox{$t\bar{t}\rightarrow \ell^+\ell^-\nu\bar{\nu}b\bar{b}$}}

\newcommand{\at}{\mbox{$a_T$}}
\newcommand{\al}{\mbox{$a_L$}}

\newcommand{\pt}{\mbox{$p_T$}}

\newcommand{\zll}{\mbox{$Z \rightarrow \ell^+\ell^-$}}
\newcommand{\zee}{\mbox{$Z \rightarrow e^+e^-$}}
\newcommand{\zmm}{\mbox{$Z \rightarrow \mu^+\mu^-$}}
\newcommand{\ztt}{\mbox{$Z \rightarrow \tau^+\tau^-$}}

\newcommand{\wenu}{\mbox{$W \rightarrow e\nu$}}
\newcommand{\wmnu}{\mbox{$W \rightarrow \mu\nu$}}

\newcommand{\met}{\mbox{$E\kern -0.6em/_{T}$}}
\newcommand{\metconstr}{\mbox{$E\kern -0.6em/_{T}^{\ \prime}$}}

\newcommand{\invfb}{fb\mbox{$^{-1}$}}

\newcommand{\pTmissTITLE}{\mbox{$p\kern -0.4em/_T$}}
\newcommand{\aTmissTITLE}{\mbox{$a\kern -0.4em/_T$}}
\newcommand{\aLmissTITLE}{\mbox{$a\kern -0.4em/_L$}}

\newcommand{\pTmissPrime}{\mbox{$p\kern -0.4em/_T'$}}
\newcommand{\aTmissPrime}{\mbox{$a\kern -0.5em/_T'$}}
\newcommand{\aLmissPrime}{\mbox{$a\kern -0.5em/_L'$}}

\newcommand{\aTmiss}{\mbox{$a\kern -0.6em/_T$}}

\newcommand{\aTdilep}{\mbox{$a_T^{\ell \ell}$}}
\newcommand{\aTjets}{\mbox{$a_T^{\rm jets}$}}

\newcommand{\aTrecoil}{\mbox{$a_T^{\rm recoil}$}}

\newcommand{\aTtrkjets}{\mbox{$a_T^{\rm trkjets}$}}

\newcommand{\aTdelta}{\mbox{$a_T^{\rm \delta}$}}

\newcommand{\aLmiss}{\mbox{$a\kern -0.6em/_L$}}
\newcommand{\aLdilep}{\mbox{$a_L^{\ell \ell}$}}
\newcommand{\aLrecoil}{\mbox{$a_L^{\rm recoil}$}}

\newcommand{\aLtrkjets}{\mbox{$a_L^{\rm trkjets}$}}

\newcommand{\aLdelta}{\mbox{$a_L^{\rm \delta}$}}

\newcommand{\pTmiss}{\mbox{$p\kern -0.4em/_T$}}
\newcommand{\pTdilep}{\mbox{$p_T^{\ell \ell}$}}
\newcommand{\pTrecoil}{\mbox{$p_T^{\rm recoil}$}}

\newcommand{\pTtrkjets}{\mbox{$p_T^{\rm trkjets}$}}

\newcommand{\pTdelta}{\mbox{$p_T^{\rm \delta}$}}

%% file: author_list.tex
%
\affiliation{Universidad de Buenos Aires, Buenos Aires, Argentina}
\affiliation{LAFEX, Centro Brasileiro de Pesquisas F{\'\i}sicas, Rio de Janeiro, Brazil}
\affiliation{Universidade do Estado do Rio de Janeiro, Rio de Janeiro, Brazil}
\affiliation{Universidade Federal do ABC, Santo Andr\'e, Brazil}
\affiliation{Instituto de F\'{\i}sica Te\'orica, Universidade Estadual Paulista, S\~ao Paulo, Brazil}
\affiliation{University of Science and Technology of China, Hefei, People's Republic of China}
\affiliation{Universidad de los Andes, Bogot\'{a}, Colombia}
\affiliation{Charles University, Faculty of Mathematics and Physics, Center for Particle Physics, Prague, Czech Republic}
\affiliation{Czech Technical University in Prague, Prague, Czech Republic}
\affiliation{Center for Particle Physics, Institute of Physics, Academy of Sciences of the Czech Republic, Prague, Czech Republic}
\affiliation{Universidad San Francisco de Quito, Quito, Ecuador}
\affiliation{LPC, Universit\'e Blaise Pascal, CNRS/IN2P3, Clermont, France}
\affiliation{LPSC, Universit\'e Joseph Fourier Grenoble 1, CNRS/IN2P3, Institut National Polytechnique de Grenoble, Grenoble, France}
\affiliation{CPPM, Aix-Marseille Universit\'e, CNRS/IN2P3, Marseille, France}
\affiliation{LAL, Universit\'e Paris-Sud, CNRS/IN2P3, Orsay, France}
\affiliation{LPNHE, Universit\'es Paris VI and VII, CNRS/IN2P3, Paris, France}
\affiliation{CEA, Irfu, SPP, Saclay, France}
\affiliation{IPHC, Universit\'e de Strasbourg, CNRS/IN2P3, Strasbourg, France}
\affiliation{IPNL, Universit\'e Lyon 1, CNRS/IN2P3, Villeurbanne, France and Universit\'e de Lyon, Lyon, France}
\affiliation{III. Physikalisches Institut A, RWTH Aachen University, Aachen, Germany}
\affiliation{Physikalisches Institut, Universit{\"a}t Freiburg, Freiburg, Germany}
\affiliation{II. Physikalisches Institut, Georg-August-Universit{\"a}t G\"ottingen, G\"ottingen, Germany}
\affiliation{Institut f{\"u}r Physik, Universit{\"a}t Mainz, Mainz, Germany}
\affiliation{Ludwig-Maximilians-Universit{\"a}t M{\"u}nchen, M{\"u}nchen, Germany}
\affiliation{Fachbereich Physik, Bergische Universit{\"a}t Wuppertal, Wuppertal, Germany}
\affiliation{Panjab University, Chandigarh, India}
\affiliation{Delhi University, Delhi, India}
\affiliation{Tata Institute of Fundamental Research, Mumbai, India}
\affiliation{University College Dublin, Dublin, Ireland}
\affiliation{Korea Detector Laboratory, Korea University, Seoul, Korea}
\affiliation{CINVESTAV, Mexico City, Mexico}
\affiliation{Nikhef, Science Park, Amsterdam, the Netherlands}
\affiliation{Radboud University Nijmegen, Nijmegen, the Netherlands and Nikhef, Science Park, Amsterdam, the Netherlands}
\affiliation{Joint Institute for Nuclear Research, Dubna, Russia}
\affiliation{Institute for Theoretical and Experimental Physics, Moscow, Russia}
\affiliation{Moscow State University, Moscow, Russia}
\affiliation{Institute for High Energy Physics, Protvino, Russia}
\affiliation{Petersburg Nuclear Physics Institute, St. Petersburg, Russia}
\affiliation{Instituci\'{o} Catalana de Recerca i Estudis Avan\c{c}ats (ICREA) and Institut de F\'{i}sica d'Altes Energies (IFAE), Barcelona, Spain}
\affiliation{Stockholm University, Stockholm and Uppsala University, Uppsala, Sweden}
\affiliation{Lancaster University, Lancaster LA1 4YB, United Kingdom}
\affiliation{Imperial College London, London SW7 2AZ, United Kingdom}
\affiliation{The University of Manchester, Manchester M13 9PL, United Kingdom}
\affiliation{University of Arizona, Tucson, Arizona 85721, USA}
\affiliation{University of California Riverside, Riverside, California 92521, USA}
\affiliation{Florida State University, Tallahassee, Florida 32306, USA}
\affiliation{Fermi National Accelerator Laboratory, Batavia, Illinois 60510, USA}
\affiliation{University of Illinois at Chicago, Chicago, Illinois 60607, USA}
\affiliation{Northern Illinois University, DeKalb, Illinois 60115, USA}
\affiliation{Northwestern University, Evanston, Illinois 60208, USA}
\affiliation{Indiana University, Bloomington, Indiana 47405, USA}
\affiliation{Purdue University Calumet, Hammond, Indiana 46323, USA}
\affiliation{University of Notre Dame, Notre Dame, Indiana 46556, USA}
\affiliation{Iowa State University, Ames, Iowa 50011, USA}
\affiliation{University of Kansas, Lawrence, Kansas 66045, USA}
\affiliation{Kansas State University, Manhattan, Kansas 66506, USA}
\affiliation{Louisiana Tech University, Ruston, Louisiana 71272, USA}
\affiliation{Boston University, Boston, Massachusetts 02215, USA}
\affiliation{Northeastern University, Boston, Massachusetts 02115, USA}
\affiliation{University of Michigan, Ann Arbor, Michigan 48109, USA}
\affiliation{Michigan State University, East Lansing, Michigan 48824, USA}
\affiliation{University of Mississippi, University, Mississippi 38677, USA}
\affiliation{University of Nebraska, Lincoln, Nebraska 68588, USA}
\affiliation{Rutgers University, Piscataway, New Jersey 08855, USA}
\affiliation{Princeton University, Princeton, New Jersey 08544, USA}
\affiliation{State University of New York, Buffalo, New York 14260, USA}
\affiliation{Columbia University, New York, New York 10027, USA}
\affiliation{University of Rochester, Rochester, New York 14627, USA}
\affiliation{State University of New York, Stony Brook, New York 11794, USA}
\affiliation{Brookhaven National Laboratory, Upton, New York 11973, USA}
\affiliation{Langston University, Langston, Oklahoma 73050, USA}
\affiliation{University of Oklahoma, Norman, Oklahoma 73019, USA}
\affiliation{Oklahoma State University, Stillwater, Oklahoma 74078, USA}
\affiliation{Brown University, Providence, Rhode Island 02912, USA}
\affiliation{University of Texas, Arlington, Texas 76019, USA}
\affiliation{Southern Methodist University, Dallas, Texas 75275, USA}
\affiliation{Rice University, Houston, Texas 77005, USA}
\affiliation{University of Virginia, Charlottesville, Virginia 22901, USA}
\affiliation{University of Washington, Seattle, Washington 98195, USA}
\author{V.M.~Abazov} \affiliation{Joint Institute for Nuclear Research, Dubna, Russia}
\author{B.~Abbott} \affiliation{University of Oklahoma, Norman, Oklahoma 73019, USA}
\author{B.S.~Acharya} \affiliation{Tata Institute of Fundamental Research, Mumbai, India}
\author{M.~Adams} \affiliation{University of Illinois at Chicago, Chicago, Illinois 60607, USA}
\author{T.~Adams} \affiliation{Florida State University, Tallahassee, Florida 32306, USA}
\author{G.D.~Alexeev} \affiliation{Joint Institute for Nuclear Research, Dubna, Russia}
\author{G.~Alkhazov} \affiliation{Petersburg Nuclear Physics Institute, St. Petersburg, Russia}
\author{A.~Alton$^{a}$} \affiliation{University of Michigan, Ann Arbor, Michigan 48109, USA}
\author{G.~Alverson} \affiliation{Northeastern University, Boston, Massachusetts 02115, USA}
\author{M.~Aoki} \affiliation{Fermi National Accelerator Laboratory, Batavia, Illinois 60510, USA}
% ...
%
%
\collaboration{The D0 Collaboration\footnote{with visitors from
$^{a}$Augustana College, Sioux Falls, SD, USA,
$^{b}$The University of Liverpool, Liverpool, UK,
$^{c}$UPIITA-IPN, Mexico City, Mexico,
$^{d}$DESY, Hamburg, Germany,
$^{e}$SLAC, Menlo Park, CA, USA,
$^{f}$University College London, London, UK,
$^{g}$Centro de Investigacion en Computacion - IPN, Mexico City, Mexico,
$^{h}$ECFM, Universidad Autonoma de Sinaloa, Culiac\'an, Mexico,
and 
$^{i}$Universit{\"a}t Bern, Bern, Switzerland.
$^{\ddag}$Deceased.
}} \noaffiliation
\vskip 0.25cm

%% file: plots/Table1.tex
\begin{tabular}{  l  ccc  ccc  }
\hline \hline
&\multicolumn{3}{c}{$e^+e^-e^{\pm}$}&\multicolumn{3}{c}{$e^+e^-\mu^{\pm}$}\\ &Accepted&fail $\met$ & fail $\mass$ &Accepted&fail $\met$ & fail $\mass$ \\\hline \hline
\rule{0pt}{2.6ex}\rule[-1.2ex]{0pt}{0pt}\zll & 0.3 $\pm$ 0.1 & 9 $\pm$ 1 & 0.0 $\pm$ 0.0 & 3 $\pm$ 1 & 7 $\pm$ 2 & 0.1 $\pm$ 0.0\\
\zgam & 0.6 $\pm$ 0.2 & 10.1 $\pm$ 0.6 & 0.0 $\pm$ 0.0 & 0.1 $\pm$ 0.0 & 0.1 $\pm$ 0.0 & 0.0 $\pm$ 0.0\\
\zzllll & 0.6 $\pm$ 0.1 & 1.0 $\pm$ 0.1 & 0.0 $\pm$ 0.0 & 1.5 $\pm$ 0.0 & 0.7 $\pm$ 0.0 & 0.1 $\pm$ 0.0\\
\ttbar & 0.0 $\pm$ 0.0 & 0.0 $\pm$ 0.0 & 0.0 $\pm$ 0.0 & 0.0 $\pm$ 0.0 & 0.0 $\pm$ 0.0 & 0.0 $\pm$ 0.0\\
\zzllnunu & 0.0 $\pm$ 0.0 & 0.0 $\pm$ 0.0 & 0.0 $\pm$ 0.0 & 0.0 $\pm$ 0.0 & 0.0 $\pm$ 0.0 & 0.0 $\pm$ 0.0\\
\wwlnulnu & 0.0 $\pm$ 0.0 & 0.0 $\pm$ 0.0 & 0.0 $\pm$ 0.0 & 0.0 $\pm$ 0.0 & 0.0 $\pm$ 0.0 & 0.0 $\pm$ 0.0\\
Predicted background & 1.5 $\pm$ 0.4 & 20 $\pm$ 1 & 0.1 $\pm$ 0.0 & 5 $\pm$ 1 & 7 $\pm$ 2 & 0.2 $\pm$ 0.1\\[4pt]
\wzlnull & 9.9 $\pm$ 0.2 & 1.7 $\pm$ 0.1 & 0.1 $\pm$ 0.1 & 13.9 $\pm$ 0.3 & 2.2 $\pm$ 0.1 & 0.6 $\pm$ 0.1\\[4pt]
Predicted total & 11.4 $\pm$ 0.4 & 21 $\pm$ 1 & 0.2 $\pm$ 0.1 & 19 $\pm$ 1 & 9 $\pm$ 2 & 0.8 $\pm$ 0.2\\[4pt]
Observed & 17 & 32 & 0 & 17 & 6 & 1\\
\hline
\hline
\end{tabular}

%% file: plots/Table2.tex
\begin{tabular}{  l  ccc  ccc  }
\hline \hline
&\multicolumn{3}{c}{$\mu^+\mu^-e^{\pm}$}&\multicolumn{3}{c}{$\mu^+\mu^-\mu^{\pm}$}\\ &Accepted&fail $\met$ & fail $\mass$ &Accepted&fail $\met$ & fail $\mass$ \\\hline \hline
\rule{0pt}{2.6ex}\rule[-1.2ex]{0pt}{0pt}\zll & 1.5 $\pm$ 0.4 & 12 $\pm$ 2 & 0.5 $\pm$ 0.2 & 4 $\pm$ 2 & 3 $\pm$ 1 & 0.1 $\pm$ 0.5\\
\zgam & 1.6 $\pm$ 0.4 & 13.0 $\pm$ 0.5 & 0.3 $\pm$ 0.1 & 0.1 $\pm$ 0.0 & 0.1 $\pm$ 0.0 & 0.0 $\pm$ 0.0\\
\zzllll & 0.9 $\pm$ 0.2 & 1.5 $\pm$ 0.2 & 0.1 $\pm$ 0.0 & 1.6 $\pm$ 0.0 & 0.7 $\pm$ 0.0 & 0.1 $\pm$ 0.0\\
\ttbar & 0.3 $\pm$ 0.0 & 0.0 $\pm$ 0.0 & 0.1 $\pm$ 0.0 & 0.1 $\pm$ 0.0 & 0.0 $\pm$ 0.0 & 0.0 $\pm$ 0.0\\
\zzllnunu & 0.0 $\pm$ 0.0 & 0.0 $\pm$ 0.0 & 0.0 $\pm$ 0.0 & 0.0 $\pm$ 0.0 & 0.0 $\pm$ 0.0 & 0.0 $\pm$ 0.0\\
\wwlnulnu & 0.0 $\pm$ 0.0 & 0.0 $\pm$ 0.0 & 0.0 $\pm$ 0.0 & 0.0 $\pm$ 0.0 & 0.0 $\pm$ 0.0 & 0.0 $\pm$ 0.0\\
Predicted background & 4.3 $\pm$ 0.8 & 26 $\pm$ 2 & 1.0 $\pm$ 0.3 & 6 $\pm$ 2 & 4 $\pm$ 1 & 0.2 $\pm$ 0.5\\[4pt]
\wzlnull & 14.0 $\pm$ 0.2 & 2.1 $\pm$ 0.1 & 0.9 $\pm$ 0.1 & 15.1 $\pm$ 0.4 & 2.0 $\pm$ 0.1 & 0.3 $\pm$ 0.1\\[4pt]
Predicted total & 18.3 $\pm$ 0.8 & 29 $\pm$ 2 & 1.9 $\pm$ 0.4 & 21 $\pm$ 2 & 6 $\pm$ 1 & 0.5 $\pm$ 0.6\\[4pt]
Observed & 26 & 23 & 3 & 25 & 12 & 5\\
\hline
\hline
\end{tabular}

%% file: metprime.tex
\begin{figure}
\centering
\includegraphics[width=\linewidth]{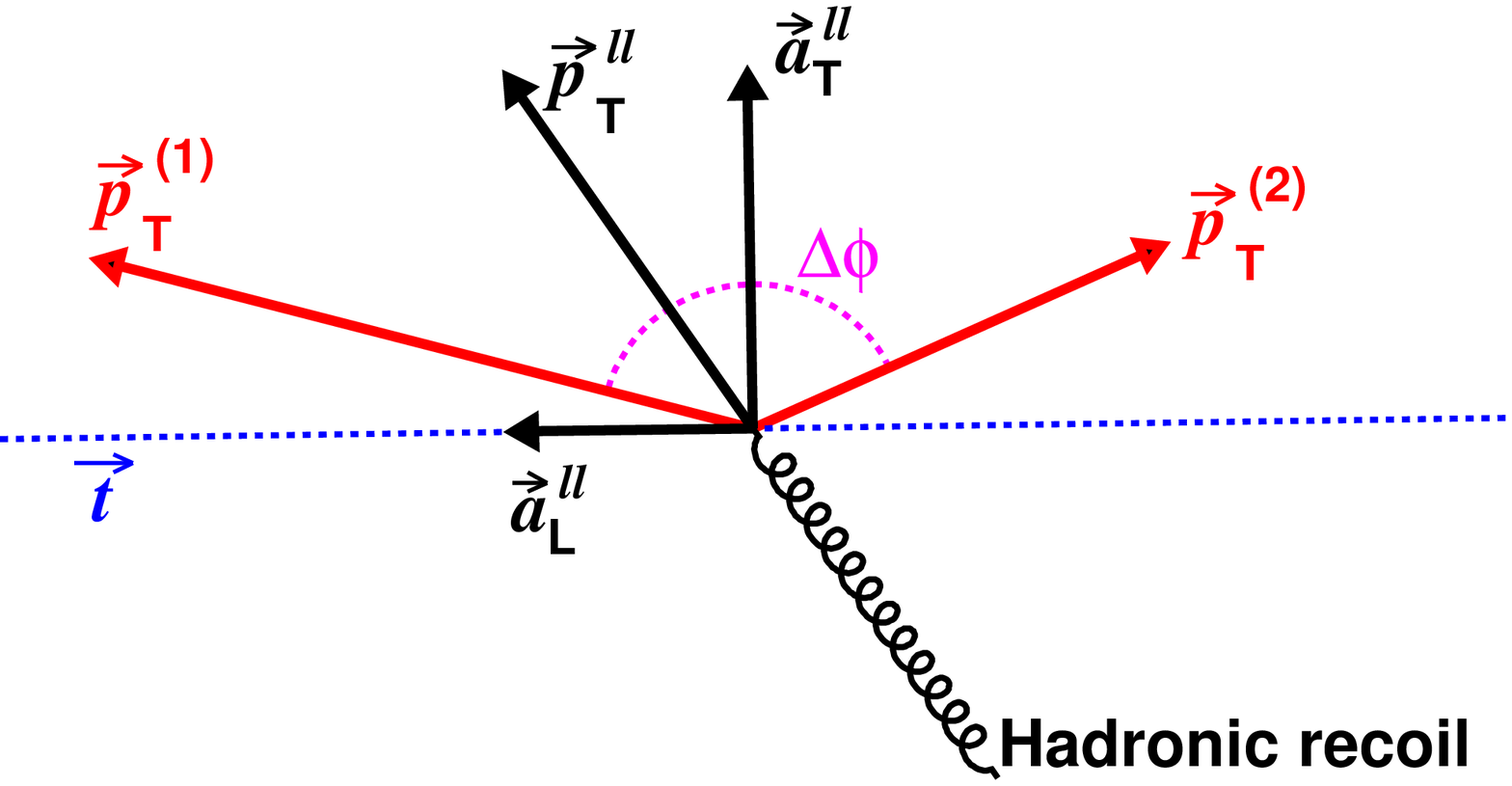}
\caption{Illustration of the decomposition of the dilepton $p_T$ into $a_T$ and $a_L$ components.}
\label{Figure:at}
\end{figure}

\begin{figure*}[htbp]\centering
\includegraphics[width=0.49\linewidth]{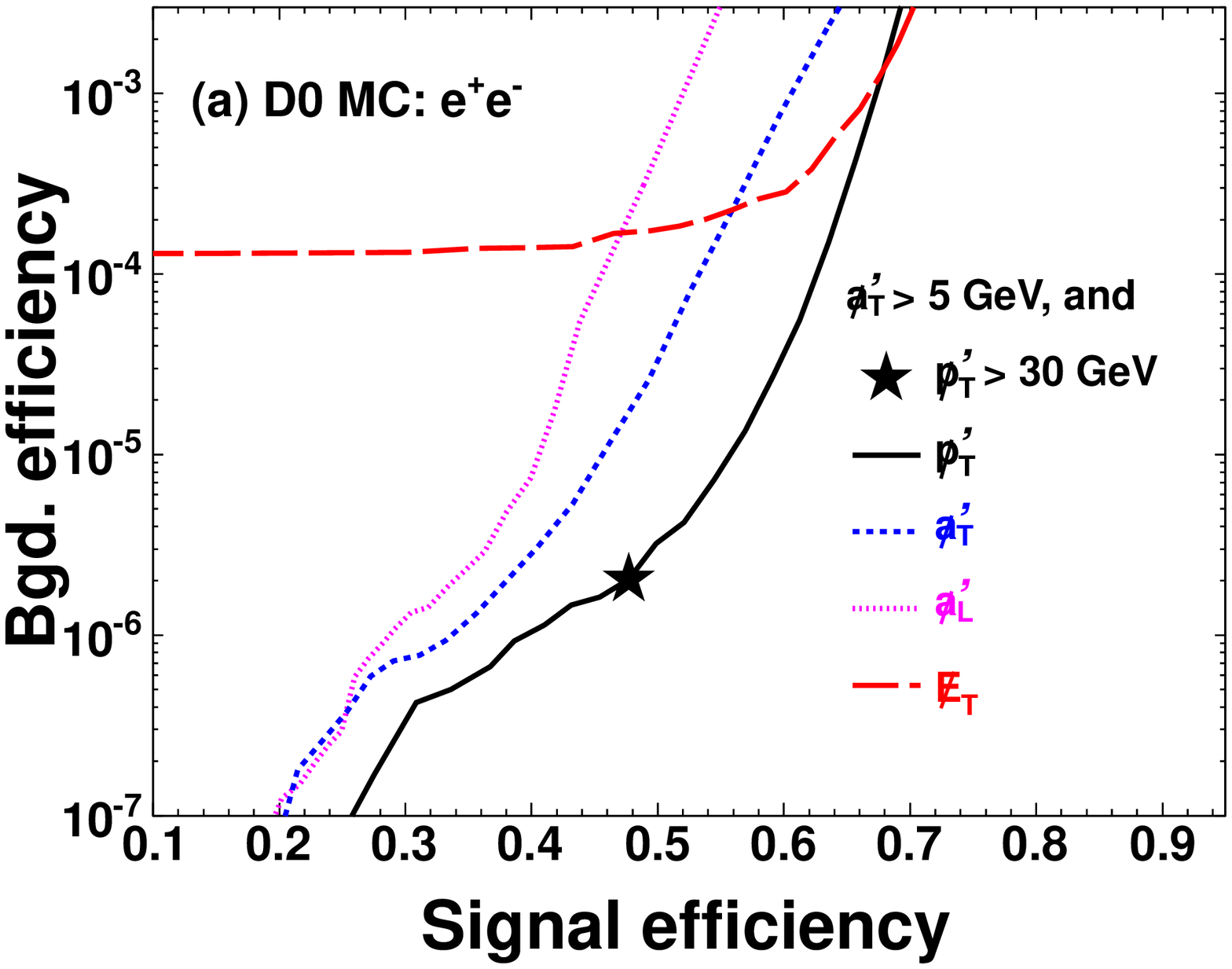}
\includegraphics[width=0.49\linewidth]{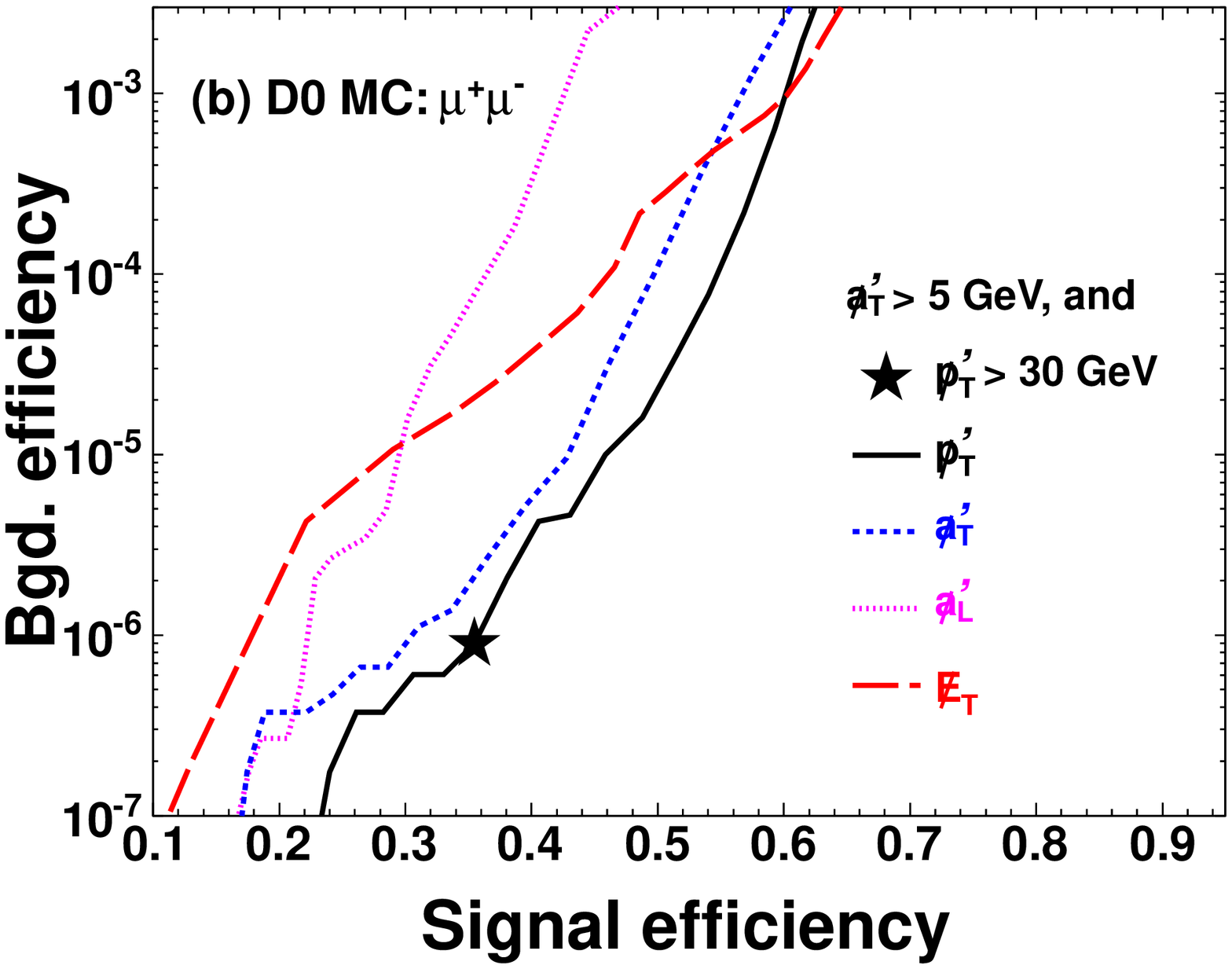}
\caption{Background vs.~signal efficiency for (a) \diem\ and (b) \dimu\ channels 
after varying the requirements on variables
that are sensitive to the missing transverse momentum.
The requirement \aTmissPrime\ $>$ 5~GeV is always applied.}
\label{Figure:metprime_eff_study}
\end{figure*}

The basic signature for the \zzllnunu\ analysis is a pair of charged
leptons with invariant mass close to $m_Z$, produced in association
with significant imbalance in transverse momentum, \pTmiss, due to the neutrinos
from the $Z\rightarrow\nu\bar{\nu}$ decay.
A substantial background corresponds to inclusive \zll\ production
in which either the leptons and/or any hadronic recoil is mis-reconstructed.
Stringent selection requirements are necessary to suppress this background,
since (i) the production cross section for $Z$ bosons
exceeds that of the signal by four orders of magnitude
and (ii) the rates for mis-reconstruction are difficult to simulate.
Rather than estimate the genuine \pTmiss\ in the event,
we use the approach of the  previous D0 analysis of this process~\cite{dzero_zzllnunu_prd},
and construct variables 
that represent the minimum \pTmiss\ consistent with the
measurement uncertainties on the leptons and the hadronic
recoil.

First, the dilepton $\vec{p}_T$ is decomposed into two components
with respect to a reference axis, $\vec{t}$, as illustrated in Fig.~\ref{Figure:at}.
We define $\vec{t} = \vec{p}_T^{\ (1)} - \vec{p}_T^{\ (2)}$,
where $\vec{p}_T^{\ (1)}$ and $\vec{p}_T^{\ (2)}$ are the $p_T$ vectors of the
two leptons.
The dilepton $p_T$ vector is defined as 
$\vec{p}_{T} = \vec{p}_T^{\ (1)} + \vec{p}_T^{\ (2)}$,
from which we define

\begin{align}
 \pTdilep\ & = |\vec{p_{T}}|, \\ 
 \aTdilep\ & = |\vec{p_{T}} \times \hat{t}|, \\ 
  \aLdilep\ & = |\vec{p_{T}} \cdot \hat{t}|, 
\label{Eq:pTdilep}
\end{align}

\noindent where $\hat{t}$ is a unit vector in the direction of $\vec{t}$.
In the region $\Delta\phi$ $>$ $\pi/2$, where $\Delta\phi$
is the azimuthal opening angle between the leptons,
the \aTdilep\ component is less sensitive to mis-measurement
in the magnitude of the individual lepton transverse momenta than is \aLdilep~\cite{dzero_zzllnunu_prd,aT_NIM}. 
For $\Delta\phi$ $<$ $\pi/2$, this decomposition is no longer
valid, and \aTdilep\ and \aLdilep\ are set equal to \pTdilep.

The missing transverse momentum estimators, \pTmissPrime, \aTmissPrime, and \aLmissPrime, are constructed as

\begin{align}
 \pTmissPrime &= \pTdilep + 2\left[\pTdelta + \pTrecoil + \pTtrkjets\right], \\
 \aTmissPrime &= \aTdilep + 2\left[\aTdelta + \aTrecoil + \aTtrkjets\right], \\
 \aLmissPrime &= \aLdilep + 2\left[\aLdelta + \aLrecoil + \aLtrkjets\right]. 
\label{Eq:pTdilep}
\end{align}

\noindent The terms, \pTdelta, \pTrecoil\ and \pTtrkjets\ (and similarly for $a_T$ and $a_L$) are corrections for 
lepton $p_T$ mis-measurement, hadronic recoil measured in the calorimeter,
and remaining hadronic recoil measured in the tracking system, respectively.
These terms are described in more detail in the following sections.
The factor of two is found to be optimal based on MC simulations.

\subsection{Dilepton mis-measurement}
A correction for possible lepton $p_T$ mis-measurement is
determined by varying each individual lepton $p_T$ within
one standard deviation of its estimated uncertainty
in order to separately minimise
\aTdilep, \aLdilep\ and \pTdilep.
Electrons that are reconstructed close to module boundaries in the CC or in the IC have relatively
poor energy resolution and are given special treatment.
The estimated uncertainty may be inflated to cover the difference between the 
calorimeter based $p_T$ measurement and the alternative measurement from the central track.
This is only allowed for the upward variation and protects against electrons for which the
calorimeter has severely under-measured the energy.
The amount by which, e.g., \aTdilep\ is reduced, is denoted \aTdelta.
These quantities are defined in such a way that they always carry
a negative sign.

\subsection{Calorimeter recoil}

Two estimates of the calorimeter recoil are made,
from the reconstructed jets and from the reconstructed \met.
Jets are reconstructed using the D0 mid-point cone algorithm~\cite{d0jets}
with a cone size of $\DR = 0.5$.
They must be separated from the leptons by at least $\DR > 0.3$ and satisfy
$p_T$ $>$ 15~GeV.
The \pt, \at, and \al\ components are calculated for each jet in the event,
e.g., 
\begin{equation}
a_T^{\rm jet(i)} = \vec{p}_{T}^{\ \rm jet(i)} \times \hat{t},
\end{equation}
where $\vec{p}_{T}^{\ \rm jet(i)}$ is the $p_T$ vector of the $i$th jet.
An individual jet that has a positive value (i.e., {\em increases} the momentum imbalance)
is ignored.
This approach ensures that jets which are not genuinely associated with the recoil system 
(e.g.,~from additional $p\bar{p}$ collisions or the underlying event)
are not allowed to generate a fake imbalance in an otherwise well reconstructed event.
The sum of contributions from the jets is denoted, e.g., for the $a_T$ component, \aTjets.

The \met\ estimate subtracts any contribution from the two leptons
and then tests how well the remaining \met\ balances the dilepton system.
Between the jet and \met\ based corrections we choose (separately for the $a_T$, $a_L$ and $p_T$ components)
the one that best balances the dilepton system.
This correction term is denoted, e.g., \aTrecoil.

\subsection{Track recoil}

As a protection against events in which at least one hadronic jet fails to be reconstructed in
the calorimeter, we attempt to recover any remaining activity in the tracker.
Track jets are reconstructed by merging together reconstructed tracks
within cones of size $\DR = 0.5$.
These tracks must satisfy $p_T$ $>$ 1~GeV.
Track jets must have at least two tracks within the cone,
and be separated by at least $\DR = 0.3$ from the leptons,
and by at least $\DR = 0.5$ from any calorimeter jets.
Corrections to each of the ($p_T$, $a_T$, and $a_L$) components 
are determined in the same way as for calorimeter jets. 

\subsection{Performance}

Figure~\ref{Figure:metprime_eff_study} shows the \zll\ background efficiency versus
the \zzllnunu\ signal efficiency for a range of requirements on each of the variables;
\aTmissPrime, \aLmissPrime, \pTmissPrime, and \met.
The decays of \ztt\ into \diem, \dimu, and \emmu\ final states produce a genuine missing $p_T$
along the $a_L$ direction.
Our \zzllnunu\ candidate selection requirements therefore include a ``soft'' requirement of \aTmissPrime\ $>$ 5~GeV.
The curves in Fig.~\ref{Figure:metprime_eff_study} correspond to the combination of this requirement
and a varying requirement on the variable under study.
The \pTmissPrime\ variable has the best performance over the range of background efficiencies
of interest.
The efficiency for the requirement \pTmissPrime\ $>$ 30~GeV (and \aTmissPrime\ $>$ 5~GeV)
is indicated explicitly by a star symbol. 
This is the requirement that is made in selecting \zzllnunu\ candidates.
The optimisation of the \pTmissPrime\ requirement is discussed later.


%% file: plots/Table3.tex
\begin{tabular}{lccccccc}
\hline
\hline
        &      &\multicolumn{6}{c}{ Rejected by requirement on }\\
Process &Accepted&\pTmissPrime&\aTmissPrime&$M_{ll}$&Extra lep.&Charge&Jets\\
\hline
\hline
\zee\rule{0pt}{2.6ex}\rule[-1.2ex]{0pt}{0pt} & 0.6 $\pm$ 0.3 & 11666 $\pm$ 1665 & 0 $\pm$ 1 & 0.3 $\pm$ 0.2 & 3 $\pm$ 2 & 0.0 $\pm$ 0.0 & 0.1 $\pm$ 0.1\\
\ztt & 0.1 $\pm$ 0.1 & 8 $\pm$ 2 & 1.4 $\pm$ 0.2 & 0.0 $\pm$ 0.0 & 0.0 $\pm$ 0.0 & 0.0 $\pm$ 0.0 & 0.0 $\pm$ 0.0\\
\wwlnulnu & 35 $\pm$ 1 & 35 $\pm$ 1 & 1.7 $\pm$ 0.1 & 33 $\pm$ 1 & 9 $\pm$ 5 & 0.3 $\pm$ 0.1 & 0.1 $\pm$ 0.1\\
\wzlnull & 2.3 $\pm$ 0.1 & 1.9 $\pm$ 0.1 & 0.2 $\pm$ 0.0 & 0.2 $\pm$ 0.1 & 14 $\pm$ 2 & 0.2 $\pm$ 0.1 & 0.0 $\pm$ 0.0\\
\wenu & 6 $\pm$ 2 & 13 $\pm$ 2 & 0.3 $\pm$ 0.1 & 5 $\pm$ 1 & 2 $\pm$ 1 & 4 $\pm$ 1 & 0.0 $\pm$ 0.0\\
$W\gamma \rightarrow e\nu\gamma$ & 3.3 $\pm$ 0.3 & 5.5 $\pm$ 0.5 & 0.0 $\pm$ 0.1 & 2.8 $\pm$ 0.5 & 0.6 $\pm$ 0.5 & 3.3 $\pm$ 0.4 & 0.0 $\pm$ 0.0\\
\zzllll & 0.0 $\pm$ 0.0 & 0.1 $\pm$ 0.0 & 0.0 $\pm$ 0.0 & 0.0 $\pm$ 0.0 & 1.3 $\pm$ 0.2 & 0.0 $\pm$ 0.0 & 0.0 $\pm$ 0.0\\
\ttbar & 1.0 $\pm$ 0.2 & 1.4 $\pm$ 0.2 & 0.4 $\pm$ 0.1 & 1.2 $\pm$ 0.1 & 7 $\pm$ 1 & 0.0 $\pm$ 0.0 & 0.2 $\pm$ 0.1\\
Predicted background & 48 $\pm$ 2 & 11749 $\pm$ 1668 & 4 $\pm$ 1 & 43 $\pm$ 2 & 37 $\pm$ 11 & 8 $\pm$ 1 & 0.4 $\pm$ 0.2\\[4pt]
\zznunull & 13.6 $\pm$ 0.4 & 7.4 $\pm$ 0.2 & 1.3 $\pm$ 0.1 & 0.6 $\pm$ 0.0 & 4 $\pm$ 2 & 0.2 $\pm$ 0.0 & 0.1 $\pm$ 0.0\\
Predicted total & 62 $\pm$ 3 & 11756 $\pm$ 1668 & 6 $\pm$ 1 & 43 $\pm$ 2 & 41 $\pm$ 13 & 8 $\pm$ 1 & 0.4 $\pm$ 0.2\\[4pt]
Observed &61 & 10560 & 12 & 50 & 63 & 12 & 1\\
\hline
\hline
\end{tabular}

%% file: plots/Table4.tex
\begin{tabular}{lccccccc}
\hline
\hline
        &      &\multicolumn{6}{c}{ Rejected by requirement on }\\
Process &Accepted&\pTmissPrime&\aTmissPrime&$M_{ll}$&Extra lep.&Charge&Jets\\
\hline
\hline
\zmm & 0.3 $\pm$ 0.5 & 8519 $\pm$ 1372 & 3 $\pm$ 6 & 2 $\pm$ 2 & 3 $\pm$ 2 & 0.4 $\pm$ 0.1 & 0.1 $\pm$ 0.1\\
\ztt & 0.0 $\pm$ 0.0 & 5 $\pm$ 3 & 1.4 $\pm$ 0.3 & 0.0 $\pm$ 0.0 & 0.1 $\pm$ 0.0 & 0.0 $\pm$ 0.0 & 0.0 $\pm$ 0.0\\
\wwlnulnu & 31 $\pm$ 2 & 48 $\pm$ 2 & 1.4 $\pm$ 0.2 & 29 $\pm$ 2 & 9 $\pm$ 5 & 0.0 $\pm$ 0.0 & 0.1 $\pm$ 0.1\\
\wzlnull & 2.0 $\pm$ 0.1 & 2.9 $\pm$ 0.2 & 0.2 $\pm$ 0.0 & 0.3 $\pm$ 0.0 & 12 $\pm$ 2 & 0.2 $\pm$ 0.1 & 0.0 $\pm$ 0.0\\
\wmnu & 2.3 $\pm$ 0.4 & 9 $\pm$ 2 & 0.0 $\pm$ 0.1 & 2.9 $\pm$ 0.7 & 1.1 $\pm$ 0.9 & 0.8 $\pm$ 0.2 & 0.0 $\pm$ 0.0\\
\zzllll & 0.0 $\pm$ 0.0 & 0.2 $\pm$ 0.0 & 0.0 $\pm$ 0.0 & 0.0 $\pm$ 0.0 & 0.9 $\pm$ 0.2 & 0.0 $\pm$ 0.0 & 0.0 $\pm$ 0.0\\
\ttbar & 0.8 $\pm$ 0.1 & 2.0 $\pm$ 0.2 & 0.4 $\pm$ 0.1 & 0.9 $\pm$ 0.1 & 6 $\pm$ 1 & 0.0 $\pm$ 0.0 & 0.2 $\pm$ 0.1\\
Predicted background & 36 $\pm$ 2 & 8602 $\pm$ 1374 & 6 $\pm$ 6 & 36 $\pm$ 2 & 32 $\pm$ 9 & 1.4 $\pm$ 0.2 & 0.4 $\pm$ 0.3\\[4pt]
\zznunull & 11.8 $\pm$ 0.3 & 11.0 $\pm$ 0.3 & 0.9 $\pm$ 0.1 & 0.8 $\pm$ 0.1 & 4 $\pm$ 2 & 0.0 $\pm$ 0.0 & 0.0 $\pm$ 0.0\\
Predicted total & 48 $\pm$ 2 & 8613 $\pm$ 1374 & 7 $\pm$ 6 & 36 $\pm$ 2 & 35 $\pm$ 11 & 1.4 $\pm$ 0.2 & 0.4 $\pm$ 0.3\\[4pt]
Observed &58 & 7416 & 7 & 42 & 60 & 4 & 1\\
\hline
\hline
\end{tabular}

%% file: plots/Table5.tex
\begin{tabular}{lccccccc}
\hline
\hline
        &      &\multicolumn{6}{c}{ Rejected by requirement on }\\
Process &Accepted&\pTmissPrime&\aTmissPrime&$M_{ll}$&Extra lep.&Charge&Jets\\
\hline
\hline
\zee\rule{0pt}{2.6ex}\rule[-1.2ex]{0pt}{0pt} & 0.0 $\pm$ 0.0 & 17 $\pm$ 7 & 0.0 $\pm$ 0.0 & 0.0 $\pm$ 0.0 & 0.0 $\pm$ 0.0 & 0.0 $\pm$ 0.0 & 0.0 $\pm$ 0.0\\
\zmm & 0.0 $\pm$ 0.0 & 6 $\pm$ 2 & 0.0 $\pm$ 0.0 & 0.3 $\pm$ 0.3 & 0.3 $\pm$ 0.1 & 0.0 $\pm$ 0.0 & 0.0 $\pm$ 0.0\\
\ztt & 0.1 $\pm$ 0.1 & 19 $\pm$ 14 & 4.5 $\pm$ 0.4 & 0.1 $\pm$ 0.1 & 0.2 $\pm$ 0.2 & 0.0 $\pm$ 0.0 & 0.0 $\pm$ 0.0\\
\wwlnulnu & 69 $\pm$ 3 & 84 $\pm$ 4 & 3.7 $\pm$ 0.2 & 67 $\pm$ 3 & 19 $\pm$ 11 & 0.4 $\pm$ 0.1 & 0.3 $\pm$ 0.2\\
\wzlnull & 0.4 $\pm$ 0.1 & 0.7 $\pm$ 0.1 & 0.0 $\pm$ 0.0 & 0.4 $\pm$ 0.1 & 3.3 $\pm$ 0.6 & 0.4 $\pm$ 0.1 & 0.0 $\pm$ 0.0\\
\wenu & 4 $\pm$ 2 & 9 $\pm$ 3 & 0.1 $\pm$ 0.2 & 5 $\pm$ 1 & 1 $\pm$ 1 & 1.2 $\pm$ 0.6 & 0.0 $\pm$ 0.0\\
\wmnu & 5 $\pm$ 4 & 12 $\pm$ 9 & 0.2 $\pm$ 0.1 & 5 $\pm$ 4 & 1 $\pm$ 2 & 3 $\pm$ 2 & 0.0 $\pm$ 0.0\\
$W\gamma \rightarrow e\nu\gamma$ & 3.4 $\pm$ 0.5 & 6.4 $\pm$ 0.4 & 0.1 $\pm$ 0.1 & 3.4 $\pm$ 0.3 & 0.7 $\pm$ 0.6 & 3.2 $\pm$ 0.3 & 0.0 $\pm$ 0.0\\
\zzllll & 0.0 $\pm$ 0.0 & 0.0 $\pm$ 0.0 & 0.0 $\pm$ 0.0 & 0.0 $\pm$ 0.0 & 0.2 $\pm$ 0.0 & 0.0 $\pm$ 0.0 & 0.0 $\pm$ 0.0\\
\ttbar & 2.3 $\pm$ 0.2 & 3.3 $\pm$ 0.2 & 1.0 $\pm$ 0.1 & 2.1 $\pm$ 0.2 & 13 $\pm$ 3 & 0.0 $\pm$ 0.0 & 0.3 $\pm$ 0.1\\
Predicted background & 84 $\pm$ 6 & 157 $\pm$ 19 & 9.6 $\pm$ 0.5 & 83 $\pm$ 6 & 39 $\pm$ 16 & 8 $\pm$ 3 & 0.5 $\pm$ 0.3\\[4pt]
\zznunull & 0.0 $\pm$ 0.0 & 0.0 $\pm$ 0.0 & 0.0 $\pm$ 0.0 & 0.1 $\pm$ 0.0 & 0.0 $\pm$ 0.0 & 0.0 $\pm$ 0.0 & 0.0 $\pm$ 0.0\\
Predicted total & 84 $\pm$ 6 & 157 $\pm$ 19 & 9.6 $\pm$ 0.5 & 83 $\pm$ 6 & 39 $\pm$ 16 & 8 $\pm$ 3 & 0.5 $\pm$ 0.3\\[4pt]
Observed &73 & 162 & 7 & 96 & 60 & 8 & 0\\
\hline
\hline
\end{tabular}

%% file: plots/Table6.tex
\begin{tabular}{ l     c    c c c   c  }
\hline
\hline
  	&$N_{\rm bgd}$	&$A_{\ell\ell}$	&$A_{\rm sig}$	&$A_{\ell\ell}/A_{\rm sig}$	&$\sigma_{\rm sig}$\\
\hline
\hline
\rule{0pt}{2.6ex}\rule[-1.2ex]{0pt}{0pt}$L_{\rm inst}$ profile	&4.0	&2.4	&3.3	&0.9	&0.2\\
Vertex $z$ profile	&1.6	&1.3	&0.9	&0.4	&0.7\\
$Z/\gamma^* p_{T}$	&0.0	&0.0	&0.0	&0.0	&0.2\\
Diboson $p_{T}$	&0.1	&0.0	&0.4	&0.4	&0.2\\
Jet energy scale	&6.0	&0.1	&0.3	&0.2	&1.3\\
Jet energy resol.	&2.2	&0.0	&0.0	&0.0	&0.2\\
IC jet treatment	&1.1	&0.0	&0.0	&0.0	&0.2\\
Electron $p_{T}$ scale	&0.3	&0.0	&0.1	&0.1	&0.2\\
Electron $p_{T}$ resol.	&1.0	&0.1	&0.0	&0.0	&0.2\\
Electron $p_{T}$ tails	&0.1	&0.0	&0.3	&0.4	&0.2\\
Muon $p_{T}$ scale	&0.1	&0.0	&0.1	&0.1	&0.2\\
Muon $p_{T}$ resol.	&0.9	&0.1	&0.1	&0.0	&0.2\\
Muon $p_{T}$ tails	&1.0	&0.2	&0.4	&0.2	&0.2\\
Track reconstr.	&0.1	&0.7	&1.1	&0.3	&0.7\\
Muon reconstr.	&0.2	&0.3	&0.5	&0.2	&0.2\\
Electron reconstr.	&0.2	&0.2	&0.2	&0.0	&0.2\\
$Z/\gamma^*$+jets model.	&17.7	&0.0	&0.0	&0.0	&2.5\\
Systematic\rule{0pt}{2.6ex}\rule[-1.2ex]{0pt}{0pt}	&19.4	&2.9	&3.7	&1.2	&3.1\\
Statistical\rule{0pt}{2.6ex}\rule[-1.2ex]{0pt}{0pt}	&--	&--	&--	&--	&13.2\\
Stat. $\oplus$ syst.\rule{0pt}{2.6ex}\rule[-1.2ex]{0pt}{0pt}	&19.4	&2.9	&3.7	&1.2	&13.6\\
\hline
\hline
\end{tabular}

%% file: plots/Table7.tex
\begin{tabular}{ l     c    c c c   c  }
\hline
\hline
  	&$N_{\rm bgd}$	&$A_{\ell\ell}$	&$A_{\rm sig}$	&$A_{\ell\ell}/A_{\rm sig}$	&$\sigma_{\rm sig}$\\
\hline
\hline
\rule{0pt}{2.6ex}\rule[-1.2ex]{0pt}{0pt}$L_{\rm inst}$ profile	&1.5	&4.5	&5.2	&0.7	&1.8\\
Vertex $z$ profile	&1.0	&1.3	&0.7	&0.6	&2.5\\
$Z/\gamma^* p_{T}$	&0.0	&0.0	&0.0	&0.0	&0.6\\
Diboson $p_{T}$	&2.6	&0.0	&1.8	&1.8	&3.7\\
Jet energy scale	&1.1	&0.8	&1.5	&0.8	&1.8\\
Jet energy resol.	&0.9	&0.1	&0.1	&0.0	&1.8\\
IC jet treatment	&0.2	&0.2	&0.4	&0.2	&0.6\\
Jet reconstr.	&0.5	&0.3	&0.0	&0.2	&0.0\\
Trkjet reconst.	&1.5	&0.0	&1.1	&1.2	&3.1\\
Electron $p_{T}$ scale	&0.4	&0.0	&0.0	&0.0	&0.6\\
Electron $p_{T}$ resol.	&1.0	&0.1	&0.5	&0.4	&1.8\\
Electron $p_{T}$ tails	&1.0	&0.0	&0.6	&0.6	&1.2\\
Muon $p_{T}$ scale	&0.1	&0.0	&0.0	&0.0	&0.0\\
Muon $p_{T}$ resol.	&0.5	&0.1	&0.5	&0.5	&0.6\\
Muon $p_{T}$ tails	&0.1	&0.1	&0.5	&0.4	&0.6\\
Lepton eff. vs $p_T$	&0.0	&0.0	&0.0	&0.0	&0.6\\
Lepton eff. vs $\eta$	&0.0	&0.0	&0.0	&0.0	&0.6\\
$W$+jets model.	&1.9	&0.0	&0.0	&0.0	&0.6\\
$W\gamma$ model.	&3.9	&0.0	&0.0	&0.0	&1.8\\
Systematic\rule{0pt}{2.6ex}\rule[-1.2ex]{0pt}{0pt}	&6.0	&4.8	&6.0	&2.6	&7.1\\
Statistical\rule{0pt}{2.6ex}\rule[-1.2ex]{0pt}{0pt}	&--	&--	&--	&--	&27.0\\
Stat. $\oplus$ syst.\rule{0pt}{2.6ex}\rule[-1.2ex]{0pt}{0pt}	&6.0	&4.8	&6.0	&2.6	&27.9\\
\hline
\hline
\end{tabular}

%% file: plots/Table8.tex
\begin{tabular}{ c c c }
\hline \hline
Sub-channel & $A_{\ell\ell}/A_{\rm sig}$ &  $N_{\ell\ell}^{\rm obs}$\\
\hline
\wzenuee & 2.242 $\pm$ 0.025 & 459336\\
\wzmnuee & 1.495 $\pm$ 0.023 & 419069\\
\wzenumm & 1.704 $\pm$ 0.027 & 493202\\
\wzmnumm & 1.443 $\pm$ 0.023 & 443869\\[4pt]
\zzeenunu & 1.638 $\pm$ 0.049 & 319797\\
\zzmmnunu & 2.052 $\pm$ 0.059 & 342603\\
\hline \hline
\end{tabular}

%% file: plots/Table9.tex
\begin{tabular}{cc}
\hline
\hline
Sub-channel & $\mathcal{R} (\times 10^{-3})$ \\
\hline
\wzenuee & 0.70~$\pm$~0.20\\
\wzmnuee & 0.40~$\pm$~0.14\\
\wzenumm & 0.66~$\pm$~0.17\\
\wzmnumm & 0.61~$\pm$~0.16\\[4pt]
\zzeenunu & 0.13~$\pm$~0.07\\
\zzmmnunu & 0.33~$\pm$~0.10\\
\hline
\hline
\end{tabular}

%% file: acknowledgement.tex
%
We thank the staffs at Fermilab and collaborating institutions,
and acknowledge support from the
DOE and NSF (USA);
CEA and CNRS/IN2P3 (France);
FASI, Rosatom and RFBR (Russia);
CNPq, FAPERJ, FAPESP and FUNDUNESP (Brazil);
DAE and DST (India);
Colciencias (Colombia);
CONACyT (Mexico);
KRF and KOSEF (Korea);
CONICET and UBACyT (Argentina);
FOM (The Netherlands);
STFC and the Royal Society (United Kingdom);
MSMT and GACR (Czech Republic);
CRC Program and NSERC (Canada);
BMBF and DFG (Germany);
SFI (Ireland);
The Swedish Research Council (Sweden);
and
CAS and CNSF (China).

%% file: prd.bbl
\begin{thebibliography}{10}
\bibitem{Note_on_Dzero_ZZ}
This cross section is corrected for the $\gamma^*$ contribution.
\bibitem{coordinates}
Rapidity is defined by $y=(1/2)\ln[(E-p_z)/(E+p_z)]$, where $E$ is the energy
  and $p_z$ is the momentum component parallel to the proton beam direction.
  Pseudorapidity is defined by $\eta = -\ln[\tan(\theta/2)]$, where $\theta$ is
  the polar angle measured relative to the center of the detector.
\bibitem{DR_ref}
We define $(\DR)^2 = (\Delta\phi)^2 + (\Delta\eta)^2$, where $\Delta\phi$ and
  $\Delta\eta$ are the differences between two objects in azimuth and
  pseudorapidity, respectively.
\end{thebibliography}
